\documentclass[useAMS,usenatbib]{mn2e}

  \usepackage{graphicx}
   \usepackage{lscape}
 \usepackage{rotating}
  \usepackage{array}
  \usepackage{flafter}
  
  \newcommand{\nosection}[1]{%
  \refstepcounter{section}%
  \addcontentsline{toc}{section}{\protect\numberline{\thesection}#1}%
  \markright{#1}}

\title{The Morphologies of Massive Galaxies at ${\bf 1<z<3}$ in the CANDELS-UDS Field: Compact Bulges, and the Rise and Fall of Massive Disks}
\author[V. A. Bruce]
{V.A. Bruce$^{1}$\thanks{E-mail: vab@roe.ac.uk}, J.S. Dunlop$^{1}$, M. Cirasuolo$^{1,2}$, R.J. McLure$^{1}$, T.A. Targett$^{1}$, E.F. Bell$^{3}$, 
\newauthor
D.J. Croton$^{4}$, A. Dekel$^{5}$, S.M. Faber$^{6}$, H.C. Ferguson$^{7}$, N.A. Grogin$^{7}$,  D.D. Kocevski$^{6}$,  
\newauthor
A.M. Koekemoer$^{7}$,  D.C. Koo$^{6}$, K. Lai$^{6}$, J.M. Lotz$^{7}$, E.J. McGrath$^{6}$, J.A. Newman$^{8}$,
\newauthor
 A. van der Wel$^{9}$\\
$^1$SUPA\thanks{Scottish Universities Physics Alliance} Institute for Astronomy, University of Edinburgh, Royal Observatory, Edinburgh EH9 3HJ\\
$^2$UK Astronomy Technology Centre, Science and Technology Facilities Council, Royal Observatory, Edinburgh EH9 3HJ\\
$^3$Department of Astronomy, University of Michigan, 500 Church St., Ann Arbor, MI 48109, USA\\
$^4$Centre for Astrophysics and Supercomputing, Swinburne University of Technology, PO Box 218, Hawthorn, VIC 3122, Australia\\
$^5$Racah Institute of Physics, The Hebrew University, Jerusalem 91904, Israel\\
$^6$UCO/Lick Observatory, University of California, Santa Cruz, CA 95064, USA\\
$^7$Space Telescope Science Institute, 3700 San Martin Drive, Baltimore, MD 21218, USA\\
$^8$University of Pittsburgh, Pittsburgh, PA 15260, USA\\ 
$^9$Max-Planck Institut f\"{u}r Astronomie, K\"{o}nigstuhl 17, D-69117 Heidelberg, Germany\\   }

\begin{document}

\date{}

\pagerange{\pageref{firstpage}--\pageref{lastpage}} \pubyear{2011}
\pagestyle{myheadings}
\markboth{V. A. Bruce et al.} {Morphologies of massive galaxies at $1 < z < 3$}

\maketitle

\label{firstpage}

\vspace*{-1.5cm}

\begin{abstract}
We have used high-resolution, {\it Hubble Space Telescope}, near-infrared imaging 
to conduct a detailed analysis of the morphological properties of the most massive galaxies at high redshift, 
modelling the Wide Field Camera 3 (WFC3/IR) $H_{160}$-band images of the $\simeq 200$ 
galaxies in the CANDELS-UDS field with photometric redshifts $1 < z < 3$, and stellar masses 
$M_* > 10^{11}\,{\rm M_{\odot}}$. We have explored the results of fitting single S\'{e}rsic and bulge+disk models, 
and have investigated the additional errors and  
potential biases introduced by uncertainties in the background and the on-image point-spread-function. 
This approach has enabled us to obtain formally-acceptable model fits to the WFC3/IR images of $> 90$\% of the galaxies. 
Our results indicate that these massive galaxies at $1 < z < 3$ lie both on and below the local size-mass relation, 
with a median effective 
radius of  $\sim 2.6$\,kpc, a factor of $\simeq 2.3$ smaller than comparably-massive local galaxies. 
Moreover, we find that bulge-dominated objects in particular show evidence for a growing bimodality in the size-mass relation 
with increasing redshift, and by $z > 2$ the compact bulges display effective radii a factor $\simeq 4$ smaller than 
local ellipticals of comparable mass. These trends also appear to extend to the bulge components of disk-dominated galaxies, and vice versa.
 In addition, we find that, while such massive galaxies at low redshift are generally bulge-dominated, 
at redshifts $1<z<2$ they are predominantly mixed bulge+disk systems, and by $z > 2$ they are mostly disk-dominated. 
The majority of the disk-dominated galaxies are actively forming stars, although this is also true for many of the 
bulge-dominated systems. Interestingly, however, while most of the quiescent galaxies are bulge-dominated, we find that a significant fraction 
($25-40$\%) of the most quiescent galaxies, with specific Star Formation Rates $sSFR<10^{-10}\,{\rm yr}^{-1}$, have disk-dominated morphologies.
Thus, while our results show that the massive galaxy population is undergoing dramatic changes at this crucial epoch, 
they also suggest that the physical mechanisms which quench star-formation activity 
are not simply connected to those responsible for the morphological transformation of massive galaxies into present-day giant ellipticals.

\end{abstract}


\newpage
\section{Introduction}

The study of the high-redshift progenitors of today's massive galaxies can provide us with invaluable insights 
into the key mechanisms that shape the evolution of galaxies in the high-mass regime, placing 
important constraints on current models of galaxy formation and evolution. In recent years the new generation 
of optical-infrared surveys have revealed that a substantial population of massive galaxies is already in 
place by $z \simeq 2$, and that the star-formation activity in a significant fraction of these objects 
largely ceases around this time, $\simeq 3$\,Gyr after the Big Bang (e.g. \citealt{Fontana2004, Glazebrook2004, Drory2005}).
These results have driven the modification of models of galaxy formation to include additional mechanisms 
for the quenching of star-formation activity in massive galaxies at early times, such as AGN feedback 
(e.g. \citealt{Granato2004, Bower2006, Croton2006, deLucia2007}).

However, explaining the number densities and ages of massive galaxies at high redshift is only part of the challenge, 
as recent advances in imaging capabilities are now providing meaningful data on their sizes and morphologies 
during the crucial cosmological epoch $1 < z < 3$, when global star-formation activity in the Universe peaked.
In particular, over the last $\simeq 5$ years, deep/high-resolution ground-based and space-based (i.e. HST) surveys have 
revealed that a significant fraction of massive galaxies at $z > 1$ are surprisingly compact 
(e.g. \citealt{Daddi2005, Trujillo2006, Trujillo2007, Cimatti2008, Franx2008, Damjanov2009, Targett2011}), 
with derived effective radii ($R_{e}<2-3$\,kpc) and stellar mass measurements which place these galaxies well below 
the local galaxy size-mass relation, as derived from the Sloan Digital Sky Survey (SDSS) \citep{Shen2003}. 
Furthermore, it appears that the largest divergence from local values arises in galaxies which exhibit very 
little sign of ongoing star-formation (e.g. \citealt{Toft2007, Kriek2009}; McLure et al. 2012). 

As befits their potential importance, these studies have been carefully scrutinized to investigate possible 
sources of bias in the measurement of galaxy size and mass  \citep{Muzzin2009, Mancini2010}. A particular concern has 
been the perceived potential for galaxy scalelengths to be under-estimated due to low signal:noise imaging (which might 
be inadequate to reveal faint extended envelopes), morphological k-corrections, or selection effects 
related to surface-brightness bias (e.g. \citealt{vanderWel2009}). 
However, the latest generation of deeper rest-frame optical morphological studies have thus far 
provided mounting evidence for the truly compact nature of many high-redshift 
galaxies (e.g. \citealt{vanDokkum2010, Ryan2012, Szomoru2012}). Moreover, several local studies have 
now clarified the relative dearth of comparably-compact systems surviving to the present day 
\citep{Trujillo2009,Taylor2010}, strengthening the argument that the compact high-redshift 
systems must undergo a period of significant size evolution with limited mass growth in order to reach 
the local galaxy size-mass relation by $z=0$ (e.g. McLure et al. 2012).

Various physical mechanisms have been suggested as the primary drivers of this process, including major or minor 
mergers \citep{Khochfar2006, Naab2007, Hopkins2009} or AGN feedback \citep{Fan2008, Fan2010}. All of these scenarios 
can potentially induce sufficient size growth, but there are problems with some of the accompanying predictions 
of these growth mechanisms. For the major-merger scenario these include reconciling 
the number of major mergers required to facilitate the required size growth with the number of 
major mergers expected since $z\sim1$ from N-body simulations \citep{Hopkins2010}, and the disparity 
between the inferred large mass growth of these (already massive) systems and the latest estimates of the 
local galaxy stellar mass function (\citealt{Baldry2012}; McLure et al. 2012). These problems, coupled with results from numerical simulations, 
which show that AGN-driven expansion occurs when the galaxy is much younger than the typical ages 
of high-redshift compact objects ($>0.5$\,Gyr)  \citep{Ragone2011}, have now led most researchers to conclude
in favour of a picture in which most size growth since $z \simeq 2$ is driven by
minor gas-poor mergers in the outer regions of galaxies, building up stellar halos around compact cores,  
with (relatively) small overall mass growth (\citealt{Bezanson2009, Naab2009, Hopkins2010, vanDokkum2010}; McLure et al. 2012; \citealt{Trujillo2012}). 

In addition to the basic question of how these compact high-redshift galaxies evolve in size, there is also still much debate 
about how these massive galaxies evolve in terms of their fundamental morphological type. Extensive studies 
of the local Universe have revealed a bimodality in the colour-morphology plane, with spheroidal galaxies typically inhabiting 
the red sequence and disk galaxies making up the blue cloud \citep{Baldry2004, Driver2006, Drory2007}. However, 
recent  studies at both low \citep{Bamford2009, Masters2010} and high redshift (\citealt{vanderWel2011}; McLure et al. 2012) have 
uncovered a significant population of passive disk-dominated galaxies, providing evidence that the physical processes 
which quench star-formation may be distinct from those responsible for driving morphological transformations. This result 
is particularly interesting in light of the latest morphological studies of high-redshift massive galaxies 
by \citet{Buitrago2011} and \citet{vanderWel2011} who find that, in contrast to the local population of massive galaxies
(which is dominated by bulge morphologies), by $z \simeq 2$ massive galaxies are predominantly disk-dominated systems.

In this paper we attempt to provide significantly improved clarity on these issues by exploiting the new 
near-infrared HST WFC3/IR imaging provided by the CANDELS survey \citep{Grogin2011, Koekemoer2011} of the central region of the UKIDSS UDS field. 
This provides the necessary combination of depth, angular resolution, and area to enable the most detailed and robust 
study to date of the rest-frame optical morphologies of massive galaxies at $1 < z < 3$. We have also taken 
this opportunity to properly explore a number of challenging technical issues in the field, investigating the extent to 
which our results are robust to the method and accuracy with which both the background and on-image PSF is determined, 
and undertaking both single and multiple-component axi-symmetric modelling (with allowance for an additional point-like 
component contribution where required). Unlike many previous studies in this area, we have placed special emphasis 
on obtaining a formally-acceptable model fit to the observed galaxy images, in order to enable meaningful 
errors to be placed on the key morphological parameters extracted from our analysis. 
   
This paper is structured as follows. First, since the CANDELS HST WFC3/IR near-infrared data have also proved crucial in the selection 
of our sample of massive galaxies, 
in Section 2 we summarize the CANDELS and associated ground-based and {\it Spitzer} datasets in the central region of the UDS field,  
and explain how these were analysed to produce the high-mass, high-redshift galaxy sample which we have then subjected 
to morphological analysis. In Section 3 we present our general morphological model-fitting technique and then, in Section 4 we 
detail our single S\'{e}rsic  model-fitting procedure, explain how meaningful errors on parameter values were determined, 
and describe our investigation of possible biases. This is followed by a description of our bulge+disk decomposition 
analysis in Sections 5 and 6. In Section 7 we present our new results on the size-mass relation, and combine our derived 
morphologies with specific star-formation rate ($sSFR$) and redshift information to explore how bulge and disk fractions 
vary as a function of star-formation activity and redshift. Finally, in Section 8 we 
discuss the implications of our results for our understanding of galaxy growth, morphological evolution, and 
the quenching of star-formation activity, before closing with a summary of our main conclusions in Section 9.
Throughout we quote magnitudes in the AB system, and calculate all physical quantities assuming a 
$\Lambda$CDM universe with $\Omega_m = 0.3$, $\Omega_{\Lambda} = 0.7$, and $H_0 = 70\,{\rm km s^{-1} Mpc^{-1}}$.

\section{Data and Sample Selection}
\subsection{HST imaging and basic sample definition}
The main aim of this paper is to present a comprehensive and robust analysis of the 
morphological properties of  a significant sample of the most massive galaxies in the redshift range $1<z<3$. 
In order to achieve this we have focussed our study on the UKIDSS Ultra Deep Survey (UDS; \citealt{Lawrence2007}), 
the central region of which has been imaged with HST WFC3/IR as part of the CANDELS multi-cycle treasury programme \citep{Grogin2011, Koekemoer2011}. 
The CANDELS near-infrared data comprise $4 \times 11$ WFC3/IR tiles covering a total area of 187\,arcmin$^2$ in both the F125W and F160W filters 
(hereafter $J_{125}$ and $H_{160}$). The integration times are 4/3-orbit per pointing in $H_{160}$ and 2/3-orbit in $J_{125}$, giving 5-$\sigma$ 
point-source depths of 27.1 and 27.0 (AB mag) respectively.
For this study we have used the catalogue from Cirasuolo et al. (in preparation) as a master sample. This sample was
constructed using {\sc sextractor} \citep{Bertin1996} version 2.8.6 run on the $H_{160}$ mosaic and then
cut at a limiting total magnitude of 24.5 (i.e. a factor of ten brighter than the 5-$\sigma$ point-source detection limit)
to ensure that a reliable morphological analysis was possible (see \citealt{Grogin2011}); in practice the subsequent stellar mass cuts described 
below result in a sample in which $>$90$\%$ of the objects under study have $H_{160}<23$ ( and so we are typically dealing with  $>$50-$\sigma$ detections) .

\subsection{Supporting multi-wavelength data}
In addition to the near-infrared imaging provided by HST, the data-sets we make use of for sample selection (i.e. photometric redshifts, stellar-mass determination, 
and star-formation rates and histories) include: deep optical imaging in the $B$, $V$, $R$, $i'$, and $z'$-band  
filters from the Subaru XMM-Newton Deep Survey (SXDS; \citealt{Sekiguchi2005}; \citealt{Furusawa2008}); 
$U$-band imaging obtained with MegaCam on CFHT; $J$, $H$, and $K$-band UKIRT WFCAM imaging from Data Release 8 (DR8) of the UKIDSS UDS; 
and {\it Spitzer} 3.6\,$\micron$, 4.5\,$\micron$, 5.8\,$\micron$, 8.0\,$\micron$ IRAC and 24\,$\micron$ MIPS imaging 
from the SpUDS legacy programme (PI Dunlop).

\subsection{Photometric redshifts}
A multi-wavelength catalogue for photometric redshift fitting was constructed for the CANDELS master sample using the dual-image mode in 
{\sc sextractor} with a ground-based PSF-matched $H_{160}$ mosaic as the detection image, and including $U$, $B$, $V$, $R$, $i'$, $z'$, 
$J$, $H$, $K$, 3.6\,$\micron$, 4.5\,$\micron$, $J_{125}$ and $H_{160}$ photometry. For full details of the catalogue extraction, 
PSF matching and treatment of source de-blending, see Cirasuolo et al. (in preparation).

Following Cirasuolo et al. (in preparation), photometric redshifts for this master sample were determined using a $\chi^{2}$ 
fitting procedure, which utilises both empirical and synthetic templates to characterise the Spectral Energy Distribution (SED) 
of galaxies. The synthetic templates used here have been generated from the stellar population synthesis models of \citet{Bruzual2003}
(BC03), 
assuming a Chabrier initial mass function (IMF). A fixed solar metallicity was used with a variety of single-component, 
exponentially decaying, star-formation histories with e-folding times in the range $0\leq \tau {\rm (Gyr)}\leq 10$, 
where the age of the galaxy at each redshift was not allowed to exceed the age of the Universe at that redshift. 
Absorption from the inter-galactic medium was accounted for using the prescriptions of \citet{Madau1995}, and 
the \citet{Calzetti2000} obscuration law was used to account for reddening due to dust within the range $0\leq A_{V} \leq 4$. 
In order to test the accuracy of the photometric redshifts they were compared with known spectroscopic estimates where possible. 
This comparison demonstrated remarkably good agreement, with a distribution of 
$(z_{spec}-z_{phot})/(1+z_{spec}$) centred on zero, with a standard deviation $\sigma=0.03$.

\subsection{Stellar masses}
Stellar masses were obtained directly from the best-fitting SED used to obtain the photometric redshift 
(for a full discussion on the stellar mass fitting procedure see Cirasuolo et al. in preparation). There is 
currently much discussion in the literature over the dependence of stellar mass estimates 
on the stellar population synthesis models employed during the fitting procedure, and more specifically 
on the treatment of thermally pulsating asymptotic giant branch (TP-AGB) stars. In particular it has been 
found that models including higher contributions from the TP-AGB population 
(\citealt{Maraston2005}; M05, Charlot \& Bruzual 2007, private communication) lead to stellar masses 
on average $\sim 0.15$\,dex smaller \citep{Pozzetti2007, Ilbert2010} than those derived using BC03 templates. 
However, the models with a strong contribution from the TP-AGB have now been essentially ruled out \citep{Kriek2010, Zibetti2012}, and 
in any case the TP-AGB contribution is only important in the specific age range $\simeq 0.5-1.0$\,Gyr.

In this work we have therefore chosen to use the BC03 models, and to define the most massive galaxies by the mass threshold 
$M_*>10^{11}\,{\rm M_{\sun}}$, as derived from single-component tau-models.
The effects of including ``double burst models'' in the SED fitting have been explored by 
\citet{Michalowski2012} and by McLure et al. (2012). However, while \citet{Michalowski2012} show that 
two-component models can produce significantly larger stellar masses than single-component models for extreme star-bursting objects 
such as sub-millimetre galaxies, McLure et al. (2012) report that the mass difference is relatively small for 
more typical $z \simeq 1-2$ galaxies ($\langle \Delta M_* \rangle \simeq 0.1$\,dex), presumably because an exponentially-decaying star-formation history 
provides a reasonable description of reality for most massive galaxies at these epochs.
Accordingly, for the present study we decided to adopt the stellar mass estimates obtained with the  
BC03 tau-models, in order to most easily facilitate direct comparison with previous studies.

In addition to inconsistencies in the stellar masses derived from various stellar population synthesis models, 
there is a further added offset in quoted stellar masses introduced by the IMF used in the fitting. 
To ease comparisons with previous studies, throughout this paper we convert stellar masses quoted in the 
literature to those that would be obtained using the BC03 models with a Chabrier IMF using the following conversions: 
$\log_{10}M_{*,M05} = \log_{10}M_{*,BC03}-0.15$ \citep{Cimatti2008}; $\log_{10}M_{*,CB07} = \log_{10} M_{*,BC03}-0.2$ \citep{Salimbeni2009}; 
$\log_{10} M_{*,Chabrier} = \log_{10} M_{*,Salpeter}-0.23$ \citep{Cimatti2008}; $\log_{10} M_{*,Chabrier} = log_{10} M_{*,Kroupa}-0.04$ \citep{Cimatti2008}.

\subsection{Final sample selection}
From the master catalogue described above we define our sample as the most massive galaxies with $M_*>10^{11}\,{\rm M_{\sun}}$ in the redshift range $1\leq z_{phot} \leq3$. 
This gives a total of 215 galaxies identified from the $H_{160}$ mosaic and provides a mass-complete sample where, for our cut at $H_{160}=24.5$, 
the mass completeness limit is $M_*<10^{11}\,{\rm M_{\sun}}$ over the full redshift range of this study (see Cirasuolo et al. in preparation, 
for a full discussion of mass completeness). 

\section{Morphologies: 2-D modelling}
We have employed the GALFIT \citep{Peng2002} morphology fitting code 
to determine the morphological properties for all 215 objects in our sample. 
GALFIT is a two-dimensional fitting routine that can be used to model the surface-brightness profile of an observed galaxy with 
pre-defined functions such as a S\'{e}rsic light profile \citep{deVaucouleurs1948, Sersic1968}.
\begin{equation}
\Sigma(r) = \Sigma_{e}exp[-\kappa(({{r}\over{r_{e}}})^{{1}\over{n}}-1)]
\end{equation}
where $\Sigma_{e}$ is the surface brightness at the effective radius $r_{e}$, $n$ (the S\'{e}rsic index) is a measure of the 
concentration of the light profile, and $\kappa$ is a a correction factor coupled to $n$ such that half of the total 
flux of the object lies with $r_e$ (obtained by numerically solving 
$\Gamma(2n) = 2\gamma (2n,\kappa)$).

It is  well established that the robustness of the GALFIT output depends heavily on the input files, such as the 
background-subtracted image, the $\sigma$ map and the PSF ( see \citealt{Haussler2007} for a full discussion of these issues). 
As a result we have conducted rigorous tests of our fitting procedure to ensure that the morphological parameters that we 
determine using the GALFIT code are not biased by the realistic uncertainties in these inputs.
Specifically, in the next section we summarize and illustrate the results of thorough tests of the robustness of the derived morphological 
parameters with respect to the accuracy of the adopted PSF, and the implementation of various background-level determinations
(further details of these tests are provided in Appendix A and Appendix B). These tests are carried out 
exclusively on the $H_{160}$ mosaic, the reddest band accessible to HST, which thus best represents the majority of 
the assembled stellar mass in our objects at useful resolution. 

We adopted a fixed set-up for the GALFIT fitting procedure. 
We first ran {\sc sextractor} on the $H_{160}$ mosaic to determine initial estimates for the centroid x,y pixel positions, 
total magnitude, axis ratio and effective radius of each object, where the total magnitude is given by {\sc mag\_auto} 
and the effective radius is taken as {\sc flux\_radius} with the fraction of total flux within this radius set as 50\%. 
{\sc sextractor} is also used to produce a segmentation map of the image.

In addition to the image and segmentation map, GALFIT also requires an input $\sigma$ map in order to conduct the 
$\chi^{2}$ fitting. To first-order this $\sigma$ map can be given by the rms map generated for the CANDELS mosaic, see \citet{Koekemoer2011}. 
This rms map contains noise from the sky, readnoise and dark current contributions from all the input exposures and 
is used as an initial input, but is adapted later in the procedure to include the Poisson noise contribution 
from the object itself, which proves to be a non-trivial contribution for the bright objects in our sample.

From the image, segmentation map and rms map we then generated  $6\times6$\,arcsec stamps for each object centred on the 
x,y pixel positions from {\sc sextractor}. These are the actual input files read into GALFIT and the code is 
allowed to use the full $6 \times 6$ arcsec area in the fit, with the exception of any pixels associated with companion 
objects in the image stamps (which are  masked out by the bad pixel map).

The method outlined here provides us with: a $\sigma$ map, a bad-pixel mask and the best-guess initial model parameters, 
which are read directly into GALFIT. This set-up procedure has been implemented in a GALFIT wrap-around script and is consistent for all following tests of the PSF and background determinations used. 

\section{Single S\'{e}rsic models}

As mentioned above, the two key elements which can significantly affect the best-fitting model parameters derived by GALFIT are, first,
the accuracy of the adopted PSF and, second, the method used to establish the sky background. We have investigated both these issues, and their impact on derived parameter values and errors.
Full details of our findings are relegated to Appendix A and Appendix B, but here we provide a summary of the most important
conclusions of this work. For simplicity, the discussion of these issues is here restricted to the single S\'{e}rsic models.

\begin{figure*}
\includegraphics[scale=0.6]{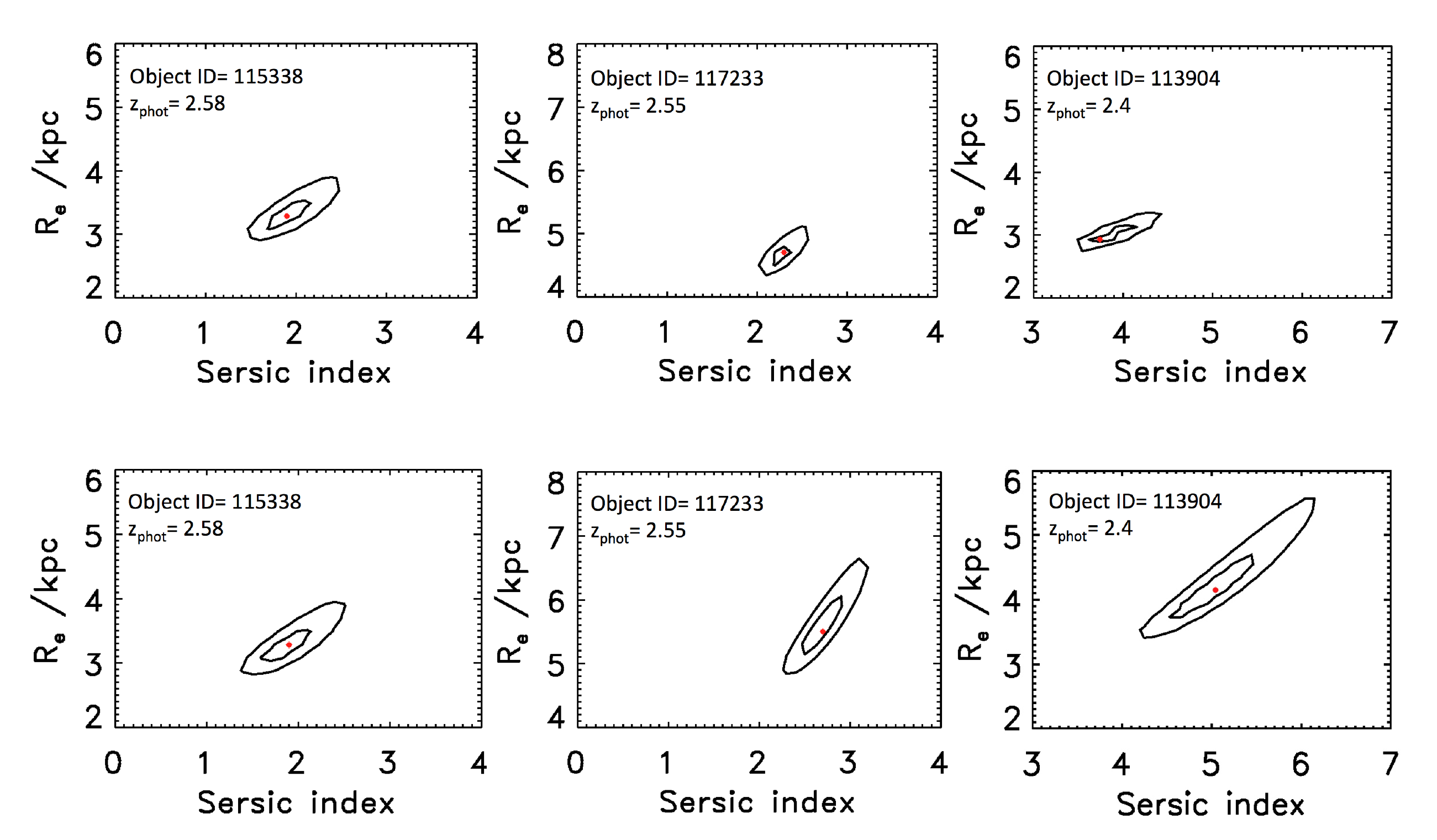}
 \caption{An illustration of the errors in, and degeneracies between, fitted effective radius $R_e$ 
and S\'{e}rsic index $n$, also showing the effect of 
allowing the background to vary during the fitting process. 
Results are shown for three example objects, with contours plotted in the $n-R_e$ plane 
at $\Delta\chi^{2} = 4$ ($\equiv 2$-$\sigma$ for 1 degree of freedom) and $\Delta\chi^{2} = 9$ ($\equiv 3$-$\sigma$ for 1 degree of freedom) above the 
the minimum $\chi^2$ value achieved by the best-fitting model (marginalising over all other fitted parameters). The location of the best-fitting model is 
indicated by the red dot in each case. 
The upper panels show the contours which result from adopting a single fixed background for each source, in this case the median background from the 
$6 \times 6$\,arcsec image stamp centred on the object in question, 
but excluding the central region of
radius 1\,arcsec (in addition to excluding pixels masked out via the segmentation map).
The lower panels show the corresponding contours which result from fitting to the same three galaxies, but in addition marginalising over a varying background 
(from the full background-grid search described in the text). 
As can be seen from these examples, allowing the background to vary during the fitting process can significantly open up the contours for some galaxies, 
increasing the errors on the fitted parameters to arguably more realistic values.
Moreover, from inspection of the third example (far right) it is clear that, for some of the largest objects in our sample, use of the $6\times6$\,arcsec median background can 
clip the wings of the galaxy and lead to an under-estimate of effective radius (note that the contours from the full background grid fitting do not in fact 
include the best-fitting solution achieved with the fixed median background, and vice versa).}
\end{figure*}

\subsection{PSF dependence}

The precision of the PSF used in the fitting procedure, especially within a 
radius of $\simeq 0.6$\,arcsec (corresponding to a physical scale of $\simeq 5$\,kpc at the redshifts of interest here), 
is crucial for the accurate determination of the scalelengths of the galaxies in our sample, as many of them transpire to have effective radii of comparable angular size.
Previous morphological studies of massive galaxies at $z>1$ have adopted both empirical and modelled PSFs in their fitting procedures, 
with modelled HST PSFs being generally determined using the Tiny Tim code \citep{Krist1995}. 
We have explored the impact of using both empirical and Tiny Tim PSFs on the resulting morphological fits. Our empirical PSF was constructed from a median stack of 
seven bright (but unsaturated) stars in the WFC3/IR $H_{160}$ image of the CANDELS-UDS field, after centroiding each stellar image.
A detailed comparison of our empirical PSF and the Tiny Tim model is presented in Appendix A. In brief, we find that 
the Tiny Tim model significantly underpredicts the emission from the real PSF around the crucial radius of $\simeq 0.6$ arcsec. 
Consequently, we found that adoption of  the Tiny Tim PSF returns fitted galaxy sizes that are on average systematically 
$5-10$\% larger than those determined using the empirical stacked PSF. As also described in Appendix A, we have confirmed that our empirical 
PSF does an excellent job of reproducing the profile of individual stars in the CANDELS $H_{160}$ image, providing 
reassurance that it is has not 
been significantly broadened or otherwise damaged by the stacking process on the angular scales of interest. Accordingly, 
for all subsequent galaxy fits presented in this paper we have adopted our empirical PSF.

\subsection{Background dependence}

The HST mosaics provided in the CANDELS data release have already been background subtracted, and so initially we attempted to use GALFIT 
on image stamps extracted from the $H_{160}$ mosaic without additional background corrections.
However, upon inspection of the radial profile plots of the fits, it became clear that additional object-by-object background corrections were required.
Moreover, the impact of background determination on the best-fitting values of, and degeneracies between, the 
fitted values of S\'{e}rsic index and effective radius is non-trivial (\citealt{Guo2009}), and merits careful exploration.

To properly explore this issue, we constructed a grid of GALFIT runs sampling the full parameter space of S\'{e}rsic index, effective radius, and plausible background values 
(see Appendix B for full details on how this grid was constructed). 
Such an analysis is computationally expensive, but it has allowed us to explicitly examine the impact of uncertainties in the background on the GALFIT results.
This problem is, of course, well known, and previous studies have attempted similar tests using different approaches (e.g. \citealt{Haussler2007,vanDokkum2010}).
However, by marginalising over the additional background subtraction value which gives the best  $\chi^{2}$ fit for each combination of S\'{e}rsic index and effective radius we
are able to properly expose the impact of background determination by constructing the $\chi^{2}$ surface in the S\'{e}rsic-index/effective-radius plane for each object.

\begin{figure*}
\includegraphics[scale=0.7]{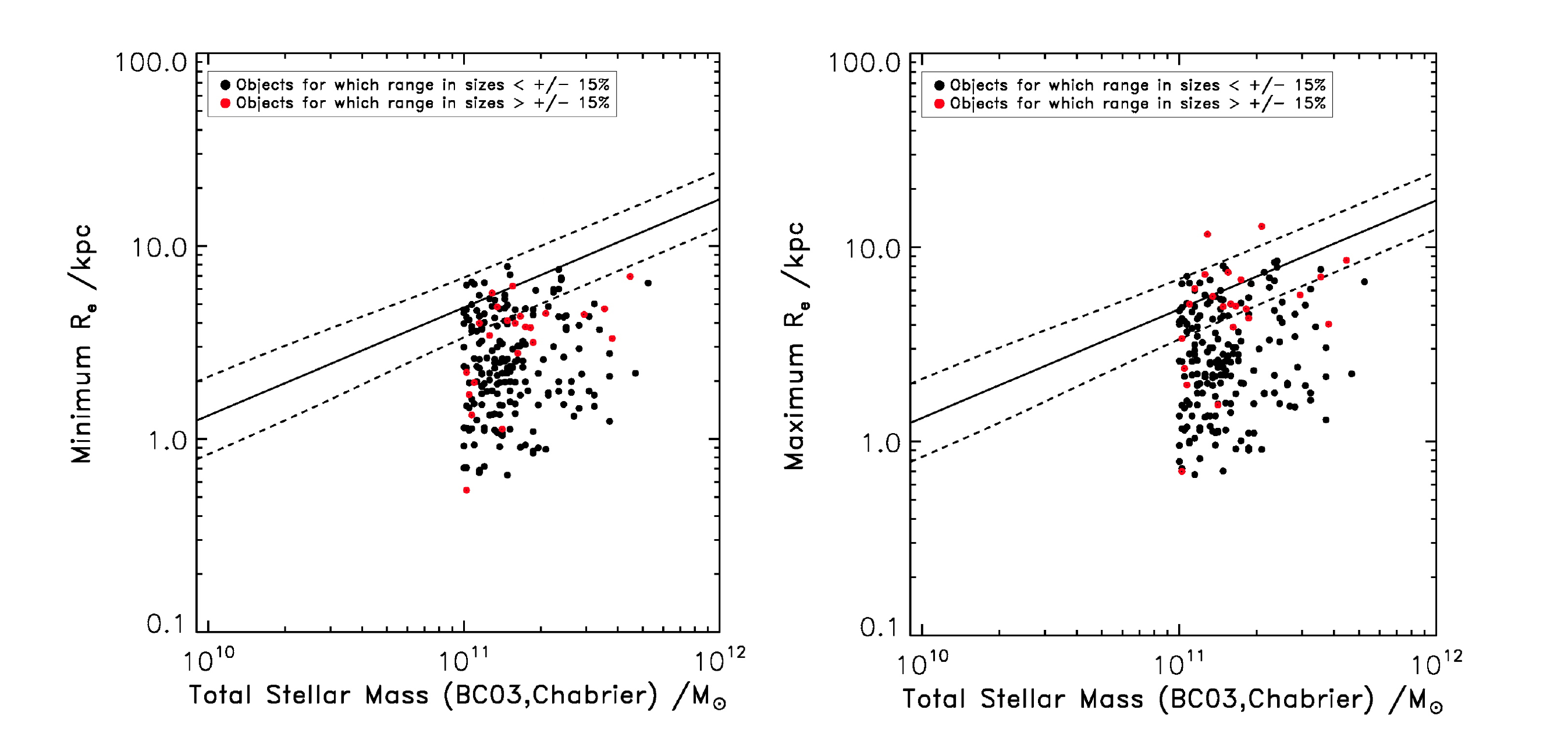}
 \caption{The size$-$stellar-mass ($R_e-M_*$, using semi-major axis $R_e$ values) relation displayed by our $M_* > 10^{11}\,{\rm M_{\odot}}$, $1 < z < 3$ galaxy sample utilising the minimum (left-hand panel) and maximum (right-hand panel) derived values of scalelength, $R_e$, as determined utilising the full range of background estimators as described in the text of Section 4.2. The objects marked in red ($\simeq 15$\% of the sample) are those for which the maximum value of $R_e$ is $>15$\% larger than the minimum, although it should be noted that the adopted range of plausible values of $R_e$ has here been chosen to be unrealistically pessimistic. Also plotted (solid line) is the local relation for early-type galaxies (ETGs) from Shen et al (2003), with its 1-$\sigma$ scatter indicated by the dashed lines. Despite our efforts 
to here exaggerate the uncertainty in $R_e$, it is clear that the size-mass relation for this sample as a whole, as derived from the high-quality CANDELS WFC3/IR imaging, is remarkably robust.}
\end{figure*}

In Fig. 1 we show the resulting $\Delta\chi^{2}$ contours in the $n-R_e$ plane for three examples of galaxies in our sample. 
The upper panels show the contours which result from adopting a single fixed background for each source, in this case the median background from the
$6 \times 6$\,arcsec image stamp centred on the object in question (but excluding the central region of
radius 1\,arcsec, in addition to excluding pixels masked out via the segmentation map).
The lower panels show the corresponding contours which result from fitting to the same three galaxies, but in addition marginalising over a varying background
(from our full background-grid search).
As can be seen from these examples, allowing the background to vary during the fitting process can significantly open up the contours for some galaxies,
increasing the errors on the fitted parameters to arguably more realistic values.
Moreover, from inspection of the third example (far right) it is clear that, for some of the largest objects in our sample, use of the $6\times6$\,arcsec median background can
clip the wings of the galaxy and lead to an under-estimate of effective radius (note that the contours from the full background-grid fitting do not in fact
include the best-fit solution achieved with the fixed median background, and vice versa). However, as discussed further below, it transpires 
that the number of such objects (i.e. objects whose scalelength is substantially 
boosted by the full background-grid search) within our sample is relatively small.

However, it should also be noted that, even with allowance for a variable background, there are a considerable number of objects within our sample for which the 
derived 1-$\sigma$ error-bars for the  S\'{e}rsic index and effective radius parameters fell below the size of the grid steps used in the 
full parameter search (0.025 in arcsec,  0.1 in  S\'{e}rsic index). Such accuracy testifies to the power of the deep, high-resolution imaging 
provided by WFC3/IR for these (relatively) bright objects. However, it does mean that it is difficult to establish a robust error for the 
parameter values in these tightly-contrained fits; to be conservative, for such objects we have simply adopted the smallest grid steps as 
the 1-$\sigma$ errors on $R_e$ and $n$.

This analysis has thus enabled us to produce more realistic errors on the S\'{e}rsic index and effective radius parameters 
for all the objects in our sample than would be inferred from the errors provided directly by GALFIT.
The error-bars produced by GALFIT are purely statistical and are determined from the covariance matrix used in the fitting, and it is well known that this 
often results in unrealistically-small uncertainties in the derived galaxy parameters. This issue is well documented in \citet{Haussler2007}, where 
they used GEMS data \citep{Rix2004} to test how well GALFIT can recover the input parameters of simulated $n=1$ (disk) and $n=4$ (bulge) galaxies. 
They found that GALFIT returns errors which are significantly smaller than the offset between the fitted and simulated input parameters,
and so concluded that the dominant contribution to the real errors in the fitting procedure arises not from statistical shot noise or read noise
(as is calculated by GALFIT), but from contamination of the fitting region by companion objects, underlying sub-structure in the sky, 
correlated pixels or potentially profile mismatching.

From our full background grid search we find that the distribution of errors is centred on $\simeq 5$\% for S\'{e}rsic index 
and $\simeq 10$\% for effective radius. This can be compared with the errors returned by GALFIT (which are often simply adopted in the literature)
where we find that, for the deep, high-quality imaging used here, 
the error distributions are centred on $\simeq 2$\% for S\'{e}rsic index and $\simeq 1$\% for effective radius.

\begin{figure*}
\includegraphics[scale=0.63]{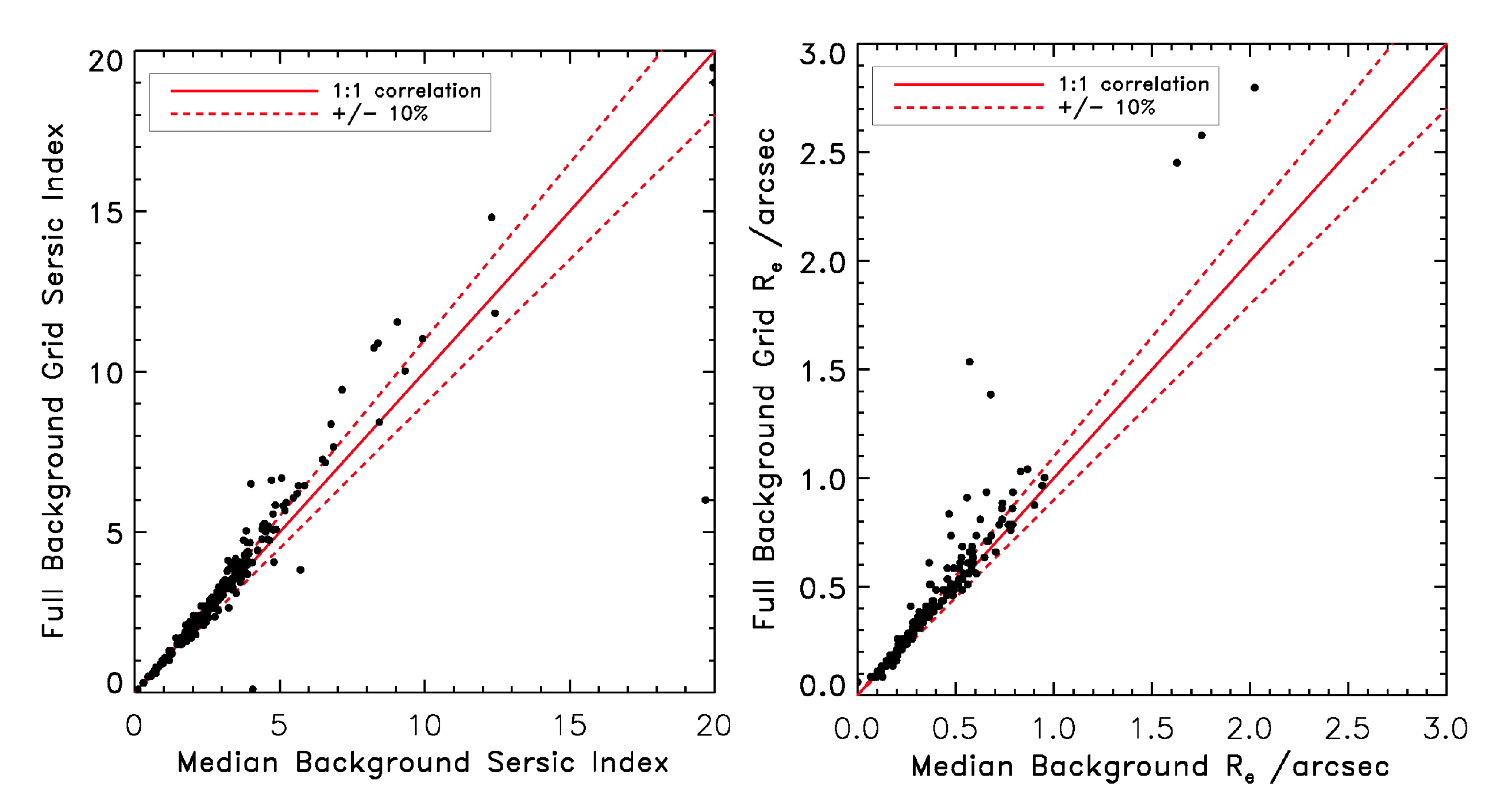}
 \caption{A comparison of the S\'{e}rsic indices and effective radii $R_e$ of the galaxies in our sample as derived 
using the background determined from the median value within a $6\times6$\,arcsec square image stamp 
(excluding pixels within 1\,arcsec of the object centroid), and 
as obtained allowing the background level to float as part of the fitting process. For $\simeq 90$\% of the objects in the sample 
the results are in excellent agreement; the full background-grid search yields significantly larger values of S\'{e}rsic index and $R_e$ for $\simeq 5$\% of the galaxies in our sample.}
\end{figure*}

To complete our analysis of the impact of background determination on derived morphological parameters we have 
considered not only the best-fitting background from the grid and the original $6 \times 6$ median background, but also an 
alternative median background determination involving exclusion of all pixels within a 
larger central aperture (i.e. the median of those pixels lying within an annulus between 3 and 5\,arcsec radius), 
and finally also zero background correction (i.e. just adopting the 
CANDELS mosaics as supplied, as we initially attempted). All four of these background values typically lie within the 
range searched within the background grid, but it is nevertheless instructive to consider these four specific alternatives
because they represent choices frequently adopted in the literature. 

Since our aim is to establish how robust our derived morphological parameters are to such choices,
we used each of these four background estimates to establish a minimum and maximum scalelength that could plausibly be 
derived for each object. The resulting extremes are almost certainly pessimistically large representations
of the uncertainty in scalelength, but nevertheless, as we show in Fig. 2, the impact on the typical 
sizes of the galaxies in our sample is still reassuringly small. Fig. 2 shows the two alternative versions of 
the size-mass relation for the galaxies in our sample which result from adopting the minimum (left-hand panel)
or maximum (right-hand panel) scalelengths as explained above. Those objects where the maximum value of
$R_e$ is $>15$\% larger than the minimum value have been highlighted in red, 
but it is clear that such objects are in a small minority ($<15\%$),
and the overall impact on the size-mass distribution exhibited by the sample as a whole can be seen to be small. 
The implications of the size-mass distribution displayed by our galaxy sample 
are discussed later, in Section 6.

In Fig. 3 we provide an additional representation of the robustness of our scalelength measurements, and also show 
that the determination of S\'{e}rsic index is extremely reliable, little affected by the alternative background determinations,
except for the very small number of objects with unusually large values of $n$.

In conclusion, therefore, our full background-grid search has enabled us to place 
realistic errors on the values of the derived parameter values such as $R_e$ and $n$, but has also shown that,
{\it for the quality of data utilised here}, our results for the sample as a whole are reassuringly robust 
to sensible alternative choices of the background level for each object.

We stress the point that the problems of systematic bias we have explored here could be much more serious for alternative datasets,
especially for ground-based observations with broader PSFs and higher backgrounds (or alternatively 
when pushing HST data closer to the detection limit).

Given the results presented in Figs 2 and 3, we did not invoke the full background search again for the multiple component modelling 
described in the next Section. For those objects which yielded robust values of $R_e$ in the single S\'{e}rsic fitting described in this section,
we have continued to simply adopt the $6 \times 6$\,arcsec (excluding the central aperture of 1\,arcsec radius) median background determination.
For the subset of $\simeq 15$\% of objects whose sizes varied by more than 15\% (i.e. those marked in red in Fig. 2) 
we found that the median background as determined in the $3-5$\,arcsec annulus returned a size centred close to the middle
of the derived range in $R_e$, and so adopted this larger annular median as the appropriate background level for this 
subset of (generally larger) objects hereafter.

\section{Multiple-component models}

Encouraged by the robustness of the single-component S\'{e}rsic fits, we decided to attempt to decompose the $H_{160}$ images of all 
the galaxies in our sample into separate bulge ($n=4$) and disk ($n=1$) sub-components. For each object we adopted the median background 
measurements as described above, and locked all sub-components at the galaxy centroid as determined from the single S\'{e}rsic fits.

To determine whether multiple components were actually merited to describe the data, we first fitted three models to each galaxy, namely
i) a bulge-only model with $n=4$ (i.e. a de Vaucouleurs spheroid), ii) an exponential disk-only model with $n=1$, and 
iii) a double-component bulge+disk model, with again the S\'{e}rsic indices locked to $n=4$ and $n=1$, but the relative amplitudes
of the components, their scale-lengths, axial ratios and position angles all allowed to vary independently.
  
It might seem that the first two of these models are simply a subset of the third (i.e. the bulge+disk model). However, 
our aim was to see if the second component was actually required (i.e. whether the more sophisticated model 
was statistically justified). In addition, inspection of the results from the double-component fits revealed that 
whenever the fainter component contributed less than $\simeq 10$\% of the $H_{160}$-band light, the parameter values for the 
fainter component were often unphysical and could not be trusted for scientific interpretation (e.g., left to its own devices, GALFIT 
will still often choose a secondary component with, for example, a completely unphysical scalelength in order to fix some unevenness in the background, even when 
such a component is not really required to achieve a formally-acceptable fit). Thus, as explained further below, whenever a secondary component
contributed less than 10\% of the flux, we simply reverted to the appropriate single-component model, designating the object as disk-only or bulge-only as appropriate.

Finally, we also explored the effect of introducing a further additional component in the form of a point-source at the galaxy centre.
This was to allow for the possibility of an AGN or central star cluster, both to quantify the evidence for such components, and to 
check whether any point sources were distorting the galaxy fits. We explored adding an additional point-source contribution 
to the single variable S\'{e}rsic, bulge-only, disk-only, and disk+bulge models. Consequently, in total we eventually fitted 
eight alternative models to each galaxy.

The exact fitting procedures implemented are detailed in the following sub-sections, while the process by which we decided 
which model to adopt for a given galaxy is described in Section 6. 

\subsection{Bulge-only and disk-only models}

The bulge-only and disk-only models are the simplest we attempted to fit to each galaxy. We constructed a GALFIT parameter file for each object 
using the best-fitting single-S\'{e}rsic parameter values as a starting point, locking the centroid position, and locking 
the S\'{e}rsic parameter at $n=4$ or $n=1$. Thus GALFIT was free to vary only the total magnitude, the effective radius, the axial ratio, and the position angle
of the forced disk or de Vaucouleurs bulge model. As with all the model fitting, great care was taken (via image masking) 
to exclude pixels which contained any significant flux from companion objects, so as not to distort the best fitting value of $\chi^2$.

\subsection{Double-component bulge+disk models}

For the bulge+disk models we again locked the centroid (of both components) at the $x,y$ position returned from the single-S\'{e}rsic fitting, and 
of course locked the S\'{e}rsic indices of the two components to $n=4$ and $n=1$. The other parameters of both components were allowed to vary independently
(i.e. allowing the bulge and disk to have very different fluxes, sizes, axial ratios and position angles if required).

When using GALFIT for this simultaneous double-component fitting, with the consequent increase in the number of degrees of freedom, we were aware 
of the increased danger of the fit becoming trapped in a local $\chi^{2}$ minimum during the minimisation routine. To tackle this issue,
and ensure that our double-component fits do indeed reflect the global minimum in $\chi^{2}$, we constructed a grid of different starting values 
for the total magnitudes and effective radii of the two components, and repeatedly restarted GALFIT from different 
positions on this grid. The grids were constructed with 11 steps in starting magnitudes for the two components, 
for each of which there were then 21 steps in initial effective radii. The grid initial magnitudes were set at 
99\% of the {\sc sextractor mag\_auto} for each object in the bulge (and hence 1\% in the disk), then 90\% bulge and 10\% disk, 80\% bulge and 20\% disk, 
continuing similarly to 10\% bulge and 90\% disk and finally 1\% bulge and 99\% disk.
Meanwhile the grid of effective radius values steps from 99\% of twice the {\sc sextractor} $r_{50}$ value for each object in the bulge and 1\%  in the disk, 
to 95\% bulge and 5\% disk, 90\% bulge and 10\% disk, and again continuing similarly to 5\% bulge and 95\% disk and finally 1\% bulge and 99\% disk.
We restarted GALFIT from each of these 231 alternative starting points in order to ensure we found the global minimum in $\chi^{2}$,
and then adopted the corresponding parameter values as our best-fitting double-component model.
After this extensive additional fitting, we found that 
the models fitted for the individual components are actually relatively robust to the initial starting conditions to an accuracy 
of $\simeq 20$\% in the fitted effective radii and magnitudes.

\subsection{Introduction of an additional point-source}
When conducting the single-S\'{e}rsic model fits (as described above in Section 4) 
we allowed the S\'{e}rsic index, which is a measure of the central concentration of 
the light profile, to range across the full 0-20 parameter space allowed by GALFIT, 
as opposed to capping it at more physical values limited to $n<8$. 
This allowed us to fully explore how $n$ and $R_{e}$ are traded off against each other 
by GALFIT when attempting to deliver model fits to some of the more unusual 
objects in the sample. 

We found that 28 out of our full sample of 215 objects 
yielded S\'{e}rsic indices in the range $5<n<20$. Upon inspection 
it appeared that these objects did indeed often have strongly-peaked central components. 
We therefore introduced the option of an additional point-source to the single-S\'{e}rsic fits, allowing GALFIT 
to vary the relative amplitude of the point-source and the single-S\'{e}rsic component.

This additional option of a point-source yielded significantly improved fits for 10 of these 28 objects, at the same time 
also yielding new, arguably more realistic, values of $n < 5$. Of the remaining 18 ``high-S\'{e}rsic objects'', 13 had 
$5 < n < 8$, and remained essentially unchanged (rejecting the additional option 
of a point-source) while the remaining five yielded only 
slightly-reduced 
values of $n$, and thus remained outside of the generally-accepted S\'{e}rsic index range.

Finally, in order to maintain a fully consistent approach across our entire sample, we decided to revisit the single-S\'{e}rsic, disk-only, bulge-only and disk+bulge 
models of every object to allow the option of an additional point-source in every case. This was done by again locking the centroid of all components at the single-S\'{e}rsic
centroid, and initially setting the brightness of the point-source at 1\% of the {\sc sextractor} {\sc mag\_auto} value.
For the bulge+disk+point-source models
we again generated a grid of initial starting parameters as detailed above in Section 5.2.

Out of the complete sample of 215 objects, 59 preferred to accept the contribution of 
a point-source comprising $>10$\% of the overall light of the galaxy (as before, we deemed unreliable/insignificant any contribution 
of $< 10$\% by any individual model sub-component). In no case did the contribution of the point-source ever exceed 43\% of the total brightness 
of the object, indicating that none of our objects is ``stellar'' or AGN dominated. Out of curiosity we checked whether those fits which 
preferred to accept a significant contribution from a point source showed any enhanced probability of yielding 
a 24\,$\micron$ detection in the SpUDS {\it Spitzer} MIPS imaging, but we did not find any significant correlation. 
However, we note that a point-source contribution might arise from a central starburst rather than an AGN. We also note that 
a preference for a point-source contribution does not necessarily mean that it is statistically required, an issue which we discuss 
further below in the general context of choosing between the array of alternative models we ultimately generated for each object.

\section{Final galaxy models}

With the inclusion of the point-source option in all models, we were left with eight alternative model fits, of varying complexity, for every 
object in the sample. In deciding which ``best-fit'' model to adopt for each source for future science analysis, we chose to  
split the models into two categories within which the models are formally nested, and thus $\chi^{2}$ statistics can be used 
to determine the ``best'' model given the appropriate number of model parameters. The first category consists of the 
single S\'{e}rsic models and the single S\'{e}rsic plus point-source models. The second category comprises the 
bulge-only and disk-only models, the bulge+disk models, and the bulge+disk+point-source models. Comparison between these two categories 
is more problematic, except in those cases where no satisfactory fit was achieved with a category-1 model, while a satisfactory 
fit was achieved with a category-2 model. As other researchers in the field may be interested in both the variable-S\'{e}rsic and bulge+disk fits, we 
have retained and present the parameter values for the best-fitting models from both categories in the tables given in Appendix D. 

\subsection{Selection of the best model}
For each object we recorded the best-fitting parameters from each of the eight models 
fitted to the data. However, before undertaking a statistical comparison of the alternative model 
options, we applied a series of criteria to reject unreasonable or physically-unrealistic models.

The first criterion imposed is the one already mentioned above, namely that we decided to throw away any model in which 
any sub-component contributed $< 10$\% of the total $H_{160}$-band light. Accordingly, any model with a very weak point-source
was rejected as unnecessary, as was any model with a very weak bulge or disk component. As discussed above, this decision was made 
after intensive inspection of the alternative model results revealed that such low-level components were 
often, in effect, artefacts of an unjustifiably complex fit (and even when physically plausible, their derived parameter 
values were too uncertain to be trusted in further analysis).

The second criterion again directly addresses how meaningful the fitted parameters are, as we decided 
to exclude any model with a sub-component whose effective radius exceeded 50 pixels (i.e. 3\,arcsec), 
the fitting radius of our image stamps. This criterion did not in fact lead to the rejection of many models, 
but those that were rejected on this basis had clearly unphysical effective radii (i.e. they substantially exceeded
the 3\,arcsec angular diameter threshold).

The third criterion, again aimed at confining our best-fitting models to those which are physically realistic,
involved the rejection of any model which contained a bulge component with an extreme axial ratio $b/a < 0.1$.
This additionally served to exclude any bulge models where the fitted effective radii were less than one pixel in size.

Having applied these criteria, it remained to consider, for each object, the relative merits of the 
surviving model alternatives within each category. First, we rejected any of the remaining models which did not 
deliver formally acceptable fits at the 3-$\sigma$ level, as judged from the absolute value of $\chi^2$ achieved, 
and the number of degrees of freedom, $\nu$ (where here, the number of degrees of freedom means the number of 
data points minus the number of fitted parameters minus 1, and is typically $7000 - 10000$ for the images and models fitted here, the precise value 
for each object depending on the degree of image masking; see Appendix C). 

A model fit was thus deemed formally acceptable if the minimum value of $\chi^2$ satisfied:

 \begin{equation}
\chi^{2}\le\nu+3\sqrt(2\nu)
\end{equation}

\noindent
and if any model failed this test it was no longer considered (although see below for model refinement).

Finally, if more than one model within each category 
survived all of the above tests, we chose between the acceptable fits of varying complexity by adopting the simplest 
acceptable model, unless a model of higher complexity satisfied:

\begin{equation}
\chi^{2}_{complex}<\chi^{2}_{simple}-\Delta\chi^{2}(\nu_{complex}-\nu_{simple})
\end{equation}

\noindent
where 
now $\nu$ represents the number of degrees of freedom in the model (in effect the number of parameters), and 
$\Delta\chi^{2}$$(\nu_{complex}-\nu_{simple})$ is the 3-$\sigma$ value for the given difference in the degrees of freedom between the two competing fits.

\begin{figure*}
\includegraphics[scale=0.7]{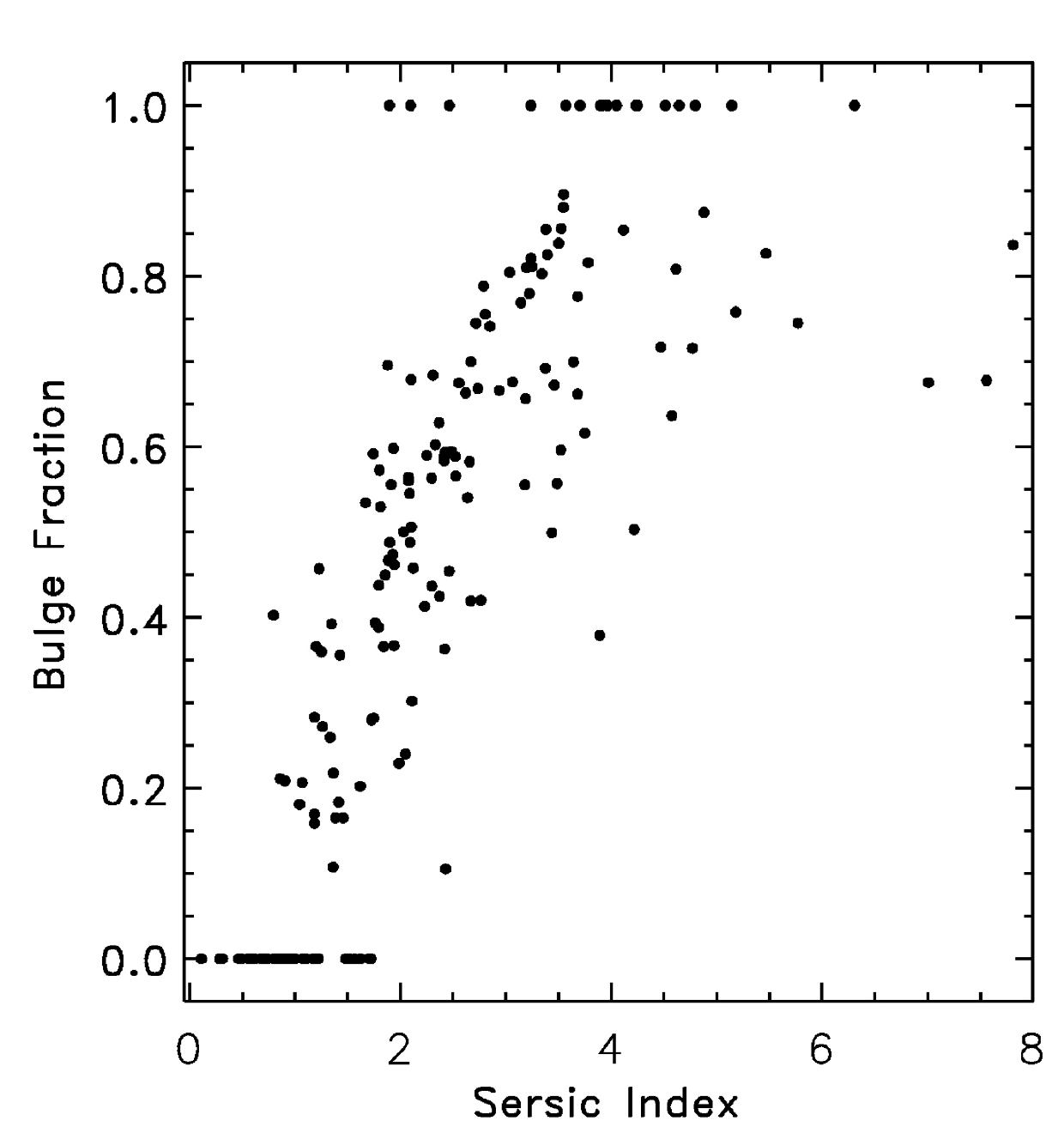}
 \caption{S\'{e}rsic index from the single-S\'{e}rsic fits, versus
bulge to total ($B/T$) fractional contribution to the $H_{160}$-band light (as determined from the multi-component modelling)
for the final sample of  192 objects used in all subsequent double-component science plots and analysis.
We have removed the 14 objects which still had formally unacceptable $\chi^{2}$ values even after the $\chi^{2}$ masking
described in the text, as well as seven objects which have unrealistically-large single S\'{e}rsic indices ($n>10$), and two unresolved
objects which may be stars. It can be seen that there is a good correlation between the two estimators
of bulge dominance, with $n \simeq 2$ corresponding to roughly equal bulge and disk contributions. 
Note that the objects at the top and bottom of the
plot are located at $B/T = 1.0$ or $B/T = 0.0$ due to our insistence (based on intensive inspection of the modelling results)
that any sub-component contributing less that 10\% of total flux is discarded as insignificant and unreliable. This of course
also leads to two artificial gaps in the distribution of bulge fraction. Reassuringly, the group of objects with $B/T$ rounded down to
zero is centred on $n=1$, while the ``pure-bulge'' objects with $B/T$ equal to unity is centred on $n=4$.}
\end{figure*}

In this way we narrowed down the alternative models to a single, final, best-fitting model within each category, and the best-fitting 
parameters for these (two) models of each object are given in Table D2 in Appendix D.

In the relatively small number of cases where no formally acceptable model survived the first of the $\chi^2$ 
tests described above, we have still applied the final {\it relative} quality-of-fit test, so as to retain 
parameter values for every galaxy in case this is required (note that very few other studies in this area have actually been concerned 
with assessing whether the best-fitting models are genuinely formally acceptable, even though a failure to achieve this renders the assessment 
of errors in parameter values problematic). The parameter values from these best-fitting, albeit formally-unacceptable 
models are also presented in Table D2 for completeness, but are flagged by an asterisk in the bulge effective-radius column. 
These unacceptable fits, and our efforts to minimize the number of such cases, are discussed further in the next sub-section and in Appendix C.

\subsection{Model fit refinement}

As a final comment on the technical aspects of the model-fitting described in this paper, we briefly consider 
the problems we encountered in achieving formally-acceptable fits to a subset of our objects, and the steps we took to minimize the
number of objects for which the modelling still proved formally inadequate. A fuller description of this work 
is provided in Appendix C for the interested reader.
 
Upon completion of our initial model fitting, we found that 70 out of our full sample of 215 objects 
had no formally-acceptable model fits as judged by the first of the two $\chi^2$ tests described above (i.e. equation 2).
To establish the cause of the excessively-high values of $\chi^2$, we visually inspected the images of all 70 objects.
We found that there were several obvious, but different, reasons for these high $\chi^2$ values, with the problematic objects including
i) $z<2$ spiral galaxies with very prominent spiral arms, ii) interacting/asymmetric systems, iii) objects in very crowded fields, and 
iv) objects with extremely close companions which had not been separately identified by {\sc sextractor}. 

We therefore included an additional round of modelling for these objects, refitting after masking out the problematic non-axisymmetric 
structures (such as spiral arms or close companions) on the basis of $\chi^{2}$ maps produced from the original attempted 
fits. Using this approach we re-ran all the model fits as described in Section 5 above, and re-selected the best-fitting models. 
Doing this delivered acceptable fits for all but 14 objects in our entire catalogue. 
The quality of the final fits achieved in this work is demonstrated by the final distribution of minimum $\chi^{2}$ 
for the full sample, which is shown in Fig. C4 and discussed further in Appendix C.

Finally, it is important to stress that, while this re-fitting was sometimes required to achieve formally-acceptable values of $\chi^2$ (and hence 
set meaningful errors on the best-fit parameters), it in fact very rarely resulted in any significant change in the best-fitting {\it values}
of these parameters. This is shown explicitly by the comparisons of the best-fitting parameter values 
(as achieved before and after this additional round of image masking) shown in Appendix C. The reason for this is simply that while 
high surface-brightness features which cannot be represented via axi-symmetric modelling can contribute significantly to $\chi^2$, they 
rarely actually dominate a sufficiently-large fraction of our object image stamps (which each contain $\simeq 10,000$ pixels) 
to significantly distort the morphological properties 
of the underlying mass-dominant galaxy as established via our modelling.

\begin{figure*}
\includegraphics[scale=0.7]{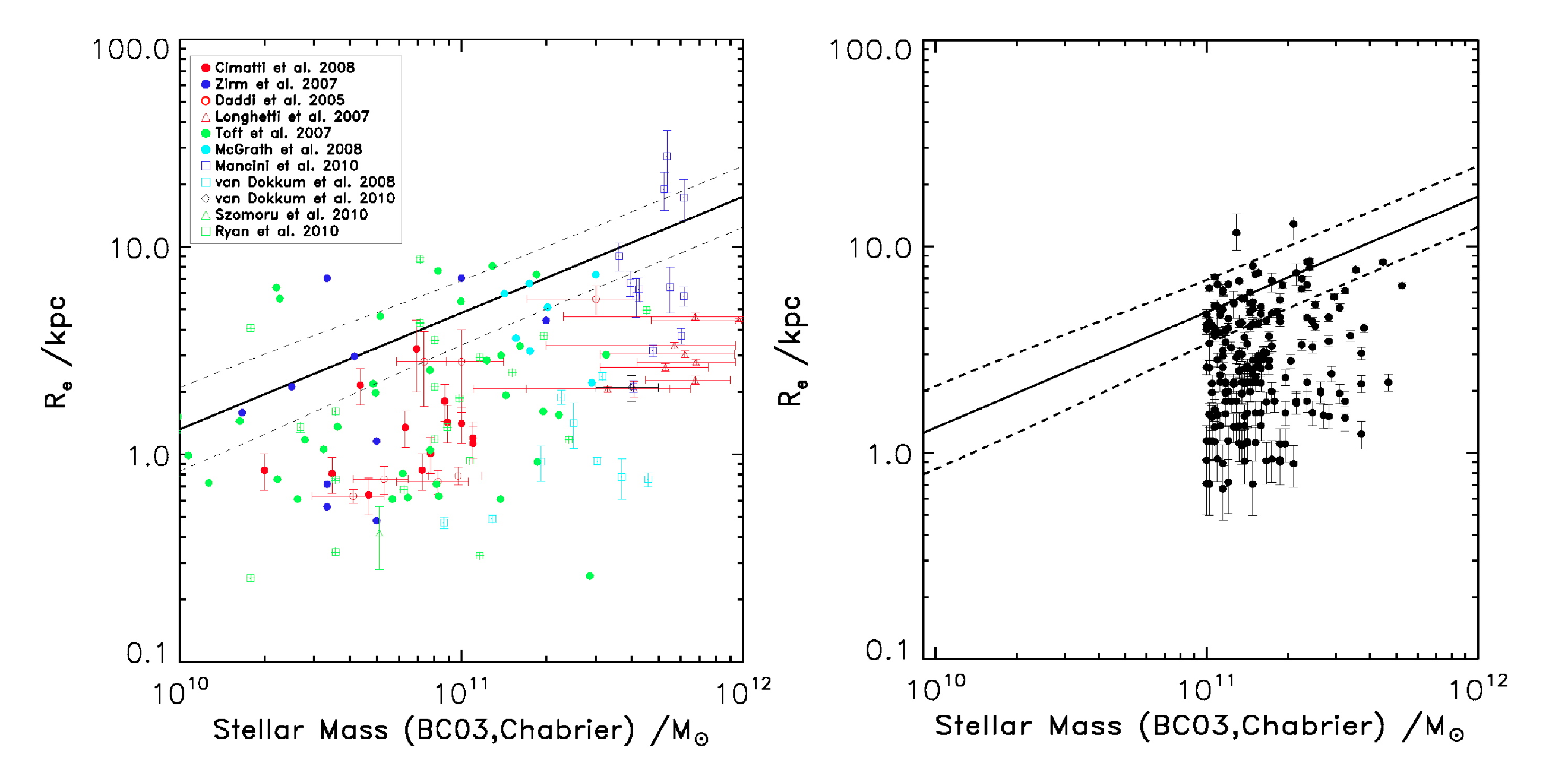}
\caption{Various determinations of galaxy-size versus stellar mass at $1 < z < 3$ from the literature are shown in the left-hand panel, 
for comparison with our new results for $M_* > 10^{11}\,{\rm M_{\odot}}$ 
galaxies over the same redshift range as shown in the right-hand panel. In order to facilitate comparison of the
semi-major axis scalelengths determined here with an appropriate low-redshift baseline we have plotted a solid line on both panels
to indicate the local early-type galaxy relation from \citet{Shen2003} (with the scatter in this relation indicated by the dashed lines).
Because the galaxy sizes determined by Shen et al. were determined by fitting 1-D surface-brightness profiles within circular apertures,
we have converted their results to reflect estimated semi-major axis sizes by dividing the circularised Shen et al. sizes
by the square root of the median axis ratio (b/a) for the $1 \times 10^{11}\,{\rm M_{\sun}} < M_* < 1 \times 10^{12}\,{\rm M_{\sun}}$ 
SDSS sample. This median axis ratio value was taken to be 0.75, following the results from \citet{Holden2012}. 
The results from the literature shown in the left-hand panel have all been converted to the masses that would 
have been derived using the \citet{Bruzual2003} models with a Chabrier IMF (see text
for details). Unfortunately the scalelengths plotted in the left-hand panel contain a mix of both circularised and semi-major axis values, but
since they come mainly from studies of early-type galaxies the correction from circularised back to semi-major axis values is generally
small. Our own points shown in the right-hand panel are all based on Chabrier BC03 masses, and semi-major axis effective radii derived
from our single-S\'{e}rsic modelling of the $H_{160}$ images. This figure serves to demonstrate the extent to which our study has advanced 
knowledge of the size-mass relation for galaxies in this crucial redshift range in this high-mass regime. It can be seen that, while the
majority of the objects in our sample lie below the local relation, a significant subset ($32 \pm 4$\%) are consistent with it within
the plotted 1-$\sigma$ errors.}
\end{figure*}

\section{Science Results}

Having determined both accurate and acceptable single-S\'{e}rsic models and bulge+disk decompositions for the vast majority
of the objects within our sample,
we are now able to proceed to explore the scientific implications of our results. First, however, it is interesting 
to consider the correlation between single S\'{e}rsic  index and $B/T$ flux-ratio delivered by our
modelling of these massive galaxies at $1<z<3$, a relation which has been extensively studied and 
debated at lower redshifts (e.g. \citealt{Ravindranath2006, Simard2011, Lackner2012}). 
This is plotted in Fig.~4, where it can be seen that, in 
contrast to some previous studies at lower redshift ($z<1$), we find that S\'{e}rsic  index and $B/T$ flux-ratio 
are generally in remarkably good agreement; from Fig.~4 it can be seen that disk-dominated systems with $B/T < 0.5$ are almost completely confined 
to the S\'{e}rsic-index range $0<n<2$, and that virtually all bulge-dominated 
galaxies with $B/T > 0.5$ have $n > 2$. These results provide further confidence in the reliability of our morphological 
analysis, and suggest that our attempt to separate the galaxies into bulge and disk components is meaningful
and, moreover, justified by the quality of the WFC3/IR data.

\subsection{The size-mass relation}

\begin{figure*}
\includegraphics[scale=0.65]{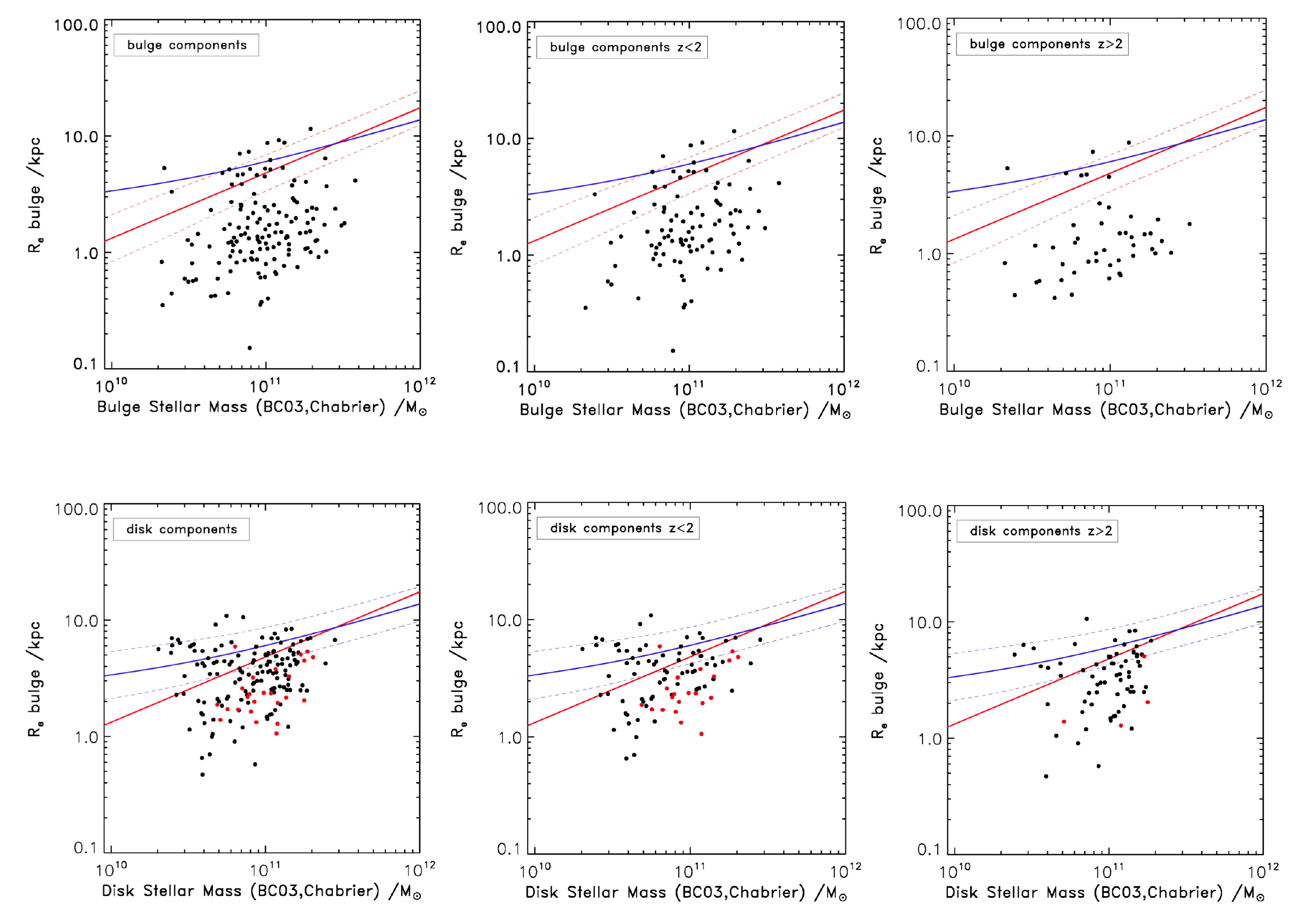}
 \caption{The size-mass relations 
displayed by the separate bulge components (upper row) and 
disk components (lower row) as produced from our bulge+disk modelling analysis of our massive galaxy sample
(shown both for $1 < z < 3$ and then subdivided into two redshift bins). The masses plotted here for the individual 
sub-components simply reflect the total mass of the ``parent'' galaxy sub-divided in proportion to the
contribution of each component to the $H_{160}$-band light. For consistency, and to avoid over-interpreting the 
location of the weakest sub-components, we have excluded nine objects whose component masses fall below 
$2 \times 10^{10}\,{\rm M_{\sun}}$. In the lower row of plots, the disk components from the passive disk-dominated galaxies 
discussed in Section 6.3 (i.e. objects with $sSFR< 10^{-10}\,{\rm yr^{-1}}$, no 24\,$\mu$m counterparts and $B/T<0/.5$) are over-plotted in red. 
In order to provide a comparison with the sizes of comparably-massive low-redshift bulge and disk counterparts, we 
have taken the local early-type, and late-type galaxy relations from \citet{Shen2003} and converted them to 
non-circularised sizes (as described in the caption to Fig.~5). 
These non-circularised relations are plotted as a solid red line for the local early-type relation,
and a solid blue line for the local late-type galaxy relation; the dashed lines indicate the typical 1-$\sigma$ scatter in these 
relations. As discussed in detail in the text, these plots reveal the more dramatic size evolution displayed by the bulges which, by $z > 2$ 
are on average a factor of $>4$ smaller than their local counterparts. Nevertheless some bulges, and a rather large fraction of disks
are still found to lie on the local relation throughout the redshift range.}   
\end{figure*}
 
We now use our modelling results to explore the size-mass 
($R_e - M_*$) relation for massive galaxies in the redshift range $1<z<3$, considering 
first the results from the single-S\'{e}rsic fits, and then the output from our bulge+disk decompositions.

\begin{figure*}
\includegraphics[scale=0.65]{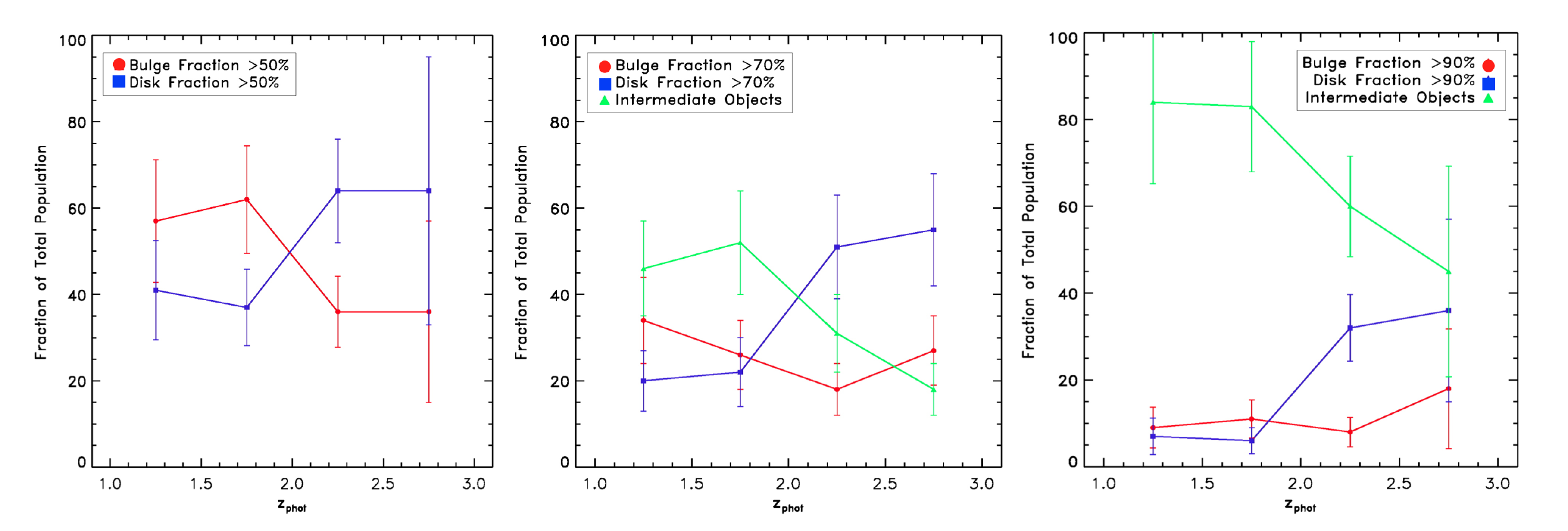}
\caption{The redshift evolution of the morphological fractions in our galaxy sample, after binning into 
redshift bins of width $\Delta z=0.5$. We show three alternative cuts in morphological classification, both to try to provide 
a complete picture, and to facilitate comparison with different categorisations in the literature. In the left-hand panel
we have simply split the sample into two categories: bulge-dominated ($B/T > 0.5$) and disk-dominated ($B/T < 0.5$). In the central 
panel we have separated the sample into three categories, with any object for which $0.3 < B/T < 0.7$ classed as ``Intermediate''. 
Finally, in the right-hand panel we have expanded this Intermediate category to encompass all objects for which $0.1 < B/T < 0.9$
(see Section 7.2 for discussion)}.
\end{figure*}

The best-fit results from our single-S\'{e}rsic analysis detailed in Section 4.2 are shown in Fig.~5, 
alongside a compilation of results from some of the previous literature at $1<z<3$. Unsurprisingly,  
previous studies have adopted a variety of different techniques for stellar-mass determinations and morphological 
modelling. However, the results plotted in the left-hand panel of Fig.~5 have been adapted to provide the fairest 
comparison with our results by ensuring that all stellar-mass estimates have been converted to those that 
would have been determined using a Chabrier IMF with a \citet{Bruzual2003} stellar population synthesis model. 
Nevertheless, the comparison remains imperfect both because the literature results are taken from imaging at a range of 
different rest-frame wavelengths, and because they comprise a mixture of both circularised and semi-major axis effective radii
(although, as the literature results come predominantly from studies of early-type galaxies, 
the correction from circularised to semi-major axis values is generally small). An additional complication arises from the  
fact that the studies in the literature have utilised a variety of different selection criteria, with most previous studies 
deliberately biased towards passive and/or early-type galaxies. By contrast our own sample is based on a relatively straightforward 
mass limit; while this inevitably limits the dynamic range of our study, it can be seen from Fig.~5 that this work represents 
a significant step forward in populating the high-mass regime of the size-mass plane at these redshifts.

Armed with our modelling results for this first substantial, complete, high-mass-limited sample, we 
find that the majority ($68 \pm 7$\%) of these galaxies have effective radii which place them well below the local relation and its 1-$\sigma$ scatter 
(where in this high mass regime the local early and late-type relations are essentially the same).
However, there is also a significant subset of $32 \pm 4$\% of our objects which, within the error-bars, are consistent with the 
sizes of similarly-massive local galaxies. Within the limited redshift range spanned by our study, we see 
no dramatic trend in these statistics with redshift; splitting the sample at $z=2$, in the redshift 
range $1<z<2$ we find $70 \pm 10$\% of objects lie below the local relation, with $30 \pm 5$\% essentially on it, while at $2<z<3$
the corresponding figures are $62 \pm 11$\% and $38 \pm 8$\% respectively.

One consequence of the majority of the galaxies lying significantly below the local relation is that  
the median size of these most massive galaxies at $1 < z < 3$ is a factor 2.25 times smaller (2.6\,kpc)  
than comparably-massive local galaxies. Again we see no really significant redshift trend in this global statistic 
within our limited redshift range, although there is a gradual trend to smaller sizes with increasing redshift; 
splitting our sample below and above $z=2$, the median size becomes 
2.73\,kpc and 2.53\,kpc respectively, corresponding to 2.18 and 2.27 times smaller than the local relation. 

Since Fig.~5 includes all objects, of whatever morphology, we next use the results from our 
bulge+disk modelling to check for any significant trends with morphological type, or indeed 
for trends with redshift within a given morphological sub-class. Since we have attempted bulge+disk 
decomposition for all galaxies in the sample, we can plot the relevant size-mass relations not just 
for bulge- or disk-dominated galaxies, but for {\it all} bulges and disks (i.e. including the 
bulges from the disk-dominated objects and vice versa). 

The size-mass relations for the separate bulge and disk components are plotted in Fig.~6, shown both for the 
full redshift range, and subdivided for $z < 2$ and $z > 2$. Because we are plotting sub-components, 
these plots contain some objects with stellar masses $M_*$ substantially smaller than our original 
mass limit. This provides additional dynamic range, but we note that the stellar-mass subdivision has been performed here solely on the basis 
of the fractional contribution of each sub-component to the $H_{160}$-band light. This is clearly not quite 
correct, but a full SED-based mass determination for each sub-component is deferred to a future paper involving
fitting of the bulge+disk models to multi-band optical-infrared imaging. It also does not mean 
that our study is in any sense mass-complete at masses substantially smaller than $M_* \simeq 10^{11}\,{\rm M_{\odot}}$.
Nonetheless, it is instructive to see whether the minor components (e.g. the bulges in disk-dominated galaxies) follow
the same trends as the dominant components (although to avoid pushing the data too far, we do not plot 
any sub-components with estimated masses $M_* < 2 \times 10^{10}\,{\rm M_{\odot}}$). In Fig.~6 we also over-plot
the local early and late-type size-mass relations as described in the figure caption.

These plots reveal a number of interesting features. First, consistent with previous studies, 
it can be seen that the size evolution is more dramatic in the bulges than in the disks, but nevertheless 
most disks are also smaller than in the local Universe; over the full redshift range $81 \pm 10$\% of the 
bulges lie significantly below their relevant local relation, while for the disks the corresponding 
figure is $58 \pm 7$\% (conversely this means that only $19 \pm 4$\% of bulges are consistent with the local relation,
but this figure rises to $42 \pm 6$\% for the disks). 

An interesting aspect of the more dramatic size evolution displayed by the bulges is that their size-mass distribution, 
especially at the highest redshifts, appears bi-modal (although the statistics are weak), with the dominant population of compact 
bulges becoming increasingly separated from the minority of objects which appear still consistent with the local relation
(see the top-right panel of Fig.~6). Interestingly 
these trends also seem to apply to the lower-mass bulges embedded in the disk-dominated galaxies, which 
display the smaller sizes as ``expected'' from a simple offset of the size-mass relation as determined 
from the more-massive bulge-dominated galaxies.

The trends with redshift shown in Fig.~6 can be quantified in terms of the fractions of bulges and disks on or significantly 
below their respective local relations at $1 < z < 2$ and $2 < z < 3$. For the bulges the relevant figures are 
$20 \pm 5$\% on and $80 \pm 12$\% below in the lower redshift bin, and $15 \pm 9$\% on, $85 \pm 18$\% below on the upper 
redshift bin. For the disks there really is no evidence for any evolution in the relevant fractions within our redshift range; 
the percentages are $41 \pm 8$\% on and $59 \pm 10$\% of disk components below the local relation 
at $1 < z < 2$, and $43 \pm 9$\% on and $57 \pm 11$\% below at $2 < z < 3$.

These trends are also reflected in the evolution of the median sizes of the bulge and disk components. Even within our limited 
redshift range the (apparent) evolution in size of the bulges is fairly dramatic, where taking the median sizes of bulges which lie below the local early-type relation
 gives an offset from the local early-type relation already a factor of 3.5 at $1 < z < 2$ rising to a factor 4.4 at $2 < z < 3$. By contrast, the offset for the disks from the local late-type relation
is more modest and apparently unchanging; a factor of 2.43 at $1 < z < 2$, and 2.55 at $2 < z < 3$.

Finally, marked in red on the lower panels of Fig.~6 are the locations of the ``passive'' disks in our sample, a
population discussed further below in Section 6.3. Interestingly, the vast majority of the passive disks lie below the local late-type size-mass relation.

\subsection{Evolution of morphological fractions}

We next consider how the relative number density of galaxies of different morphological type
changes over the redshift range probed by our sample. In Fig.~7 we illustrate this by binning our sample into four 
redshift bins of width $\Delta z = 0.5$, and consider three alternative cuts in morphological classification
as measured by $B/T$ from our disk-bulge decompositions. We present the data in this way both to try to provide
a complete picture, and to facilitate comparison with different categorisations in the literature. In the left-hand panel
of Fig.~7 we have simply split the sample into two categories: bulge-dominated ($B/T > 0.5$) and disk-dominated ($B/T < 0.5$). In the central
panel we have separated the sample into three categories, with any object for which $0.3 < B/T < 0.7$ classed as ``Intermediate''.
Finally, in the right-hand panel we have expanded this Intermediate category to encompass all objects for which $0.1 < B/T < 0.9$.

From the first panel it can be seen that disks dominate at $z > 2$ and that this situation is reversed at $z < 2$. However, the other
two panels help to emphasize that, at $z < 2$, pure bulges and disks are rare, and that the vast majority of lower-redshift objects are,
to a varying degree, disk+bulge systems. Interestingly, however, it is clear that, however the cuts are made, at $z > 2$
the population is disk-dominated, and a substantial fraction
of the sample are ``pure'' disks, which have largely disappeared by $z < 2$. Since the number density of galaxies in this high-mass
regime falls dramatically with increasing redshift at $z > 3$, these plots illustrate that the redshift range $2 < z < 3$
is the {\it era of massive disks}. 

Conversely, at the lowest redshifts probed by this study ($z\simeq1$) it is seen that, while bulge-dominated objects are on the rise,  
pure-bulge galaxies (i.e. objects comparable to present-day giant ellipticals) have yet to emerge in significant numbers, with $>90$\%
of these high-mass galaxies still retaining a significant disk component.

 \begin{figure*}
\includegraphics[scale=0.7]{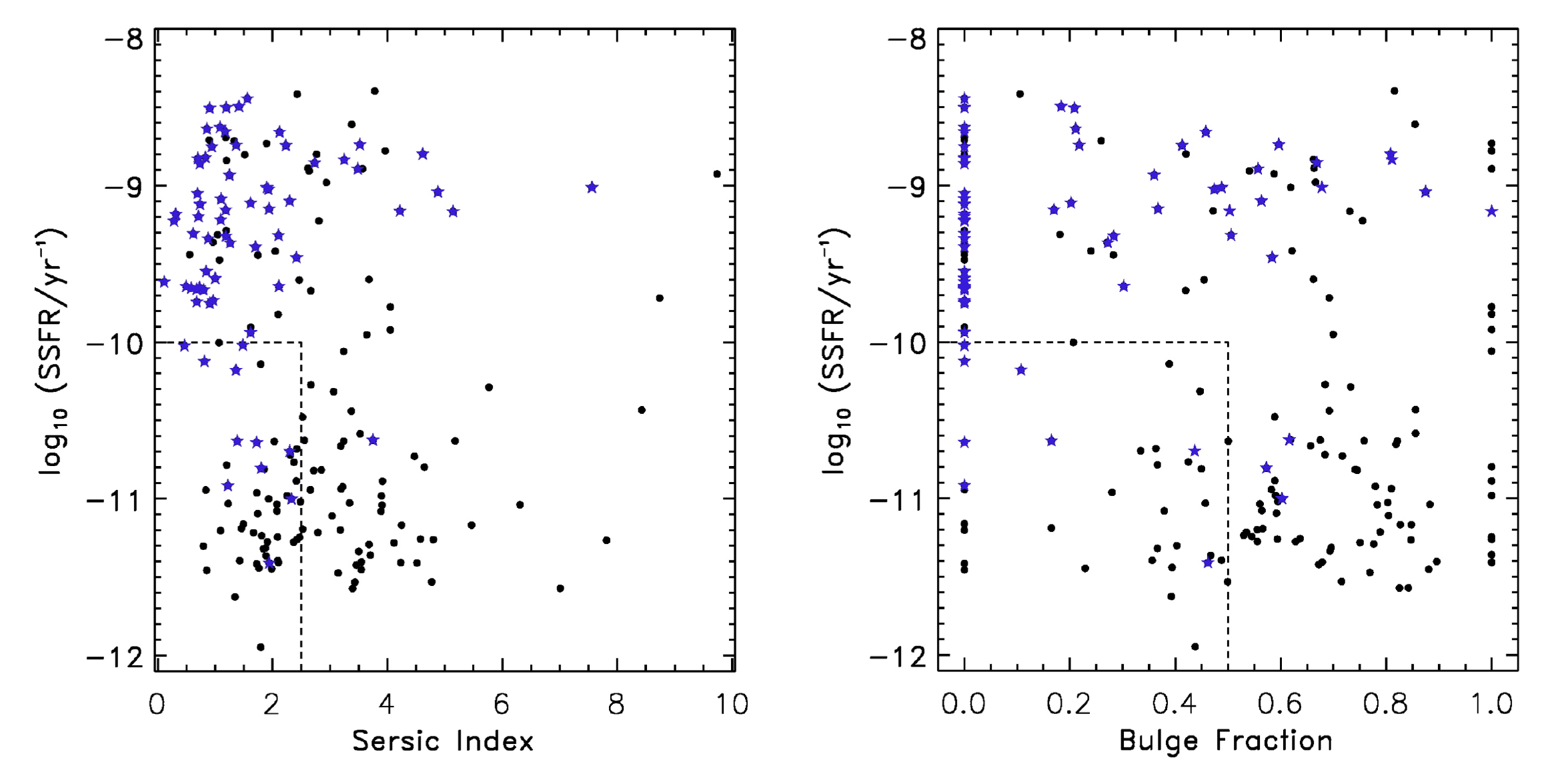}
 \caption{Plots of specific star-formation rate ($sSFR$) versus morphological type as 
judged by single S\'{e}rsic index (left-hand panel) and bulge-to-total $H_{160}$-band flux ratio ($B/T$) (right-hand panel).
The values of $sSFR$ plotted are derived from the original optical-infrared SED fits employed in the sample selection, and include 
correction for dust extinction as assessed from the best fitting value of $A_V$ derived during the SED fitting. Highlighted by blue stars 
are those galaxies which we found have a 24\,$\mu$m counterpart in the {\it Spitzer} SpUDS MIPS imaging of the UDS, indicative of 
some dust enshrouded star-formation and/or AGN activity. It is clear from these plots that the vast majority of disk-dominated galaxies are star-forming,
and the majority of bulge-dominated galaxies are not (as judged by $sSFR < 10^{-10}\,{\rm yr^{-1}}$). However, we have 
indicated by a box on both the panels the region occupied by a potentially interesting population of passive disk-dominated objects; in the left-hand panel 
disk-dominated is defined as $n < 2.5$, and $52 \pm 9$\% of the quiescent galaxies lie within this box, while in the right-hand panel 
disk-dominated is defined by $B/T < 0.5$, in which case $34 \pm 7$\% of the quiescent objects lie within this region.}
\end{figure*}

\subsection{Star-forming and passive disks.}

The primary aim of this paper is to focus on the morphological analysis of the $H_{160}$ images, with 
a full treatment of the SEDS, including dependence of morphology on wavelength, deferred to a future paper.
Nevertheless, in Fig. 8 we make use of the SED fitting already employed in the sample selection to explore 
the relationship between star-formation activity and morphological type. 

Fig. 8 shows specific star-formation rate ($sSFR$) versus morphological type for the massive galaxies in our sample, where morphology 
is quantified  by single S\'{e}rsic index in the left-hand panel, and by bulge-to-total $H_{160}$-band flux ratio ($B/T$) in the right-hand panel.
The values of $sSFR$ plotted are derived from the original optical-infrared SED fits employed in the sample selection, and include
correction for dust extinction as assessed from the best fitting value of $A_V$ derived
during the SED fitting. As a check of the potential failure of this approach to correctly identify reddened dusty star-forming 
galaxies, we have also searched for 24\,$\mu$m counterparts in the {\it Spitzer} SpUDS MIPS imaging of the UDS, 
and have highlighted in blue those objects which yielded a MIPS counterpart within a search radius of $<2$\,arcsec.
Reassuringly, relatively few 24\,$\mu$m detections have been uncovered in the lower regions
of the panels shown in Fig. 8, while the vast majority of star-forming objects are confirmed via MIPS counterparts.
 This shows that the determination of $sSFR$ as deduced from the optical--mid-infrared
SED fitting has been (perhaps surprisingly) good at cleanly separating the star-forming galaxies from the more quiescent objects.

It is clear from these plots that the vast majority of disk-dominated galaxies are star-forming,
whereas the majority of bulge-dominated galaxies are not (as judged by $sSFR < 10^{-10}\,{\rm yr^{-1}}$). Nonetheless, the sample
also undoubtedly contains a few star-forming bulge-dominated galaxies and, perhaps more interestingly, a significant 
population of apparently-quiescent disk-dominated objects, which we quantify and discuss further below.

First, though, we note that the most obvious 
feature of Fig.~8 is the prominent group of pure-disk galaxies which dominate the star-forming population. 
Since we already emphasized in Fig. 7 that the pure-disk population is largely confined to $2 < z < 3$, it becomes clear 
that, at $z > 2$, our massive galaxy sample is dominated by disk-dominated/pure-disk star-forming galaxies. 
As we discuss in a related CANDELS paper, this population of massive star-forming disks at $2 < z < 3$ is, to first-order, 
the same as the population of sub-millimetre galaxies revealed by continuum sub-millimetre and millimetre 
wavelength surveys over the last decade \citep{Targett2012}.

Equally interesting, however, is the apparently-significant population of quiescent disks revealed on these plots. To highlight and quantify  
this population we have indicated by a box on both the panels the region occupied by objects with disk-dominated morphologies and 
$sSFR < 10^{-10}\,{\rm yr^{-1}}$. In the left-hand panel, disk-dominated is defined as $n < 2.5$, and $52 \pm 9$\% of the quiescent galaxies 
lie within this box ($40 \pm 7$\% if we exclude the 24\,$\mu$m detections as indicating obscured star-formation activity), 
while in the right-hand panel, disk-dominated is defined by $B/T < 0.5$, 
in which case $34 \pm 7$\% of the quiescent objects lie within this region ($25 \pm 6$\% if we exclude the 24\,$\mu$m detections).

As discussed further in Sections 8.3 and 8.4, quiescent disk galaxies are of particular interest because they suggest 
that the quenching or exhaustion of star-formation activity need not be simply linked to a process (e.g. major merging) 
which is also directly associated with inducing morphological transformations. We re-emphasize that it is clear the 
majority of disk-dominated galaxies in our sample are star-forming, and that this is true for an even clearer 
majority of the pure disks. However, our sample does appear to include a significant population of quiescent 
disk-dominated objects, including $\simeq 5$ pure disks (ten pure disks lie in the box, but the upper five of these 
possess 24\,$\mu$m detections indicating that they may be reddened star-forming disks, or contain obscured AGN; note that at the 
depth of the {\it Spitzer} SpUDS MIPS imaging, and the redshifts and masses of interest here, a significant detection at 
24\,$\mu$m always corresponds to an $sSFR$ above our adopted threshold of $sSFR = 10^{-10}\,{\rm yr^{-1}}$ 
if the mid-infrared emission is interpreted as due to star-formation activity). 

We have double-checked that none of the quiescent disk-dominated objects not already marked by the blue stars in Fig. 8
(indicating a counterpart in the MIPS catalogue) have even marginal detections in the 24\,$\mu$m imaging. We have also checked 
that this population is not biased towards higher redshift, which might make MIPS detections more challenging. 
We thus conclude that this population really is quiescent as judged by $sSFR$, and needs to be explained in any viable 
model of galaxy formation/evolution.

\begin{figure*}
\includegraphics[scale=0.5]{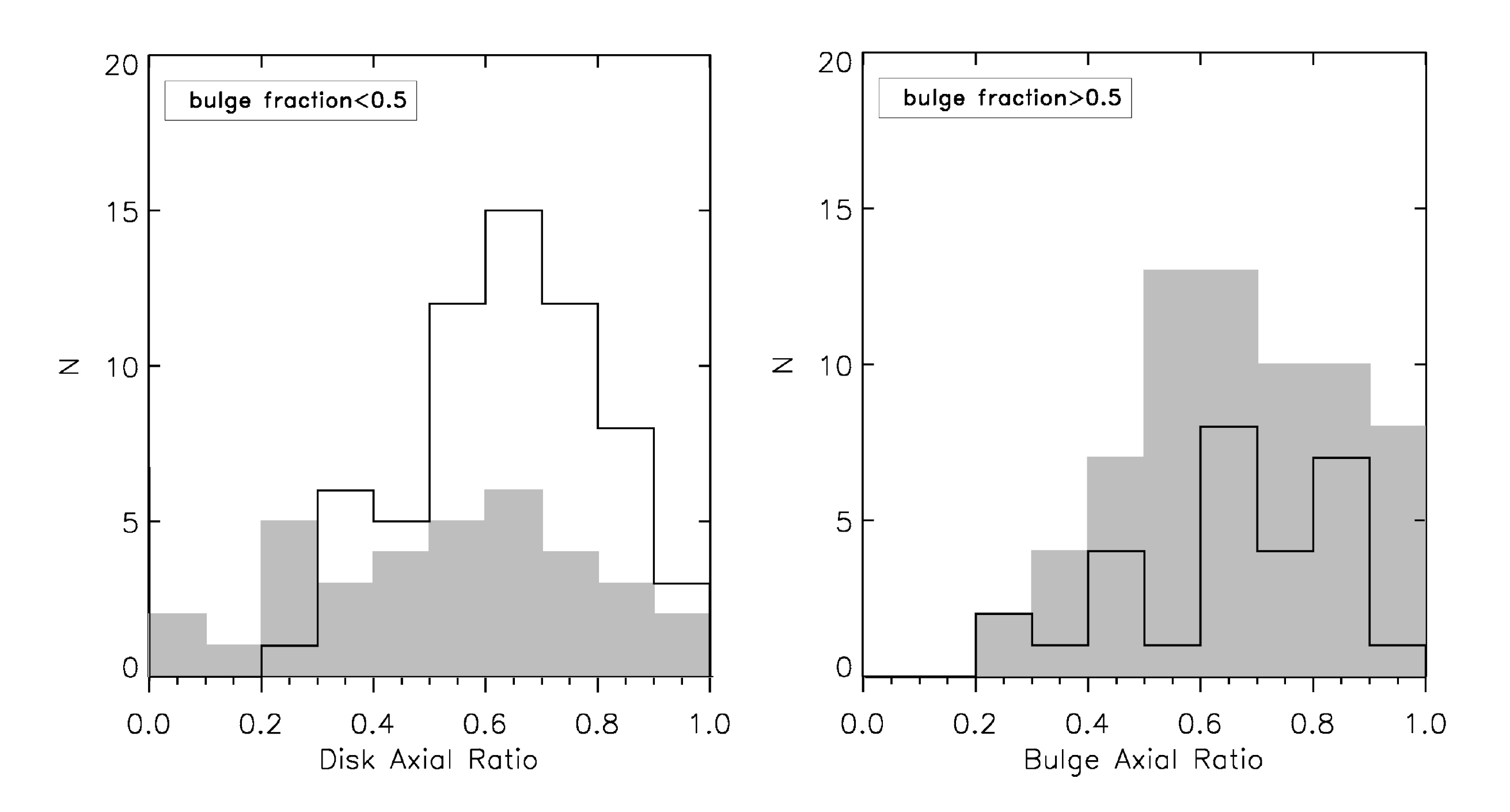}
 \caption{Axial-ratio distributions displayed by the dominant disk components in the disk-dominated galaxies ($B/T < 0.5$; left-hand panel)
and by the dominant bulge components in the bulge-dominated galaxies ($B/T > 0.5$; 
right-hand panel). These sub-samples have been further split into
star-forming objects ($sSFR > 10^{-10}\,{\rm yr^{-1}}$; black outlined histograms) and quiescent objects ($sSFR > 10^{-10}\,{\rm yr^{-1}}$; grey shaded histograms).
Both the star-forming and quiescent bulge populations show similar distributions peaked, as broadly expected, at $b/a \simeq 0.7$. However,
the active and passive disk populations are marginally different, with the passive disks showing a relatively flat distribution as seen for low-redshift disks (see also Fig. 10),
while the star-forming disks display a peaked distribution more comparable to that displayed by the bulges (see text for details and K-S statistics).}
\end{figure*}

\subsection{Axial ratio distributions}

Some additional (and independent) information on the morphologies of the galaxies in our sample can potentially 
be gained from examining the distribution of their axial ratios. In Fig.~9 we have split our 
sample into disk-dominated ($B/T < 0.5$) and bulge-dominated ($B/T > 0.5$) galaxies, and then plotted the axial-ratio 
distributions of the disk components in the disk-dominated galaxies (left-hand panel), and of the bulge components 
in the bulge-dominated galaxies (right-hand panel) (we do this to avoid potential contamination 
of these plots by poorly-constrained axial ratios from weak sub-components; Fig. 9 thus displays the axial ratio distributions
of our more robustly measured disks and bulges). In addition, in each panel we split the sub-samples further into star-forming (black outlined histogram) 
or quiescent (shaded grey histogram) 
objects, as again defined by whether a given galaxy lies above or below our adopted specific star-formation threshold 
$sSFR = 10^{-10}\,{\rm yr^{-1}}$.

From the right-hand panel of Fig. 9 it can be seen that the axial-ratio distributions of the star-forming and quiescent bulges are indistinguishable, both peaking 
around $b/a \simeq 0.7$ (the K-S test yields $p=0.71$ for the null hypothesis that they are drawn from the same distribution).
This result is consistent with previous studies of bulge-dominated objects, both at low \citep{Padilla2008} and high redshifts \citep{Ravindranath2006}.

Perhaps of more interest are the axial distributions of the disk components as plotted in the left-hand panel of Fig.~9. 
Here the two distributions look markedly different (although the statistical significance of the difference is marginal; $p=0.09$). Specifically,
it appears that the passive disks display a fairly flat distribution (as expected for a set of randomly-oriented thin disks) 
whereas the star-forming disks display a significantly more peaked distribution, in fact indistinguishable from the 
axial-ratio distributions displayed by the bulges. 

The flat axial-ratio distribution found for the passive disk-dominated galaxies 
lends some additional support to our conclusion that we have uncovered a 
genuine population of passive disk-dominated galaxies, but the peaked 
distribution of the star-forming disks might be viewed as surprising. However, these results 
agree well with other recent studies of star-forming disk-dominated galaxies at comparable redshifts, as we illustrate in Fig.~10. 
The left-hand panel of Fig.~10 shows again the axial-ratio distribution of our star-forming disks (simply taken from the left-hand panel of Fig.~9), 
but this time over-plotted with results from \citet{Law2012}, who utilised a larger sample of galaxies at $z\sim1.5-3.6$, but plot only the 
single-S\'{e}rsic model 
axis ratios of $n \simeq 1$ galaxies. It can be seen that the two distributions are in good agreement, both peaked around $b/a \simeq 0.6 - 0.7$, and displaying a deficit
of objects with $b/a < 0.3$; these results are also consistent with those obtained by
\citet{Ravindranath2006} who used HST ACS optical data to model the rest-frame UV morphologies of galaxies at $z \sim 3-4$, and with \citet{Yuma2011} who conduct a similar analysis at $z \simeq 2$.
The implications of these peaked axial-ratio distributions are discussed further in Section 8.4.

\begin{figure*}
\includegraphics[scale=0.5]{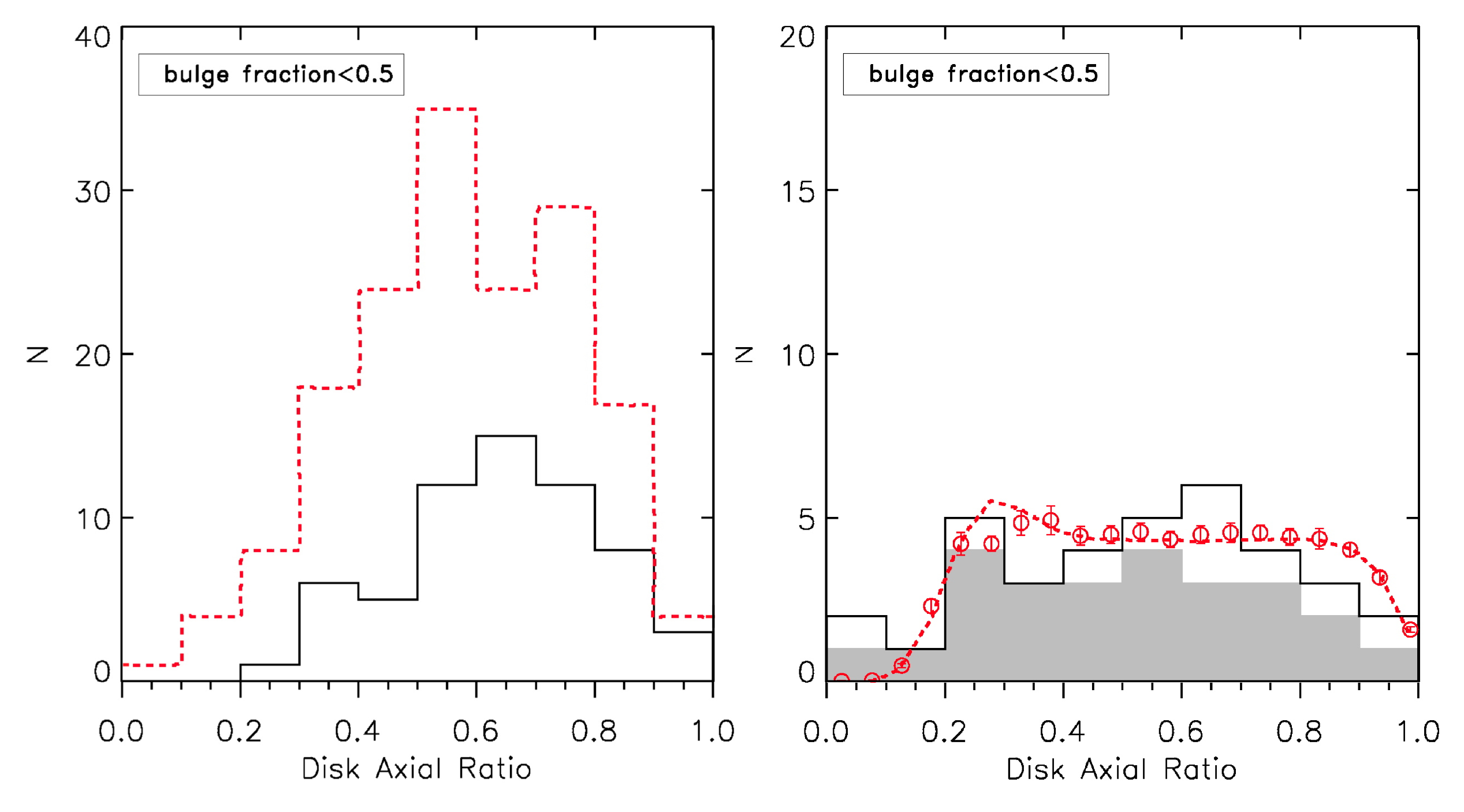}
 \caption{Comparison of our disk-galaxy axial-ratio distributions with other relevant recent results from the literature. In the left-hand panel
we again plot the axial-ratio distribution of our {\it star-forming} disks (in black solid outline), but also over-plot (in the red/dashed histogram)
the axial-ratio distribution of $n \simeq 1$,  $z \simeq 1.5-3.6$ star-forming disk-galaxies from \citet{Law2012}; the two distributions 
are indistinguishable. In the right-hand panel we plot (in black solid outline) the axial-ratio distribution of our {\it passive} disks as judged 
by $sSFR = 10^{-10}\,{\rm yr^{-1}}$ from SED fitting and over-plot (in grey shaded regions) the corresponding distribution after excluding 
the apparently-passive disks which appear to have  24\,$\mu$m counterparts. These are compared with local results
in the form of the best-fit model axial-ratio distribution (red dashed line) and the actual measured distribution of axial ratios from a fitted single-component model 
(red points with corresponding error-bars) of local SDSS spiral galaxies from \citet{Padilla2008} 
(where here we plot their normalised frequency 
scaled appropriately for direct comparison with our results). This comparison illustrates that the relatively flat axial-ratio distribution 
displayed by our sample of passive disks at $1 < z < 3$ is consistent with results from local disks, 
whilst the peaked distribution of star-forming galaxies is in good agreement with previous studies of similar galaxies conducted at $z \simeq 2$.}
\end{figure*}

Finally, in the right-hand panel of Fig.~10 we confirm that the axial-ratio distribution 
displayed by our {\it passive} disks at $1<z<3$ is indeed consistent with that displayed
by the disk-galaxy population at low redshift as deduced from the SDSS. Tha axial-ratio distribution for our passive disks is shown
here both with and without inclusion of the 24\,$\mu$m-detected objects, to demonstrate that its shape is unchanged by this 
extra level of caution in excluding potential star-forming objects.
These histograms have been over-plotted with the data points and best-fit model from \citet{Padilla2008}; their 
normalised frequencies have simply been re-scaled here by the area under our 
solid histogram to ease direct comparison with our results. 
As can be seen, our distribution agrees well with the relatively 
flat distribution displayed by present-day disk-dominated galaxies. 
We also compared our results with the 
axial-ratio distribution presented by \citet{vanderWel2011} for a sample of 14 $z\simeq 2$ disk-dominated passive galaxies,
and found them to be consistent, although the statistics are weak given the limited size of both samples ($p=0.15$).

\vspace*{0.5cm}

\section{Discussion}

We now discuss the implications of our results in the context of other recent studies of massive 
galaxies at comparable redshifts, and current models of galaxy formation and evolution.

\subsection{Galaxy Growth}

Based on a complete, mass-selected sample of $\simeq 200$ galaxies
with $M_* > 10^{11}\,{\rm M_{\odot}}$, our HST WFC3/IR study
provides the most detailed information to date on the sizes of the
most massive galaxies at $1 < z < 3$. Considering the sample as a whole,
our most basic statistical measurement is that the median size of these
galaxies is 2.6\,kpc, a factor 2.25 smaller than the size of
comparably-massive galaxies today.
Splitting the sample into $z < 2$ and
$z > 2$ subsamples yields a gentle trend with redshift, with the median
descending from 2.73\,kpc at $1 < z < 2$, to 2.53\,kpc at $2 < z < 3$,
corresponding to factors of 2.18 and 2.27 below the local size-mass relation.
These figures are somewhat ($\simeq 20$\%) smaller than the
results reported for a comparable mass-selected sample of
galaxies at comparable redshifts by \citet{vanDokkum2010}
($R_e \simeq 4 \pm 0.3$ at $z \simeq 1.6$, $R_e \simeq 3 \pm 0.3$
at $z \simeq 2.0$), but their results were based on stacks of
ground-based images taken in 1.1 arcsec seeing, and are thus
superseded by the results presented here.

However, these basic statistics conceal a number of potentially important
details. First, the scatter in size is large, spanning $\simeq 1$\,dex
(see Fig. 5b) and, due to our relatively small errors on $R_e$ ($<10$\%;
e.g. Fig. 1)
and our exploration of systematic effects (e.g. Fig. 2), we can say with
confidence that this scatter is real. Our analysis reveals that massive
galaxies display half-light radii which range from $R_e \simeq 8$\,kpc, fully
consistent with comparably massive local galaxies, to $R_e \simeq 1$\,kpc,
consistent with the very small sizes previously reported for the most extreme
examples of compact galaxies at these redshifts (e.g. \citealt{Kriek2009}).

Second, when our galaxies are split into their bulge and disk components,
it is clear that the bulges display more rapid evolution to small sizes, both
in terms of median size, and in terms of the relative numbers of objects
which lie on and below the present-day size-mass relation. For the disks,
we find that, throughout our redshift range, $\simeq 40$\% lie on the local
relation, with $\simeq 60$\% below, while for the bulges the percentage of
objects which lie significantly below the local relation rises from an already
high 80\% at $1 < z < 2$ to 85\% at $2 < z < 3$. Clearly bulges consistent
with the local size-mass relation are rare at these redshifts and, moreover,
the compact bulge population appears to become increasingly compact
with increasing lookback time,
lying a factor $\simeq 3.5$ below the local relation at
$1 < z < 2$ but a factor $> 4$ below at at $2 < z < 3$ (the corresponding
figures for the subset of compact disks are more modest, 2.43 and 2.55
respectively). Here, our results for bulges match very well those recently
reported by \citet{Szomoru2012}, who used the CANDELS
imaging in GOODS-South to deduce that quiescent
galaxies at $1.5 < z < 2.5$ with median S\'{e}rsic indices $n \simeq 3.7$
lie a factor of $\simeq 4$ in size below the local size-mass relation.
A related issue is the morphological mix of the objects selected as compact.
For example, it has recently been suggested by \citet{vanderWel2011} (albeit based on a sample
 of only 14
objects) that the ``majority'' of compact galaxies at $z \simeq 2$ are disk-dominated.
Fig. 6 illustrates that such a statement is not straightforward, as it depends on
what one defines as compact and what mass range is to be considered.
Certainly it is clear from Fig. 6 that the most massive and compact objects
(i.e. the galaxies with $M_* > 2 \times 10^{11}\,{\rm M_{\odot}}$, and $R_e < 3$\,kpc)
are {\it all} bulge-dominated, but at more moderate masses the situation
is certainly more mixed. A direct comparison is limited by the somewhat complex
mix of criteria used by \citet{vanderWel2011} to classify an object
as disk-dominated (as compared to our straightforward use of $B/T < 0.5$) but
clearly Fig. 6 does reveal a substantial population of compact disks as
quantified above, and we confirm that essentially all the {\it passive}
disks are comparably compact to their spheroidal counterparts.

Third, while our sample is clearly somewhat limited in terms of dynamic
range in stellar mass, we find evidence for a lower envelope in size which
tracks the slope of the present-day size-mass relation. This trend is
strengthened by the results of our bulge+disk decomposition, which extends
the size mass relation down to estimated sub-component masses
$M_* \simeq 2 \times 10^{10}\,{\rm M_{\odot}}$. Thus,
for $M_* > 2 \times 10^{11}\,{\rm M_{\odot}}$ we find no objects significantly
smaller than $R_e \simeq 1$\,kpc,
while at $M_* < 1 \times 10^{11}\,{\rm M_{\odot}}$ we
start to see examples of even smaller bulges and disks, with some bulges as
small as $R_e \simeq 0.4$\,kpc. These details, including
the trend of minimum size with stellar mass are
important when comparing with previous studies; for example,
\citet{Szomoru2010} have reported a very small scalelength
of $R_e = 0.42 \pm 0.14$\,kpc from WFC3/IR imaging of a compact
bulge-dominated galaxy at $z = 1.91$, but with an estimated stellar mass
of $M_* \simeq 5 \times 10^{10}\,{\rm M_{\odot}}$ \citep{Wuyts2008}, it
is clear that this object lands perfectly on the lower envelope
of the size-mass relation displayed by our bulge components in Fig.~6.
The single object studied by \citet{Szomoru2010}
was the most massive, quiescent $z \simeq 2$ galaxy available for study
in the Hubble Ultra Deep Field. A comparably-detailed study of the
brightest galaxy at $z > 1.5$ in the ten-times-larger ERS field by
\citet{vanDokkuma2010} again yielded a S\'{e}rsic index $n \simeq 4$,
but this time an effective radius $R_e \simeq 2.1 \pm 0.3$\,kpc and a much
larger galaxy mass $M_* \simeq 4 \times 10^{11}\,{\rm M_{\odot}}$; again,
comparison with the results shown in Fig. 6 shows that this is perfectly
consistent with the size-mass locus for bulges uncovered here.
We also note that within Fig. 6 we see no real evidence in support
of the claim advanced by \citet{Ryan2012} that the required size growth of
galaxies from $z \simeq 1.5$ to the present is a strong function of stellar
mass. A direct comparison is difficult because our extension
to lower masses is primarily based on bulge+disk decomposition, but we note
here that  \citet{Cimatti2012} also find no evidence
for any stellar mass dependence in the redshift growth-rate of
early-type galaxies.

Fourth, it is also clear that the objects which remain on the local relation,
even out to the highest redshifts, are star-forming disks, with the passive
galaxies, including the passive disk components, confined to the more compact
population. This result mirrors that recently reported by McLure et al. (2012)
who found, for spectroscopically-confirmed
galaxies of comparably high mass at $z \simeq 1.4$, that all objects with
low $sSFR$ (i.e. $sSFR < 10^{-10}\,{\rm Gyr^{-1}}$)
lie below the present-day size-mass relation, irrespective
of morphological classification. At $z \simeq 2.3$ a comparable trend
for star-forming objects to be $2 - 3$ times larger then their
quiescent counterparts has been reported by \citet{Kriek2009}
for a sample of 28 galaxies with $M_* \simeq 3 \times 10^{10}\,{\rm M_{\odot}}$,
a result confirmed as extending to even lower masses
by \citet{Szomoru2011}, who also found star-forming
galaxies at $z \simeq 2$ to be larger than their quiescent counterparts
in the mass range $M_* \simeq 1 - 10 \times 10^{10}\,{\rm M_{\odot}}$.

In summary, our results confirm and clarify a number of trends in the galaxy
size-mass relation previously reported from detailed studies of
small numbers of objects with HST,
or larger samples studied via ground-based imaging. Within the high-mass
regime our study provides significantly improved statistics on the
scatter in size, and how the size-mass relation evolves
differently for bulges and disks
in the redshift range $1 < z < 3$. Our bulge+disk decomposition
is the most extensive attempted to date, and
suggests that these trends extend to the bulge components of disk-dominated galaxies, and
to the disk components of bulge-dominated galaxies. We also provide the first clear evidence
for a lower envelope in size
which our bulge+disk decomposition suggests
extends from our high-mass sample down to lower masses
($M_* \simeq 2 \times 10^{10}\,{\rm M_{\odot}}$),
tracking the slope of the present-day size-mass relation.

Many authors have discussed the theoretical challenge of
explaining the growth in the size of massive galaxies from $z \simeq 2$ to the present.
Various arguments, based on $\Lambda$CDM simulations, clustering analyses
(e.g. \citealt{Quadri2007, Hartley2010}) and
simple comoving number density comparisons (e.g. \citealt{vanDokkum2010}) indicate that the $M_* \simeq 10^{11}\,{\rm M_{\odot}}$
galaxies studied here at $1 < z < 3$ must evolve into galaxies with stellar masses
$M_* \simeq 3 \times 10^{11}\,{\rm M_{\odot}}$ which are
essentially
all giant elliptical galaxies on the high-mass end of the local
early-type size-mass relation plotted in red in Fig. 6. The issue of
what happens to the disk components so evident in the high-redshift population (but essentially absent in the present-day descendants)
is discussed further below. But in terms of size evolution, the challenge
is to explain how such compact massive galaxies (especially the extremely
compact bulges at $z > 2$ which lie a factor $\simeq 4$ below the present-day relation) can evolve onto the present-day size-mass relation without
simultaneously attaining excessively high masses which violate constraints
imposed by the measured present-day mass function \citep{Baldry2012}.

As pointed out by various authors (e.g. McLure et al. 2012),
major mergers do not provide a sufficiently vertical evolutionary track
on the size-mass plane to lift the compact high-redshift galaxies
onto the present-day relation without yielding excessively high masses.
In any case, size growth driven primarily by major mergers would require
many more major mergers since $z \simeq 2$ than appears plausible from
N-body simulations (which suggest $< 2$ per massive galaxy by the present day;
e.g. \citealt{Hopkins2010}), or indeed from observed merger rates
(e.g. \citealt{Robaina2010}).

Thus while the rare major mergers may be
responsible for the relaxation process which at some stage destroys the disk
component (although a series of minor mergers may also achieve this;
\citealt{Naab1999,Bournaud2007})
it appears that the bulk of the size growth must be attributed
to minor mergers which are much more effective at adding stars and dark
matter in the outer regions of galaxies, increasing observed size with
relatively limited increase in stellar mass.
It is also
worth noting that minor mergers are more effective than
major mergers at raising the dark-matter
to stellar mass ratio to the levels observed for the most massive galaxies
today, are better able to add mass
while leaving the age and metallicity gradients
in the central regions of massive galaxies unscrambled, and may provide
a natural explanation for the kinematically decoupled
cores frequently observed in present-day ellipticals
(e.g. \citealt{vandenBosch2008})

As illustrated
by McLure et al. (2012), a combination of five minor (mass ratio 1:10) mergers
and a single major merger (mass ratio 1:3) appears sufficient to achieve
the required evolution since $z \simeq 1.4$. Recent
simulations analysed by \citet{Oser2012} also support the idea that
minor mergers can produce the required size evolution at $z < 2$. However,
whether this sort of
evolutionary path can also solve the problem for the most compact spheroids
at $z > 2$ is still a matter of some debate \citep{Newman2012,
Cimatti2012}.

Finally, it is worth emphasizing that despite the ongoing debate of how
such compact high-redshift galaxies can climb onto the present-day size-mass
relation, the existence of such compact objects at early times, while perhaps
initially unexpected, is in fact a natural prediction of modern
galaxy-formation simulations (e.g. \citealt{Khochfar2006,
Obreschkow2009,Wuyts2010}).

\begin{figure*}
\includegraphics[scale=0.5]{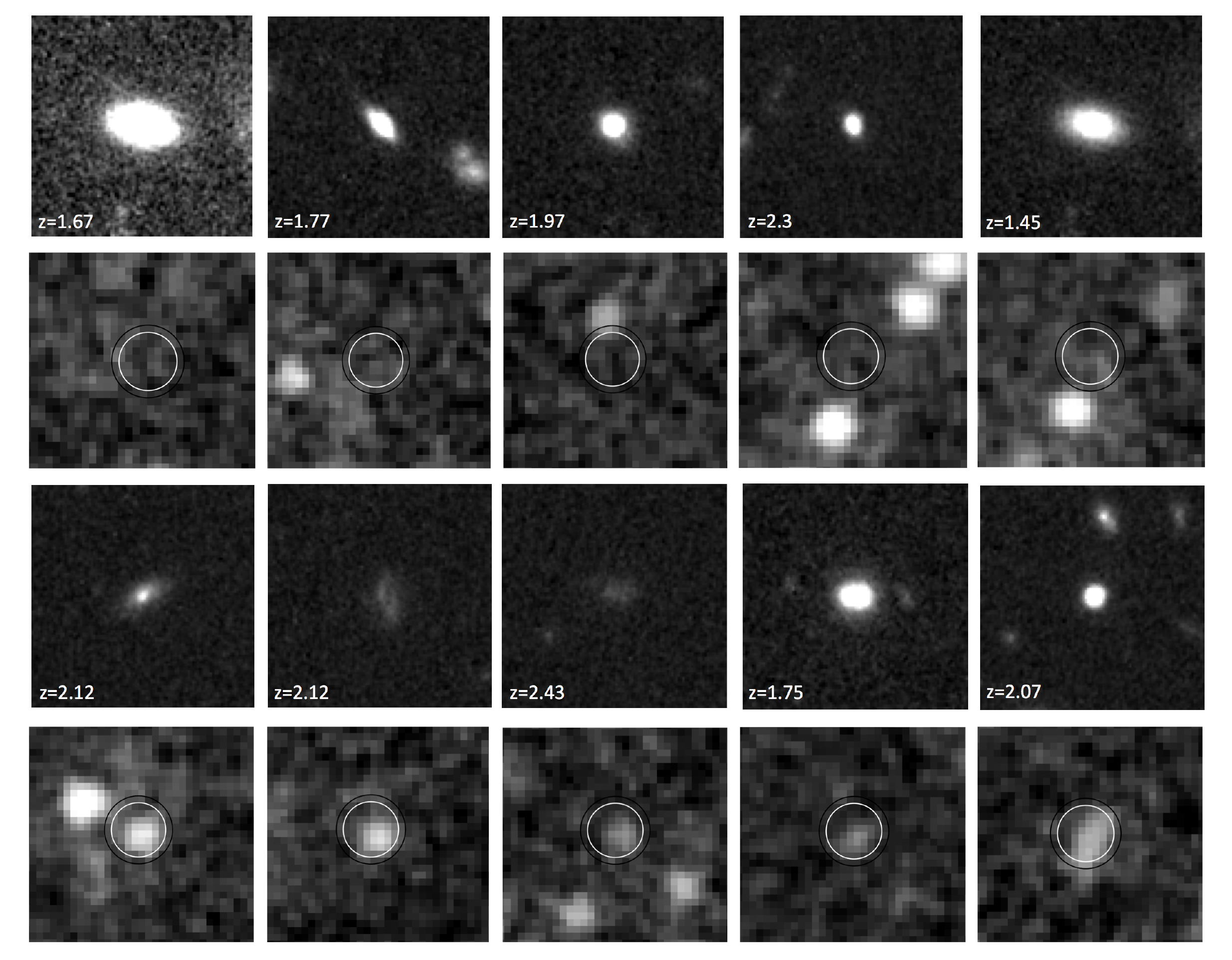}
 \caption{The WFC3/IR $H_{160}$ and {\it Spitzer} 24\,$\mu$m images of the ten apparently bulgeless pure-disk objects
in our sample which the optical--near-infrared SED fitting suggests are passive (i.e. $sSFR < 10^{-10}\,{\rm yr^{-1}}$).
The top row shows $6 \times 6$\,arcsec images of the five objects which have no significant 24\,$\mu$m counterpart, as shown 
in the $20 \times 20$\,arcsec MIPS image stamps in the second row (the circle indicates a 5 arcsec radius aperture, which is a very 
generous search radius). The third and fourth rows show the same information for the five objects which do have 24\,$\mu$m counterparts.}
\end{figure*}

\subsection{Morphological Evolution}

As the bulge components decline in size with increasing redshift, we also
find a clear trend for the massive galaxies in our sample to become
increasingly disk-dominated. As shown in Fig. 7, $z \simeq 2$ appears to
mark a morphological transition epoch, at least for our
chosen galaxy mass range; crudely speaking, the majority of our galaxies are
bulge-dominated ($B/T > 0.5$)
at $z < 2$, while the situation is reversed at $z > 2$. Moreover, at the
highest redshifts ($z \simeq 2.5$), over half the galaxies have $B/T < 0.3$
and over half of these (i.e. $\simeq 35$\% of all objects in the
relevant redshift range) are ``pure disks'' as judged by $B/T < 0.1$ (which we
cannot distinguish from $B/T = 0$). Such highly disk-dominated objects are
virtually absent in our high-mass sample by $z \simeq 1.5$, although it
is still true that the vast majority of objects contain some
detectable disk component,
with ``pure de Vaucouleurs bulges'' (i.e. $B/T > 0.9$)
still largely absent until $z < 1$.

The relative lack of pure de Vaucouleurs bulges at $z > 1$ appears broadly
consistent with the findings of \citet{Buitrago2011} who reported
that ellipticals have been the dominant morphological class for massive
galaxies only since $z \sim 1$, although a direct
comparison of our results is difficult as \citet{Buitrago2011}
did not attempt bulge+disk decomposition and relied on a combination
of single S\'{e}rsic fitting and visual classification.

The presence of a significant fraction
of disk-dominated objects, even among the apparently passive subsample,
has already been reported at $z \simeq 1.5$ for masses
$M_* > 10^{11}\,{\rm M_{\odot}}$ by McLure et al. (2012) ($44 \pm 12$\%)
and at $z \simeq 2$ for masses $M_* > 6 \times 10^{10}\,{\rm M_{\odot}}$ by
\citet{vanderWel2011} ($40 - 65$\%).  However, these studies
do not extend to high enough redshift to capture
the full extent to which disk-dominated galaxies, primarily star-forming,
come to dominate the massive galaxy population at $z > 2$ as illustrated
in Figs 7 and 8.

Given the axial ratio distributions plotted in  Figs 9 and 10, it might
be argued that, while the more passive disks may indeed be disks, the
star-forming disk-like objects might be more tri-axial in nature, given
their more peaked (i.e. typical rounder) axial ratios. However, as discussed
further in the next subsection, visual inspection of both the
active and passive disk dominated objects supports the view
that they are indeed
disks; the only mystery is the lack of any very thin edge-on disks
in the star-forming population which we return to at the end.

It is worth again bearing in mind that virtually all the objects in this study
are destined to evolve into today's very
massive $M_* > 3 \times 10^{11}\,{\rm M_{\odot}}$
giant elliptical galaxies
which display, at most, very low-level disk components.
This alone means it may be
naive to expect the properties of many of these disks to correspond closely
to those of $M_* \simeq 1 \times 10^{11}\,{\rm M_{\odot}}$ disk galaxies
in the present-day Universe. Indeed it has been argued that the stellar
densities of these high-redshift massive disks are comparable to those
found in the cores of massive present-day bulges \citep{Bezanson2009, vanDokkum2010},
 consistent with the inside-out
model of massive galaxy growth discussed above.

\subsection{Star-forming and Quiescent Galaxies}

\begin{figure*}
\includegraphics[scale=1.15]{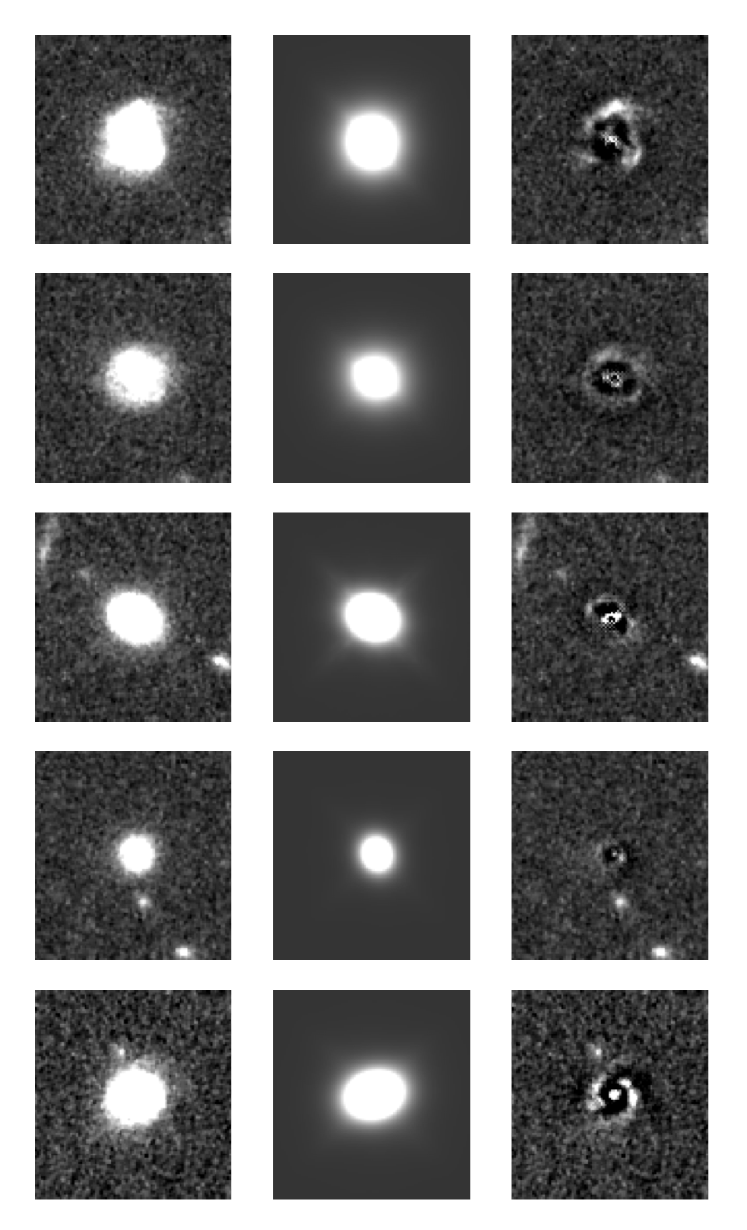}
 \caption{$H_{160}$ images (left), models (centre) and model-data residual (right) (all $6 \times 6$\,arcsec) 
for a subset of five of our star-forming ($sSFR > 10^{-10}\,{\rm yr^{-1}}$) disk-dominated ($B/T < 0.5$) galaxies. 
The five galaxies shown here have been chosen to have single S\'{e}rsic indices in the range $\simeq 0.9-1.1$ 
thus demonstrating that, despite the bulge-like axial ratio distributions for our sample of star-forming galaxies 
(as discussed in Section 7.4), the galaxies with single S\'{e}rsic index consistent with 
traditional disk-like ($n=1$) values show clear face-on disk morphologies, 
and are not especially disturbed systems. In addition our residual image stamps highlight the clumpy structure within 
these disks, as expected for violently star-forming disks at high redshift (see Section 8.3).}
\end{figure*}

This paper is deliberately focussed on $H_{160}$ morphologies, with a detailed
analysis of the colours of the bulge and disk components deferred to a future
paper. Nevertheless, as explained in Section 7.3, the SED fitting undertaken
to deduce the photometric redshifts also yielded dust-corrected star-formation
rates and stellar masses, from which we can derive an estimate of
$sSFR$ for each galaxy in our sample. As illustrated in Fig. 8, we have
then followed \citet{Bell2011} by also searching for 24\,$\mu$m detections
to try to ensure against misinterpreting dust-reddened star-forming galaxies
as quiescent objects. In general the results of this latter test are reassuring,
with the vast majority of star-forming objects (defined as $sSFR > 10^{-10}\,{\rm                   
yr^{-1}}$) yielding 24\,$\mu$m detections, as compared to relatively few
of the objects with UV $sSFR > 10^{-10}\,{\rm                                                       
yr^{-1}}$ being detected by the SpUDS MIPS imaging. As already summarized in
Section 7.3, the vast majority of the disk-dominated
galaxies are star-forming,
while the majority of the bulge-dominated objects are quiescent, but yet
our sample contains a significant number of star-forming bulges
and a significant number of ``quiescent'' disks; $25-50$\% of the passive
subsample are disk-dominated, depending on whether
one splits by S\'{e}rsic index or $B/T$, and on
whether the few 24\,$\mu$m detections of the supposedly passive objects
are deemed symptomatic of star-formation or buried AGN.

Thus, to first order, our results show that the well-documented bimodality in the
colour-morphology plane seen at low redshift, where spheroidal
galaxies inhabit the red sequence, while disk galaxies occupy the blue cloud
\citep{Baldry2004, Driver2006, Drory2007} is at least partly already in place by $z \simeq 2$.
However, the colour-morphology division
is undoubtedly much less clean than in the nearby Universe, and a key challenge
is to determine the prevalence and physical significance of the
passive disks and the active bulges.

Recent studies have produced apparently conflicting results over the
prevalence or otherwise of massive passive disks at these redshifts. Specifically,
while \citet{vanderWel2011} and McLure et al. (2012) both conclude that
$\simeq 50$\% of passive objects at these redshifts are disk dominated,
\citet{Bell2011} find that the key parameter which correlates best with
quiescence at these redshifts is still S\'{e}rsic index, with the presence
of a substantial bulge a necessary (but not necessarily sufficient) condition
for the termination of star-formation activity. This confusion may be partly a matter
of definition; it is not clear what a ``substantial'' bulge component means, or
how comparable the morphological criteria applied in these studies really are.
Nevertheless, given the controversy over this issue, and its potential
importance, we have carefully revisited the passive disk-dominated objects
in our sample, motivated in part by the fact that five of the ten apparently
passive ``pure disks'' (i.e. $B/T < 0.1$)  originally
isolated on the basis of optical/near-infrared photometry
in Fig. 8 transpired to have 24\,$\mu$m detections.

In Fig. 11 we show the $H_{160}$ image stamps for these ten interesting objects,
along with their 24\,$\mu$m MIPS imaging. The 24\,$\mu$m detections
of the five MIPS catalogue-matched objects (shown in the bottom row) are clear,
but equally clear is the fact the the top five objects do not possess even marginal
mid-infrared detections at the depth of the SpUDS imaging. We note
that the 24\,$\mu$m-detected objects in the bottom row of Fig. 11
have fluxes which,
if interpreted as arising from
star formation, imply typical values of
$sSFR \simeq 10^{-9}\,{\rm yr^{-1}}$, and that
the SpUDS MIPS detection limit conveniently
corresponds rather closely to the adopted passive/active $sSFR$ threshold
of $sSFR \simeq                                                                                    
10^{-10}\,{\rm yr^{-1}}$ (for galaxies in this redshift and mass range).
Thus, since we have no real reason to assign the MIPS detections to AGN activity
(other than the fact that several of these objects prefer a small contribution
from a point-source rather than a resolved
bulge in the multi-component $H_{160}$ modelling) we have taken a
conservative approach, and
have classified the lower five objects in Fig. 11 as star-forming, which reduces the number
of passive ``pure'' disks by half, to five. This represents less than 15\% of the
``pure disk'' sample, and so clearly the vast majority of
{\it apparently bulgeless} disks are actively star-forming
galaxies on the main-sequence. Nevertheless, this still means that
a substantial fraction of the passive galaxy subsample ($25-40$\%) is
disk-dominated, and it is as yet unclear whether the relative rarity
of completely
bulgeless quiescent disks reflects an important causal link between
bulge growth and passivity at these redshifts, or is simply an inevitable
symptom of the dimming of star-forming disks as star-formation activity dries up
(for whatever reason). These issues, and the prospects
for further progress, are discussed further in Section 8.4.

Moving now to consider the active disks, we attempt to investigate a little
further the apparent contradiction between the results of our S\'{e}rsic fitting
and the axial-ratio distribution displayed by these supposedly disk-like
star-forming objects. As already mention in Section 7.4 (and see Figs 9 and 10)
while the axial distribution for the
passive disk components is as flat as that
displayed by low-redshift disk galaxies, that displayed by the star-forming
``disks'' does not extend to such low values, and peaks at $b/a \simeq 0.7$.
This is essentially identical to the distribution found by \citet{Law2012},
who also commented that such an axial-ratio distribution was more in line with that expected
from a population of tri-axial objects.

We have therefore tried to check whether our active disk-dominated objects
do indeed look like star-forming disks. This is somewhat against
the spirit of our analysis which seeks to deliberately avoid the pitfalls
of visual classification. Nevertheless, image inspection can still offer an
interesting sanity check on the interpretation of modelling results.
In Fig. 12, we therefore show, for illustrative purposes, the images,
model-fits, and residual data-model images of the five star-forming
galaxies which we find to have S\'{e}rsic indices closest to unity (in practice,
$n \simeq 0.9 - 1.1$). By (possible) coincidence all five of these objects
are in fact fairly round, but it is visually obvious that they are not spheroidal galaxies, but
rather face-on disks with spiral arms and/or star-forming clumps. We are thus left to conclude
that we have no reason to really doubt the disk-like nature of these objects just because
of their axial ratio distribution. Perhaps it is simply the case that very few of
the (violently) star-forming disks at these epochs
are genuinely thin enough to display low axial ratios, or alternatively such disks may be so
dusty that near edge-on examples have in fact evaded our detection limit (this might
seem unlikely, but see \citealt{Targett2012}).

A full review of the already extensive observational and theoretical literature on the
nature and importance of clumps in star-forming disk galaxies at $z \simeq 2$ is beyond the scope
of this paper. Suffice
to say that, given the above-mentioned lack of evidence for major mergers being the primary
driver of elliptical galaxy evolution, it has now been suggested that the progenitors
of todays giant ellipticals are these high velocity dispersion, clumpy disks, in which
star formation is fed by cold streams and minor mergers (e.g. \citealt{Dekel2009, Ceverino2010,Ceverino2012})
 with the clumps eventually coalescing to form a spheroid. However this view
of the potential importance of the observed clumps in building bulges
has been challenged observationally (e.g.
\citealt{Wuyts2012}) and theoretically (e.g. \citealt{Genel2012}). Nevertheless, whether or not the
clumps are the direct ancestors of bulges, what is clear from our study is that the majority
of progenitors of todays most massive elliptical galaxies are indeed, at least at $2 < z < 3$,
clumpy, and fairly extended, star-forming disk galaxies (a result reinforced by
the properties of the extreme star-forming galaxies as deduced from the CANDELS imaging
of sub-millimetre galaxies by \citealt{Targett2012}).

Finally, we note that the presence of at least some star-forming spheroids in our $1 < z < 3$ 
sample
is unsurprising. Various authors have observed this before at comparable redshifts, 
including \citet{Bell2011} who,
while arguing that bulge formation was a potentially necessary condition for the quenching 
of star-formation, also concluded that it was not sufficient to ensure this, given the presence 
of star-forming galaxies in their sample with $n > 2.5$ (although see also \citealt{Wang2012}).

\begin{figure*}
\includegraphics[scale=0.71]{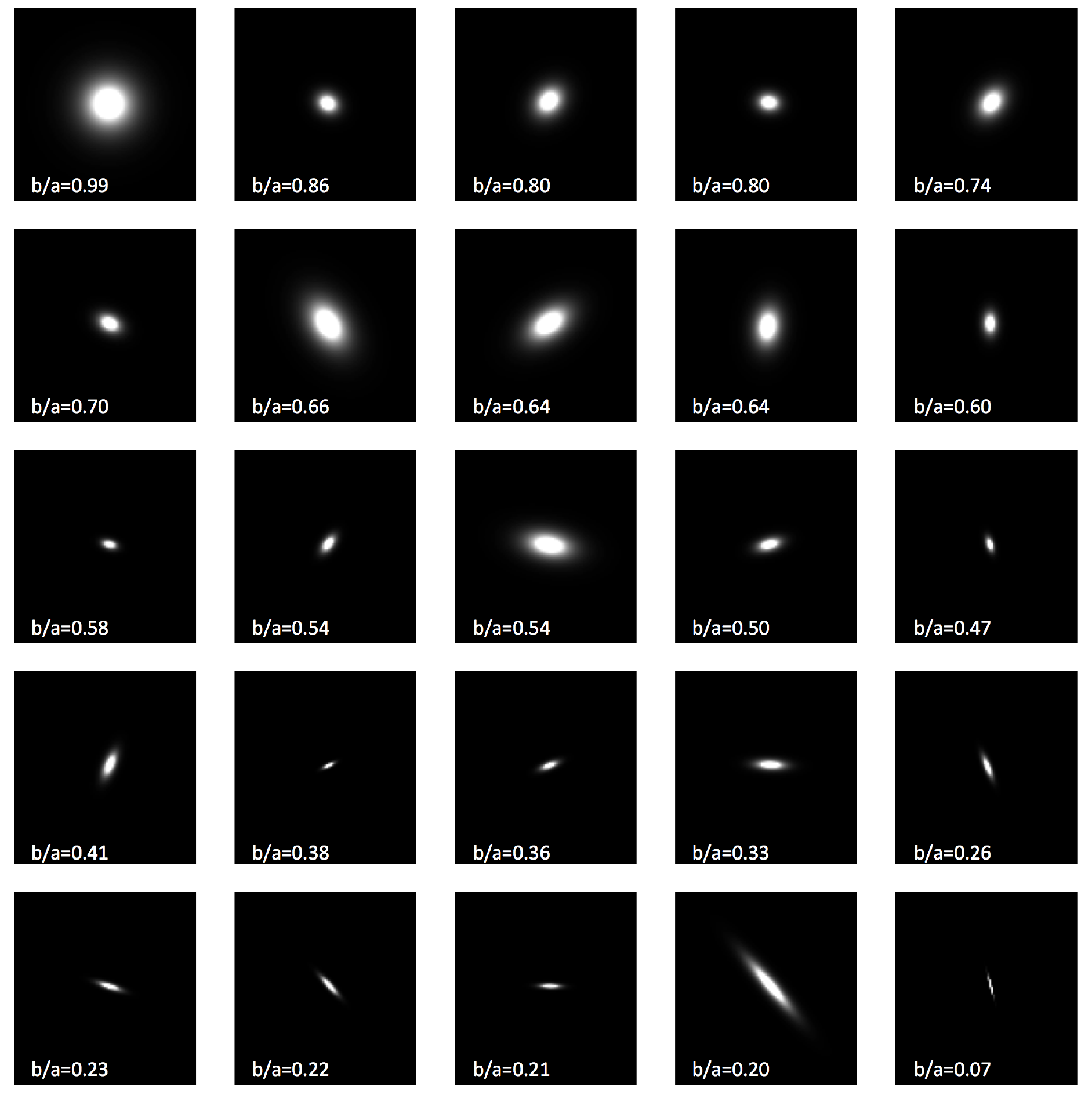}
 \caption{The model disk components of the 
25 disk-dominated ($B/T < 0.5$) galaxies within our sample which show no evidence for star-formation 
from either SED fitting ($sSFR < 10^{-10}\,{\rm yr^{-1}}$) or 24\,$\mu$m counterparts. 
The models have been constructed from the best-fit disk parameters from our 
double-component analysis and have been convolved with a model PSF generated from a Gaussian of FWHM=0.05\,arcsec, 
providing artificial imaging comparable to that achievable by HST at the bluest optical wavelengths.
Each stamp is again $6 \times 6$\,arcsec in size, and the grey-scale for these images is 
set at black=0 and white=1/3 of the maximum pixel value of each image, so as to provide 
consistent brightness cuts for each stamp at an appropriate level. 
The models have been ranked by descending axial ratio from the top left to the bottom right
(the value of axial ratio, $b/a$ is given in the corner of each stamp). These are the 25 models
which were used to produce the axial-ratio distribution of passive disks shown in Fig. 10 (in the grey histogram 
of Fig. 10b), and this illustration 
shows that there is no reason to doubt that they are genuine disks (i.e. no disk displays an unreasonable 
scalelength, and disks covering the full range of fitted sizes are apparently visible over the full 
range of inclination angles). This provides further evidence of the genuine 
disk-like morphologies of these passive systems, 
the implications of which are discussed in Sections 8.3 and 8.4.}
\end{figure*}

\subsection{Passive disks and quenching}

We conclude this discussion by exploring further the nature of the apparently passive disk-dominated 
objects in our sample, and considering briefly the potential implications for the 
the connection, if any, between termination of star-formation activity and morphological 
transformation.

As already noted, the axial-ratio distributions presented in Figs 9 and 10 suggest that the
passive disks in our sample have similar intrinsic shapes to low-redshift disks, while,
on average, the star-forming disks do not. As a final check on the nature of the passive disks
we show, in Fig. 13, images of the model disks fitted to all 25 of the confirmed passive
disks in  our sample (i.e. those which also have no 24\,$\mu$m detection). In this plot
the disks are shown at high resolution (i.e. FWHM 0.05\,arcsec) and scaled to comparable
surface brightness levels, making it easier to see the full range of axial ratios found.
This figure demonstrates that the flat axial-ratio distribution of passive disks is not a result
of strange, excessively-elongated or otherwise unphysical disks which GALFIT has attempted to
fit to deal with other peculiarities in the data. In addition, the full range of fitted sizes
can be seen at a wide range of axial ratios (i.e. viewing angles). We thus have no reason to
doubt that these are, as suggested by the S\'{e}rsic and double-component fits, genuine
passive disks.

The presence of a significant population of passive disks among the massive galaxy population
at these redshifts indicates that star-formation activity can cease without a disk galaxy
being turned directly into a disk-free spheroid, as generally previously expected if the process that
quenches star formation is a major merger. Thus, while some fraction of the substantial
population of star-forming disks may indeed suffer a major merger (possibly transforming
rapidly into a compact passive spheroid) our results argue that another process must
exist which is capable of terminating star-formation activity while leaving a substantial disk intact.

One possibility arises from the latest generation of hydrodynamical simulations \citep{Keres2005, Dekel2009a} 
and analytic theories \citep{Birnboim2003, Dekel2006}, which suggest a formation scenario whereby at high redshift 
star-formation in massive disks takes place through inflows of cold gas until the dark-matter halos in which 
the galaxies reside reach a critical mass ($> 10^{12}\,{\rm M_{\odot}}$) below $z=2$. 
At this point the virial temperature of the halos is high enough to prevent efficient cooling 
such that pressure can be built to support a stable extended virial shock, which can be triggered 
by minor mergers. This results in the galaxy residing in a hot medium and below $z=2$ a stable 
shock can also be sustained in the cold streams, which stops cold gas inflowing and 
quenches star-formation, but does not cause any accompanying change in underlying morphology.

The idea that star-formation quenching and morphological transformation are distinct processes is also 
consistent with the empirical description of \citet{Peng2010}, who suggest that, in this high-mass regime, 
the star-formation quenching of galaxies is driven by a process governed by ``mass-quenching'', 
where the rate of star-formation suppression is proportional to the star-formation rate of the galaxy (although 
Peng et al. do not attempt to posit a physical mechanism responsible for this observed relation). 

Another scenario which can account for star-formation quenching, whilst still being consistent with 
the existence of passive disks, is the model of violent disk instabilities 
\citep{Dekel2009, Ceverino2010, Cacciato2012}. This model suggests that, as the disk evolves, there is 
an inflow of mass to the centre of the disk, which gradually builds to form a massive bulge. 
This mass inflow can quench star-formation whilst still retaining a massive disk in a process 
known as ``morphology quenching" \citep{Martig2009}. In addition to this, it also agrees 
with the observed trend in morphologies with redshift observed in this study, i.e. the 
transition from predominantly bulge systems in the local Universe, to the increase in mixed bulge+disk 
morphologies between $1 < z < 2$, and then the dominance of disk-dominated objects beyond $z=2$.

Finally, returning to the data, in considering the possible evolutionary links between the active
and passive disks in our sample, we must remember that there are important observational differences 
between these populations. First, while the
passive disks are not especially compact (median disk-component $r_e = 2.37$\,kpc), they are, on average,
significantly smaller that the active disks (median disk-component $r_e = 4.08$\,kpc).
However, it is not clear that this is a serious problem; Fig. 6 shows a significant
fraction of the active disks are also reasonably compact and, in any case, some scenarios (e.g. the model of morphology quenching
described above) might naturally lead to a disk reducing in size somewhat as star-formation activity
turns off. Second, of course, we still need to explain how the relatively thin disks in the passive population
emerge from a star-forming population which apparently lacks objects with low axial ratios. Again, it is hard to know
if this is a real problem. It seems entirely plausible that a maximally-unstable, violently star-forming disk will settle
down into a significantly flatter configuration once the fuelling source of, and violent feedback from star-formation
activity ceases, but (to our knowledge) this has yet to be convincingly and quantitatively demonstrated by
simulations. There are also still potential issues of selection effects which might mean that edge-on star-forming disks 
are unrepresented in flux-limited optical-UV selected samples (due, possibly, to dust obscuration). Interestingly, the axial-ratio 
distribution presented by \citet{Targett2012} for the extreme population of star-forming disks selected 
via sub-millimetre emission is relatively flat, and statistically indistinguishable from the axial-ratio distribution 
displayed by the passive disks in the current study.

\section{Conclusions}
We have isolated a sample of $\simeq 200$ galaxies in the CANDELS UDS field for which we 
have determined stellar masses $M_* > 10^{11}\,{\rm M_{\odot}}$, and photometric redshifts in the 
range $1 < z < 3$. These objects are relatively bright, being selected from a parent 
sample with $H_{160} < 24.5$ (a factor of 10 brighter than the CANDELS WFC3/IR 5-$\sigma$ 
detection limit of $H_{160} < 27$), and in practice virtually all objects have $H_{160} < 23$ 
(equivalent to 100-$\sigma$ detections). Consequently, we have been able to exploit the exquisite CANDELS imaging
to undertake a detailed analysis of their rest-frame optical morphologies, 
and how these vary as a function of redshift, mass and star-formation rate. 

Crucial to this work is proper control 
of both the random and systematic errors. We have undertaken a detailed study of the 
form of the adopted PSF, constructing and justifying the use of an empirical on-image 
PSF over that produced by the Tiny Tim modelling software. We have also 
explored in detail the effect of errors in background determination 
on both the best-fitting values of, and errors in, the derived physical parameters 
such as S\'{e}rsic index and effective radius. In addition we have 
placed a high premium on the importance of obtaining formally acceptable model fits to as many 
objects as possible, in order to enable realistic error estimation. In the end, via careful 
object-by-object masking, and the use of models ranging from single-S\'{e}rsic fits to 
disk+bulge+point-source combinations, we achieved satisfactory models for $\simeq 95$\% 
of the massive galaxies in our complete $1 < z < 3$ sample.

Armed with the resulting unparalleled, robust morphological information on massive galaxies during this
key epoch in cosmic history, we have been able to reach the following conclusions.

 \begin{enumerate}
 \renewcommand{\theenumi}{(\arabic{enumi})}
 
 \item Our single S\'{e}rsic results indicate that these massive galaxies at $1 < z < 3$ lie both on and below the local size-mass relation, with a median effective 
radius of  $\sim 2.6$\,kpc, a factor of $\simeq 2.25$ smaller than comparably-massive local galaxies. 

\item Our study is the first to attempt bulge+disk decomposition on such a large sample at these redshifts.
 We find that bulges in particular show evidence for a growing bimodality in the size-mass relation 
with increasing redshift; the fraction of bulges consistent with the 
local size-mass relation is $20 \pm 5$\% at $1 < z < 2$, and $15 \pm 9$\% at $2 < z < 3$, while the offset in size of the (dominant) 
compact population from the the local early-type relation is already a factor of 3.5 at $1 < z < 2$ and rises to a factor 4.4 at $2 < z < 3$.
These trends appear to extend to the bulge components we have isolated from the disk-dominated galaxies, and 
we find evidence that the lower envelope of galaxy size is a function of mass which broadly parallels the local relation; no galaxies more 
compact than $R_e = 1$\,kpc are found at masses $M_* > 2 \times 10^{11}\,{\rm M_{\odot}}$, while bulges as small as 
$R_e < 0.5$\, kpc are found at lower stellar masses $M_* \simeq 5 \times 10^{10}\,{\rm M_{\odot}}$.

\item The statistics for disks are less dramatic, with $\simeq 40 \pm 8$\% of disks still consistent with the relevant local size 
mass relation over our full redshift range, and the offset of the compact population from the local late-type relation 
growing gently from a factor 2.43 at $1 < z < 2$, to 2.55 at $2 < z < 3$. We do, however, find that the objects which remain 
consistent with the present-day size mass relation are virtually all active star-forming disks, with the population 
of apparently passive disks confined to the more compact subset.

\item Even within the relatively limited redshift range of our study, we find evidence 
for dramatic changes in the morphologies of massive galaxies with redshift, with $z \simeq 2$ apparently marking a key transition epoch. 
While similarly massive galaxies at low redshift are generally bulge-dominated (and the expected more massive 
$M_* \simeq 3 \times 10^{11}\,{\rm M_{\odot}}$ descendants of our high-redshift galaxies are virtually all 
giant ellipticals today), by a redshift 
of $1 < z < 2$ they are predominantly mixed bulge+disk systems, and by $z > 2$ they are mostly disk-dominated. 
Furthermore, at the lowest redshifts covered by this study while bulge-dominated objects are on the rise,  
pure-bulge galaxies (i.e. objects comparable to present-day giant ellipticals) have yet to emerge in significant numbers, with $> 90$\%
of these high-mass galaxies still retaining a significant disk component.

\item We find that the majority of the disk-dominated galaxies are actively forming stars, although this is also true for many of the 
bulge-dominated systems. Interestingly, however, while most of the quiescent galaxies are bulge-dominated (indicating early
emergence of the red sequence), we find that a significant fraction 
($25 \pm 6$\% using a disk-dominated definition of $B/T < 0.5$, and $40 \pm 7$\% using a disk-dominated definition of $n < 2.5$) of the most quiescent galaxies, 
with specific Star Formation Rates $sSFR<10^{-10}\,{\rm yr}^{-1}$, have disk-dominated morphologies 
(including a small number (five) of ``pure disk'' galaxies with $B/T < 0.1$). We show that these passive disks appear to be ``normal''
disks in the sense that they display an axial-ratio distribution comparable to that displayed by present-day disks, while the 
more prevalent actively star-forming disks seem, on average, rounder and clumpier. We consider various possible 
reasons for this, including selection effects, and briefly discuss the theoretical implications.

\smallskip
Our results challenge theoretical models of galaxy formation to {\bf i)} include a mode in which star-formation quenching is not simply connected 
to morphological transformation, {\bf ii)} explain the relationship between active and passive disks, {\bf iii)} predict the relatively
rapid demise of massive star-forming disks, but the relatively gradual emergence of genuinely bulge-dominated morphologies, and {\bf iv)} provide 
the necessary dramatic size evolution (but with limited mass increase) to lift the compact bulges we see at $z \simeq 2$ onto 
the local size-mass relation $\simeq 10$\,Gyr later.

\end{enumerate}

\nocite{Zirm2007,Longhetti2007,vanDokkum2008,Szomoru2010,McLure2012}

\section*{Acknowledgments}

VAB and MC acknowledge the support of the Science and Technology Facilities
Council (STFC) via the award of an STFC Studentship and an STFC Advanced 
Fellowship respectively. JSD and TAT acknowledge the support of the 
European Research Council via the award of an Advanced Grant. 
JSD and RJM acknowledge the support 
of the Royal Society via a Wolfson Research Merit Award and a University 
Research Fellowship respectively.
RJM acknowledges the support of the Leverhulme Trust via the award of a Philip Leverhulme Research Prize.
DC acknowledges the support of an Australian Research Council QEII Fellowship.
The work of AD has been partly supported by the ISF through grant 6/08, by GIF through grant G-1052-104.7/2009, 
by the DFG via DIP grant STE1869/1-1.GE625/15-1, and by an NSF grant AST-1010033 at UCSC.

This work is based in part on observations made with the NASA/ESA {\it Hubble Space Telescope}, which is operated by the Association 
of Universities for Research in Astronomy, Inc, under NASA contract NAS5-26555.
This work is based in part on observations made with the {\it Spitzer Space Telescope}, which is operated by the Jet Propulsion Laboratory, 
California Institute of Technology under NASA contract 1407. 

\bibliographystyle{mn2e}

\bibliography{mybibtex}

\newpage

\appendix

\section{PSF Dependence}

We show in Fig. A1 radial profile plots of the two PSFs tested here, 
the empirical stellar stack and the Tiny Tim model, along with the residuals 
between them and a magnified plot between 0.5 and 0.8 arcsec, the 
range encompassing a physical size comparable to the fitted sizes of the objects
(which more clearly demonstrates the difference between the PSFs). 
In the top-left plot we show the difference between the empirical stacked PSF and the 
Tiny Tim model. The other plots are included to emphasize the uniformity of the 
individual stars that were included in the stack as they compare the stack with 
three out of the seven stars that comprise the stack.

This figure clearly highlights that the Tiny Tim model under-predicts the flux 
in the PSF at this critical radius and thus explains why the 
fitted sizes using this PSF are $5-10$\% larger than those from the empirical stacked PSF.

In order to ascertain the reason for this discrepancy between the modelled and 
empirical PSFs we constructed a difference image of the empirical stack - Tiny Tim PSF, 
show in Fig. A2. The offsets at the very centre of the image are due to centroiding 
issues but it is clear that further out, beyond 0.5 arcseconds, 
there is a distinct halo in the empirical PSF which is not present in the Tiny Tim model. 
This unequivocally shows that the empirical stacked PSF contains a much stronger  
contribution from the airy rings, which is not properly modelled fails by the Tiny Tim PSF. 
In addition to this, 
the Tiny Tim model does not accurately reproduce the diffraction spikes.

As a result of these tests, we adopted the empirical stacked PSF for all the model fitting 
and testing undertaken in this work. Consequently we have generally derived 
fitted sizes which are systematically a factor of $5-10$\% smaller than those 
which would have been determined using a Tiny Tim PSF.

\begin{figure*}
\includegraphics[scale=0.5]{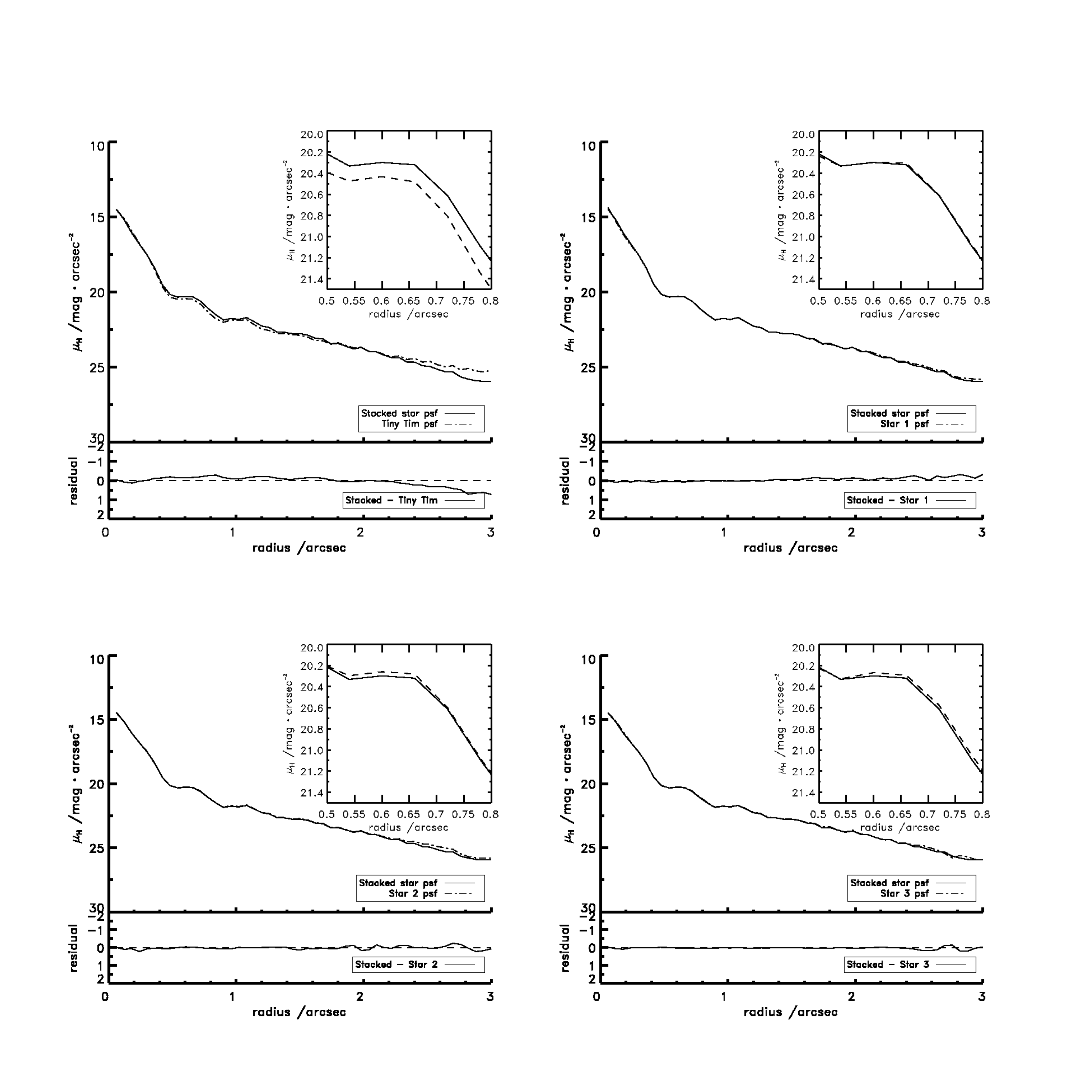} 
 \caption{Comparison of the radial surface-brightness profiles of alternative $H_{160}$ PSFs. The top-left panel
compares the empirical PSF we obtained from stacking stars taken from the real $H_{160}$ CANDELS mosaic (solid-line)
with the PSF produced by the Tiny Tim model (dashed line) (with the residuals given below). The inset shows a magnified 
view of the crucial region around $\simeq 6$\,arcsec, which corresponds to a physical size of $\simeq 5$ kpc at $1 < z < 3$,
comparable to the typical effective radii of the galaxies in our sample (the surface-brightness scale in the inset 
has been expanded to demonstrate more clearly the level of the offset between the Tiny Tim model 
and the empirical stack at these important scales). The remaining three panels simply show how well the empirical PSF matches 
the profiles of three of the seven individual stars which went into it, demonstrating that our empirical PSF has not been 
significantly broadened or otherwise damaged by the stacking process at any angular scales of interest.}
\end{figure*}

\begin{figure}
\begin{center}
\includegraphics[scale=0.58]{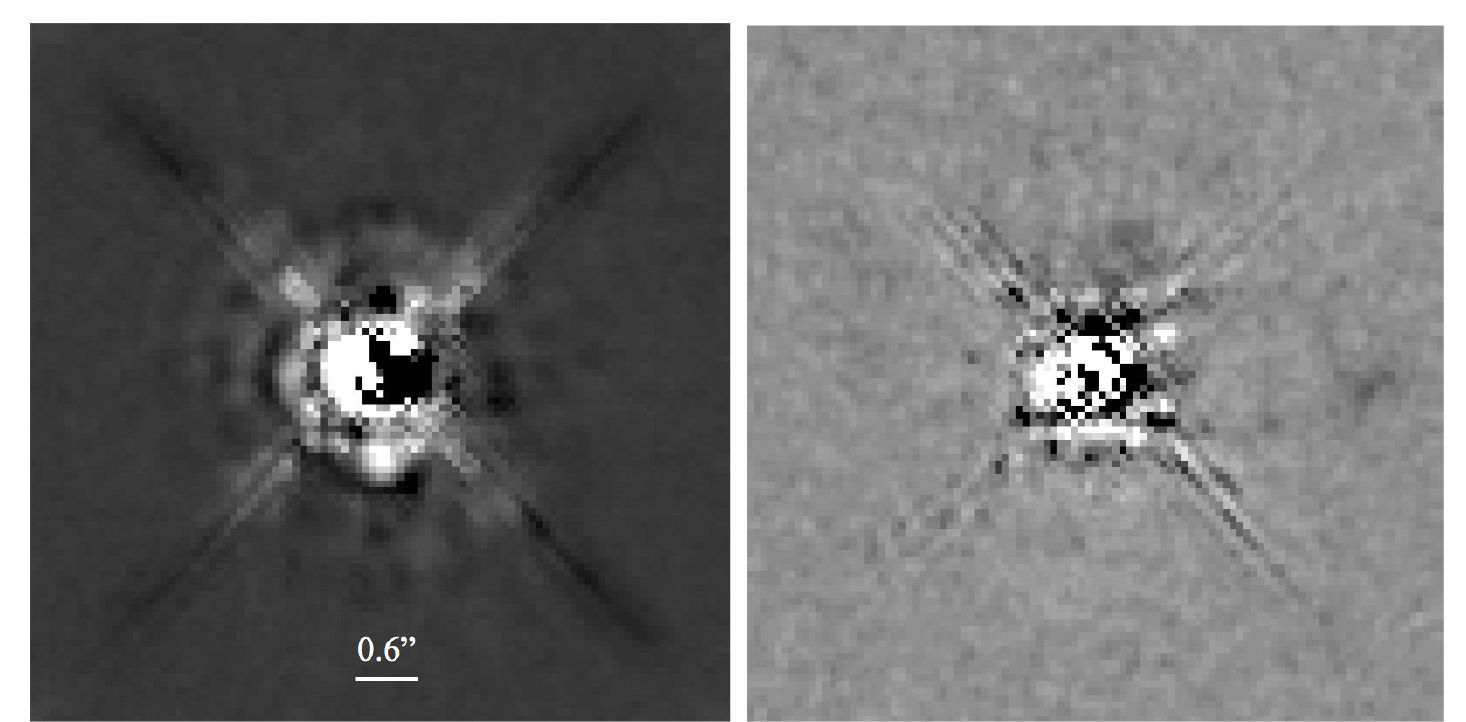}
 \caption{ Left: Difference image of the stacked empirical PSF -- the Tiny Tim model. 
The image is $6 \times 6$\,arcsec with a pixel scale of 0.06\,arcsec (an illustrative 0.6 arcsec 
line has been added for clarity). The grey scale shows negative pixels as darker and positive pixels as whiter. 
The discrepancy between the two PSFs at the centre is due to minor mismatching during centroiding, 
but a real positive halo can be clearly seen at a radius of 0.5\,arcsec and greater. 
This is due to the empirical PSF including a stronger airy disk pattern than is modelled by the Tiny Tim PSF,
and perhaps also containing additional scattered light. 
Inconsistencies in the contribution from diffraction spikes are also visible in the image. 
Right: A difference image of the stacked PSF -- one of the component star PSFs is given for comparison, 
where the images have been constructed using the same cut in the brightness level.}
 \end{center}
\end{figure}

\section{Background Dependence}

In Section 4.2 we discuss the additional level of background subtraction needed before the image stamps 
taken from the CANDELS mosaics can be fitted with GALFIT, and how the fitting procedure trades off the 
treatment of background light with the fitting of the the degenerate S\'{e}rsic index and effective 
radius parameters. We fully explored this issue by constructing a grid of GALFIT runs throughout 
the full parameter space of  S\'{e}rsic index and effective radius parameters, and background subtraction values.

This grid contains a set of additional background values to be subtracted from the image. This is done by 
determining two initial estimates of the additional background light. The first is done by masking out an 
aperture of radius 1\,arcsec around the object centroid position and calculating the median background 
value in the remaining $6 \times 6$\,arcsec image stamp. This method provides a reliable estimate of the 
median background light in most cases, with the exception of those for the largest objects in our sample. 
These largest objects are particularly susceptible to biased size estimates as careful consideration 
must be given to their extended wings. For these cases it is clear that the masking of a 1 arcsec radius 
aperture may not be sufficient to mask out the full extent of the wings, therefore for every object we adopt 
a second median background estimator. This secondary method expands the image stamp of each object to 
$12 \times 12$\,arcsec, generates an annular aperture centred on each source with an inner radius of 
3\,arcsec and an outer radius of 5\,arcsec, and measures the median background light within this aperture. 
By adopting this second technique, although our median background estimate is conducted further from the source, 
it ensures we have not biased our median background estimate too high by failing to account properly 
for the extended wings of the largest objects.

For each object we therefore have two estimates of the local median background, where comparison of these 
estimates gives us an indication of the error associated with determining median background estimates 
from the CANDELS images. We find that the offset between these two estimates for each object is well 
described by a Gaussian distribution centred on 0 with a $2 \times$ FWHM value of 0.001 electrons/s. 
We subsequently use this $2 \times$ FWHM as the error associated with any median background estimate.

From our inspection of the individual sources we are aware that sources can be equally subject to 
background over-subtraction from the first order analysis performed on the images, as well as under-subtraction. 
Thus, for each object, we generate a grid of additional background subtraction values to be used in 
the fitting procedure, which is taken to be the range $-0.001$ to $+0.001$ electrons/s (where $-0.001$ is the 
upper limit of background light that will be added back into the image, accounting for original over-subtraction, 
and $+0.001$ is the upper limit to the amount of background light that will be additionally 
subtracted off the images, accounting for original under-subtraction). 

For each of the points in additional background subtraction space we then construct a loop 
over S\'{e}rsic index and effective radius parameters allowed in the fit. We run an initial fit 
on every object using the median additional background subtraction value determined above 
using a 1 arcsec masking aperture. The S\'{e}rsic index and effective radius parameters 
returned for these fits are used as the centroid points for the S\'{e}rsic index and effective 
radius loops. For S\'{e}rsic index we construct a loop of steps of 0.1 in size, and for 
effective radius we make steps of size 0.025\,arcsec. These step sizes have been determined to 
incorporate the full range of generally accepted realistic S\'{e}rsic index and effective radius 
values (i.e. $0.1-10$ in S\'{e}rsic index  and $0.025-2$ arcsec in angular effective radius).

For each point in the effective radius and S\'{e}rsic index grid we lock these values during the 
GALFIT fit and step through a range of different additional background subtraction values 
to find the best-fit background subtraction value at that grid point, using the $\chi^{2}$ values of each background fit.

\section{Model Fit Refinement}

\begin{figure*}
\includegraphics[scale=0.7]{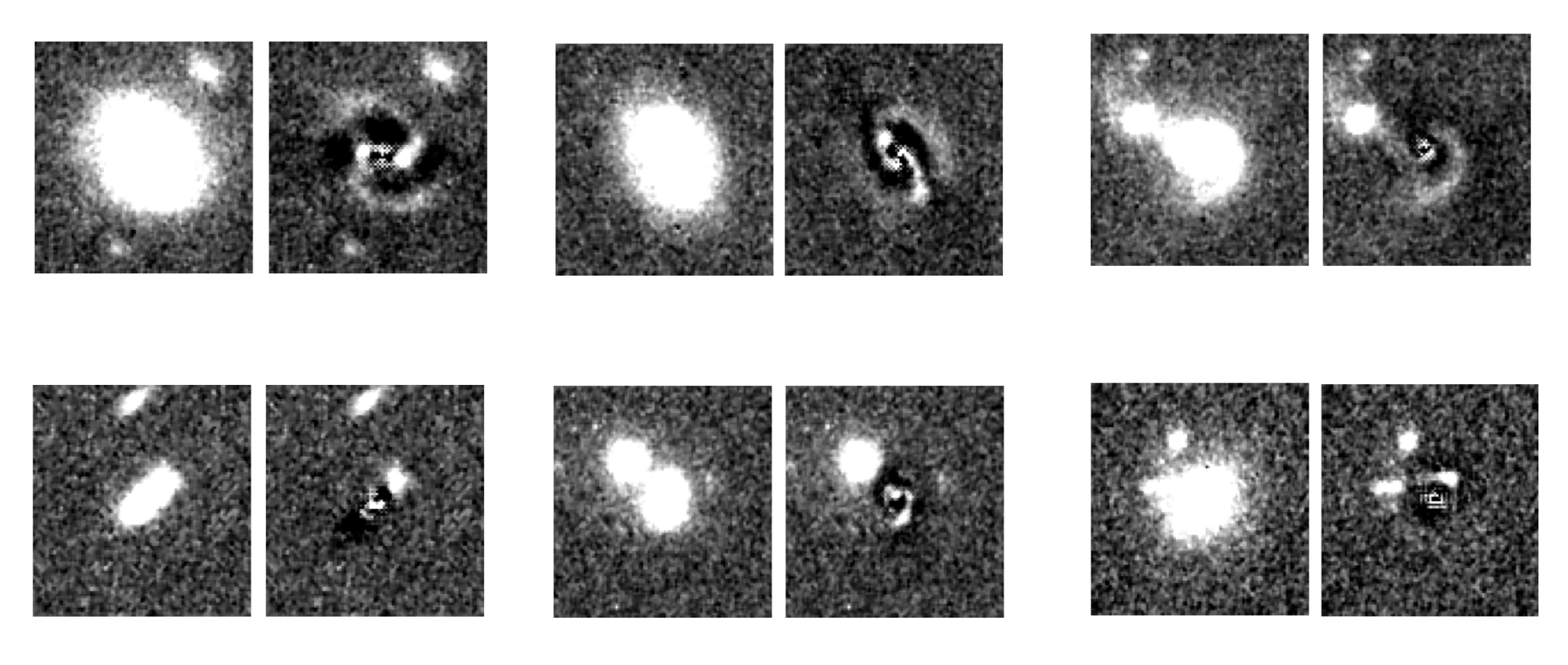}
 \caption{Six examples of objects where our initial modelling failed our $\chi^{2}$ acceptability test due to additional 
structure which could not be properly accounted for by the smooth models. 
For each object we show the $6 \times 6$\,arcsec image stamp on the left, and the data--model 
residual image on the right at the same grey-scale (as produced by the best-fitting double-component model).
The top row shows some clear examples of spiral structures and interacting systems, 
whereas the bottom row shows objects where the fits have been influenced by close companions the light from which 
has not been adequately masked out.}
\end{figure*}

\begin{figure*}
\includegraphics[scale=0.7]{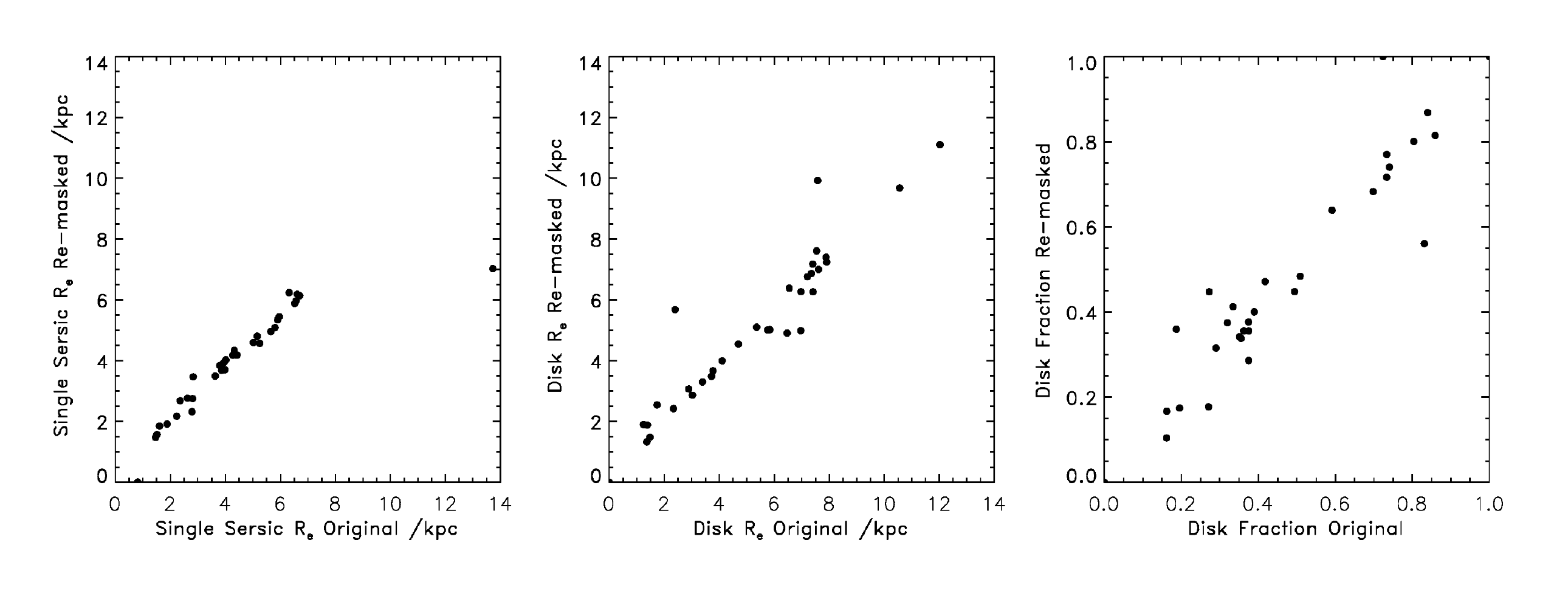}
 \caption{These plots demonstrate the excellent agreement between the key derived galaxy physical parameter 
values obtained with the original model-fitting and with the first set of re-masked/refined fits. 
Left: comparison between single S\'{e}rsic model effective radii, middle: comparison between disk effective radii, 
right: comparison between disk fractions. 
These plots clearly illustrate that the underlying structure of these more complicated systems has in fact 
been accurately fitted by our procedure and has not been significantly influenced by the 
high surface brightness features, such as spiral arms, etc. }
\end{figure*}

As detailed in Section 6.2 a significant fraction of our sample ($\sim 30$\%) were 
initially found to have statistically unacceptable model fits, as judged by:

\begin{equation}
\chi^{2}>\nu+3(\sqrt(2\nu)
\end{equation}

\noindent
However, from close visual inspection of these objects it was found that they display additional levels of 
complex structure such as $z < 2$ grand design spirals with clear spiral arms, interacting systems, objects 
in very crowded fields and objects with extremely close companions, which have not been separately 
identified by {\sc sextractor} despite the high level of de-blending employed in our 
catalogue generation ({\sc deblend\_mincont}=0.0008). Examples of these systems are shown below in Fig. C1, and they contain
some of the best examples of prominent spiral structure.

\begin{figure*}
\includegraphics[scale=0.7]{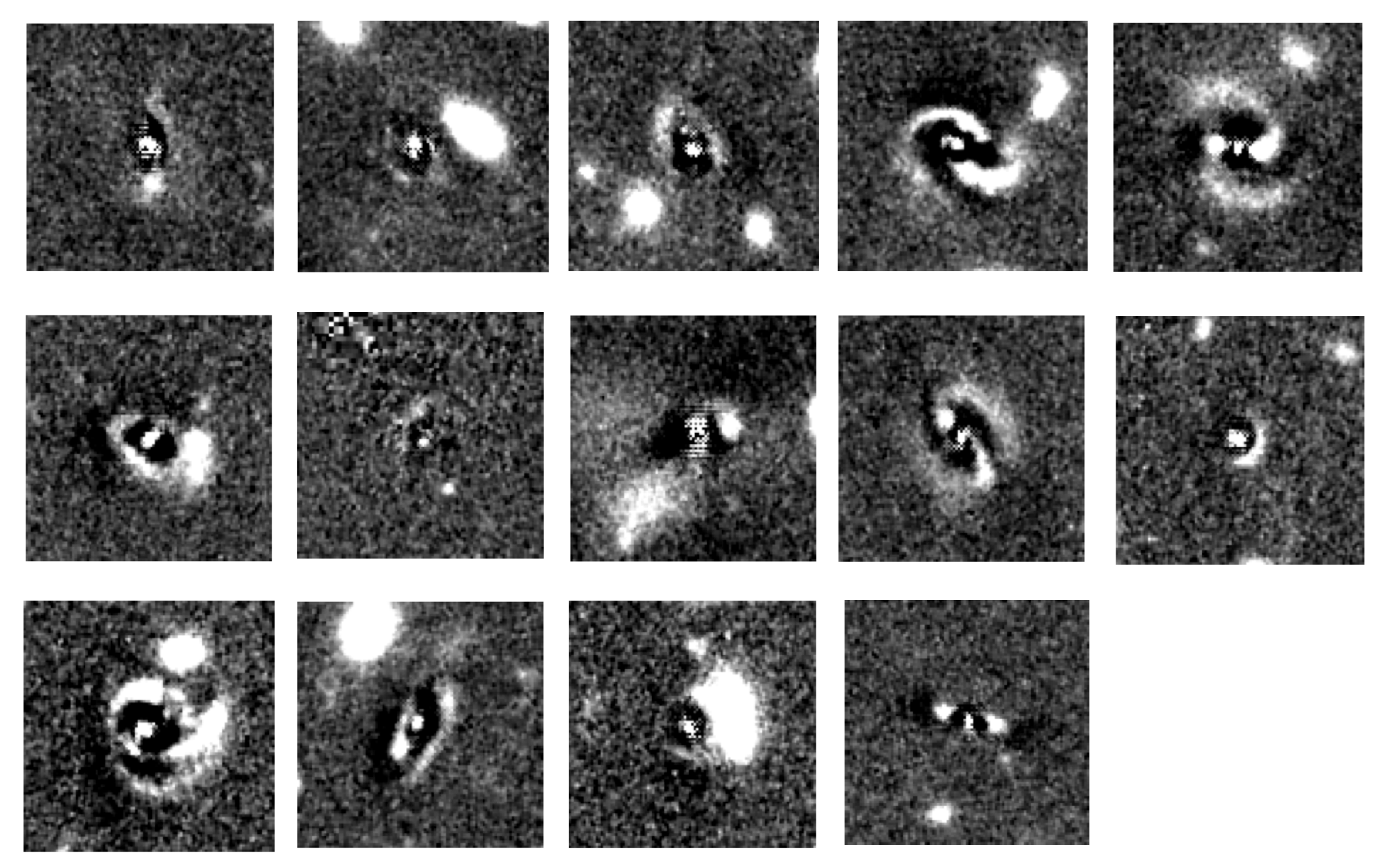}
 \caption{Residual map image stamps for the 14 objects which continued to fail the formal model-fitting acceptability criteria,
even after additional masking. These image stamps have been constructed in the same way as in Fig. C1, 
with the same brightness level and pixel scale.}
\end{figure*}

\begin{figure*}
\includegraphics[scale=0.6]{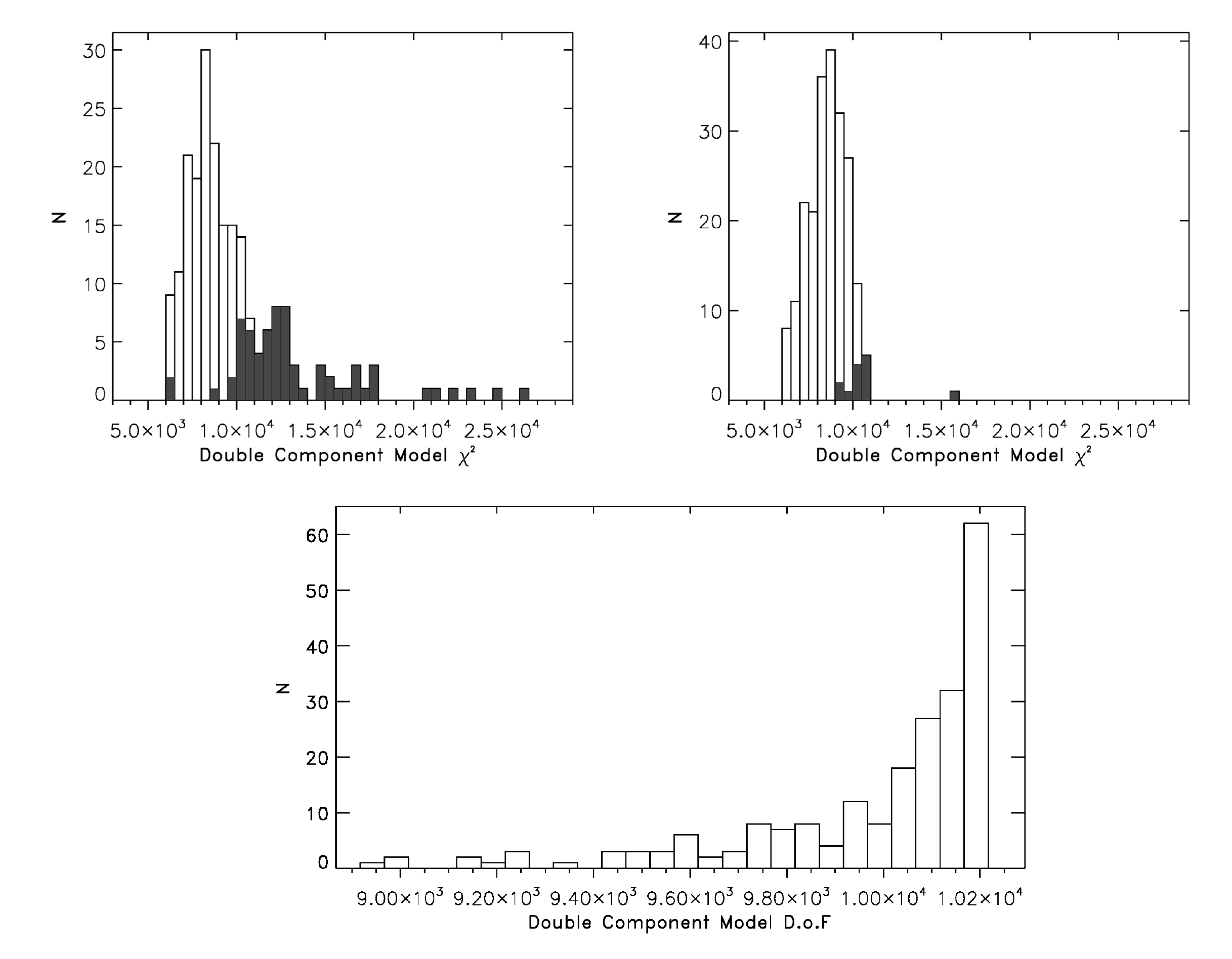}
 \caption{Distributions of minimum $\chi^{2}$ achieved by the modelling 
of all objects in the sample. The upper-left panel shows this distribution as it resulted
from the first pass of modelling, with the shaded region indicating those objects 
which failed to pass the acceptability criterion as defined in equation C1, given the number of degrees 
of freedom (which is typical $\simeq 10,000$, but 
varies on an object by object basis depending on the level of local pixel masking, as illustrated 
in the lower panel). The upper-right panel shows the final distribution achieved after the model refinement including additional
pixel masking of high surface-brightness features as described in the text. Here the remaining shaded region indicates
the 14 objects for which we still failed to achieve an acceptable model fit (and whose residual images are shown in Fig. C3).
In practice, equation C1 means that a formally acceptable model has to typically have $\chi^{2} < 1.05 \times 10^{4}$.
 As a result of the careful treatment given to modelling of all the objects in our sample,
this have been achieved for 94\% of the galaxies studied here.}
\end{figure*}

By additional masking based on closer examination of the residual images of the model fits to these complex systems 
(and refinements to our fitting procedure), 
we have been able to achieve formally-acceptable model fits to the vast majority of these objects. 
Furthermore, from comparison of the morphological 
parameters fitted by our general procedure and those from the refined procedure employed on this subset of  systems, 
we find that, despite the unacceptable $\chi^{2}$ statistics produced by the initial attempt to 
model these objects, we did in fact successfully recover their key morphological parameters (even if errors on these quantities would
have been under-estimated on the basis of $\delta \chi^2$) despite the presence of additional high-surface brightness 
features which cannot be reproduced by our smooth models. This is clearly illustrated in Fig. C2, which shows the tight 
correlation between the underlying physical properties determined from our initial general fitting procedure 
and the first stage of additional modelling refinement.
 
Our refinement procedure is outlined in Section 6.2 and, in brief, incorporates masking of pixels for which the model fit to our 
data exceeds a certain $\chi^{2}$ threshold. This serves to mask out any additional structure, 
which is not modelled by our symmetric  S\'{e}rsic profiles, by ensuring that such pixels 
are not considered during the fitting process, and so do not contribute to the $\chi^{2}$ returned for the overall fit.

Our first refinement involved setting the $\chi^{2}$ threshold for each pixel at 9, the point 
at which secondary structure became clearly visible in the $\chi^{2}$ maps of these objects, and 
the point in the $\chi^{2}$ distribution for all pixels for these objects where the 
distribution has peaked and begins to fall into the tail. Applying the refinement with 
this threshold improved the fits of 32 objects to within statistically acceptable levels, 
but we were still left with a further 37 objects which still failed to meet the acceptability criterion.

Accordingly, we re-ran our modelling with a lower $\chi^{2}$ threshold for a second refinement in the fitting. 
This second pass used a $\chi^{2}$ threshold of 5, a value cutting further into the 
main distribution of the $\chi^{2}$ values for each pixel (from inspection of the $\chi^{2}$ maps 
of these complex objects it became apparent that spiral structure could be present and significant enough to 
influence the fits even at this low level.

This second level of refinement resulted in formally 
example model fits for all but 14 objects. Residual image stamps of these 14 objects are shown in Fig. C3.

Throughout the analyses presented in this paper it is the parameters derived from the best-fit refined models 
which have been utilised. For the 14 remaining unacceptable fits, we report the morphological 
properties from the second refinement in Table D2 with an asterisks marked in the column for 
the bulge effective radius in order to clearly  distinguish them from the acceptable models. 
These 14 unacceptable models have been removed from all further results presented 
in Section 7 onwards so as not to potentially bias any science results.

The statistical quality of our final model fits is illustrated in Fig. C4, which shows the distribution of 
minimum $\chi^{2}$ achieved from the modelling of the 69 ``troublesome'' galaxies in our sample 
both before and after the refinement in the model fitting as described above. The figure  
also shows the distribution of degrees of freedom for all objects, which is typically $\simeq 10,000$ but 
varies on an object-by-object basis depending on the degree of object masking employed. As can be seen, our model fits 
have reduced $\chi^{2}$ values centred exceptionally close to unity with very little spread
(as detailed in the caption to Fig. C4, in practice equation C1 dictates that an acceptable model has to have typically
$\chi^{2} < 1.05 \times 10^{4}$ given the number of degrees of freedom involved in the fit).

\nosection{}

 \begin{table*}

\begin{minipage}{10in}

\textbf{APPENDIX D: TABLES OF SAMPLE PROPERTIES AND BEST-FIT PARAMETERS.}

\smallskip

\smallskip

\end{minipage}

\begin{minipage}{5in}

\begin{center}

\textbf{Table D1.}

Table of Galaxy Properties

\smallskip

\end{center}

\end{minipage}

 \begin{tabular}{ m{1.5cm}m{2.0cm}m{2.0cm}m{2.8cm}m{1.2cm}m{2.8cm}}

 \hline

 ID  &

 RA &

 DEC &

 H$_{160}$ Total Mag &

 $z_{phot}$ &

 Mass /$10^{11} M_{\sun}$  \\

 \hline

104291&

02:18:19.44&

-05:14:45.9&

20.78&

1.00&

1.17\\

107814&

02:17:27.75&

-05:13:30.3&

20.68&

1.00&

1.02\\

110641&

02:17:21.17&

-05:12:24.0&

19.90&

1.00&

1.91\\

117875&

02:17:29.65&

-05:09:47.6&

20.51&

1.00&

1.62\\

104128&

02:17:28.87&

-05:14:48.8&

20.20&

1.02&

1.17\\

109330&

02:17:24.39&

-05:12:52.2&

19.17&

1.02&

4.47\\

120725&

02:18:11.26&

-05:08:49.5&

20.75&

1.02&

1.29\\

121549&

02:17:21.81&

-05:08:23.2&

20.44&

1.02&

1.58\\

121600&

02:17:21.56&

-05:08:28.8&

21.44&

1.02&

1.05\\

107906&

02:17:19.33&

-05:13:25.6&

20.63&

1.05&

1.32\\

115478&

02:17:22.30&

-05:10:38.5&

20.29&

1.05&

2.24\\

107886&

02:17:41.12&

-05:13:30.8&

20.63&

1.07&

1.38\\

111163&

02:17:32.53&

-05:12:18.0&

20.84&

1.07&

1.51\\

116097&

02:17:39.01&

-05:10:32.3&

21.39&

1.07&

1.15\\

116189&

02:17:16.44&

-05:10:28.2&

20.17&

1.07&

1.95\\

117976&

02:17:30.84&

-05:09:43.8&

20.13&

1.07&

2.24\\

108718&

02:17:15.63&

-05:13:07.7&

20.37&

1.10&

1.66\\

109018&

02:17:31.36&

-05:13:04.4&

21.01&

1.10&

1.15\\

105061&

02:18:05.72&

-05:14:33.8&

20.61&

1.15&

1.55\\

116928&

02:18:09.98&

-05:10:08.9&

20.05&

1.15&

3.09\\

117116&

02:17:05.02&

-05:10:07.1&

20.43&

1.15&

1.82\\

120336&

02:18:12.03&

-05:08:56.8&

20.73&

1.15&

1.07\\

102534&

02:18:09.13&

-05:15:30.2&

20.06&

1.17&

1.07\\

103000&

02:18:15.05&

-05:15:20.6&

20.11&

1.17&

3.09\\

108988&

02:18:13.70&

-05:13:06.4&

20.50&

1.17&

1.35\\

113554&

02:17:06.12&

-05:11:23.0&

20.91&

1.17&

1.26\\

102857&

02:17:04.77&

-05:15:18.1&

19.97&

1.20&

3.80\\

113491&

02:17:06.45&

-05:11:23.1&

20.14&

1.20&

2.95\\

116852&

02:17:17.49&

-05:10:03.3&

19.87&

1.20&

1.48\\

118791&

02:17:41.09&

-05:09:26.3&

20.17&

1.20&

3.24\\

104282&

02:17:43.91&

-05:14:50.7&

21.93&

1.25&

1.02\\

120093&

02:17:13.51&

-05:09:03.2&

21.43&

1.25&

1.20\\

120134&

02:18:07.80&

-05:09:00.9&

21.75&

1.27&

1.20\\

112575&

02:17:15.55&

-05:11:47.9&

21.58&

1.30&

1.07\\

105017&

02:18:06.10&

-05:14:33.7&

21.03&

1.32&

1.23\\

109704&

02:18:06.16&

-05:12:44.9&

20.23&

1.32&

1.32\\

113419&

02:17:05.68&

-05:11:31.6&

20.70&

1.32&

1.58\\

113972&

02:17:40.35&

-05:11:16.8&

20.77&

1.35&

1.45\\

122843&

02:18:16.99&

-05:07:55.7&

21.20&

1.35&

1.55\\

102704&

02:18:20.21&

-05:15:30.5&

21.52&

1.37&

1.00\\

104371&

02:17:30.82&

-05:14:47.5&

21.03&

1.37&

1.41\\

113151&

02:17:15.83&

-05:11:37.2&

20.67&

1.37&

1.70\\

107026&

02:17:12.46&

-05:13:48.5&

21.14&

1.40&

1.70\\

107210&

02:17:17.69&

-05:13:47.3&

21.53&

1.40&

1.74\\

100222&

02:17:51.22&

-05:16:21.8&

19.79&

1.42&

4.68\\

107573&

02:18:07.15&

-05:13:39.8&

21.57&

1.42&

1.45\\

119019&

02:17:46.64&

-05:09:26.3&

21.06&

1.42&

1.48\\

102613&

02:17:38.51&

-05:15:33.3&

21.64&

1.45&

1.20\\

104240&

02:17:55.30&

-05:14:52.1&

21.53&

1.45&

1.07\\

105024&

02:17:17.95&

-05:14:34.1&

21.58&

1.45&

1.23\\

112384&

02:17:01.03&

-05:11:54.3&

21.60&

1.45&

1.17\\

114942&

02:17:55.53&

-05:10:54.0&

20.28&

1.45&

1.45\\

 \end{tabular}

 \end{table*}

 \newpage

  \begin{table*}

\begin{minipage}{5in}

\begin{center}

\textbf{Table D1. Continued}

\smallskip

\end{center}

\end{minipage}

 \begin{tabular}{ m{1.5cm}m{2.0cm}m{2.0cm}m{2.8cm}m{1.2cm}m{2.8cm}}

 \hline

 ID  &

 RA &

 DEC &

 H$_{160}$ Total Mag &

 $z_{phot}$ &

 Mass /$10^{11} M_{\sun}$  \\

 \hline

122919&

02:18:20.92&

-05:07:59.2&

21.02&

1.45&

2.34\\

105818&

02:17:48.93&

-05:14:19.0&

22.56&

1.47&

1.51\\

118545&

02:17:47.44&

-05:09:35.7&

20.68&

1.47&

2.45\\

112149&

02:16:54.03&

-05:11:59.2&

21.83&

1.50&

1.00\\

110670&

02:18:01.56&

-05:12:30.6&

21.53&

1.52&

1.35\\

100855&

02:16:56.99&

-05:16:13.6&

20.84&

1.55&

2.63\\

111783&

02:17:20.48&

-05:12:06.1&

20.99&

1.55&

2.34\\

118417&

02:17:35.31&

-05:09:43.6&

22.10&

1.55&

1.78\\

123058&

02:17:22.66&

-05:07:56.8&

21.45&

1.55&

1.20\\

109795&

02:16:59.40&

-05:12:50.7&

21.77&

1.57&

2.45\\

110261&

02:17:23.83&

-05:12:39.3&

21.62&

1.57&

1.38\\

121157&

02:17:31.87&

-05:08:37.1&

20.47&

1.57&

5.25\\

102712&

02:17:56.97&

-05:15:31.8&

21.04&

1.60&

1.20\\

121682&

02:17:08.19&

-05:08:25.5&

21.52&

1.60&

1.86\\

102967&

02:17:19.38&

-05:15:10.6&

20.58&

1.62&

2.82\\

117838&

02:17:25.02&

-05:09:52.4&

21.38&

1.62&

1.66\\

118244&

02:17:13.62&

-05:09:39.8&

20.63&

1.62&

2.51\\

113309&

02:17:07.58&

-05:11:33.0&

22.02&

1.65&

1.74\\

115725&

02:18:21.09&

-05:10:33.0&

20.52&

1.65&

3.39\\

116275&

02:18:10.69&

-05:10:29.5&

21.39&

1.65&

1.10\\

121585&

02:17:08.63&

-05:08:26.1&

20.94&

1.65&

2.88\\

101385&

02:18:11.91&

-05:16:04.3&

21.78&

1.67&

1.10\\

109022&

02:17:47.08&

-05:13:05.3&

21.73&

1.67&

1.48\\

110839&

02:17:53.86&

-05:12:26.0&

21.09&

1.67&

1.35\\

113066&

02:17:47.22&

-05:11:38.6&

21.45&

1.67&

1.29\\

105503&

02:18:04.97&

-05:14:24.6&

21.94&

1.70&

1.48\\

105929&

02:17:16.44&

-05:14:14.6&

21.40&

1.70&

1.12\\

110901&

02:17:23.65&

-05:12:24.9&

21.84&

1.70&

1.15\\

113549&

02:17:58.64&

-05:11:32.4&

22.49&

1.70&

1.12\\

117922&

02:18:17.10&

-05:09:52.6&

21.63&

1.70&

1.45\\

119123&

02:18:18.96&

-05:09:24.9&

21.53&

1.70&

1.86\\

120201&

02:17:58.08&

-05:09:01.7&

21.63&

1.70&

1.51\\

100592&

02:17:19.34&

-05:16:22.3&

21.96&

1.72&

1.05\\

111966&

02:17:47.26&

-05:12:02.5&

21.17&

1.72&

2.45\\

115630&

02:17:35.41&

-05:10:42.9&

21.54&

1.72&

1.58\\

116508&

02:18:21.54&

-05:10:19.8&

20.71&

1.72&

3.55\\

120574&

02:18:14.44&

-05:08:51.1&

21.24&

1.72&

2.24\\

120940&

02:17:15.07&

-05:08:40.5&

20.76&

1.72&

2.40\\

121595&

02:17:11.97&

-05:08:23.2&

20.64&

1.72&

2.51\\

104918&

02:17:15.54&

-05:14:35.7&

20.95&

1.75&

2.82\\

110317&

02:16:53.99&

-05:12:39.8&

21.63&

1.75&

1.26\\

114574&

02:18:17.20&

-05:11:05.7&

21.42&

1.75&

1.41\\

115661&

02:17:52.70&

-05:10:42.9&

21.63&

1.75&

1.74\\

117377&

02:18:17.61&

-05:10:04.1&

21.05&

1.75&

3.24\\

101558&

02:17:35.20&

-05:15:57.7&

21.40&

1.77&

1.38\\

102867&

02:17:33.59&

-05:15:28.7&

21.81&

1.77&

1.58\\

110152&

02:17:33.37&

-05:12:41.4&

20.71&

1.77&

1.15\\

113470&

02:17:29.40&

-05:11:29.6&

20.86&

1.77&

1.48\\

117047&

02:18:19.32&

-05:10:13.4&

21.99&

1.77&

1.20\\

117332&

02:18:10.51&

-05:10:06.9&

22.17&

1.77&

1.41\\

122623&

02:17:46.36&

-05:08:03.9&

21.40&

1.77&

1.62\\

103252&

02:17:36.13&

-05:15:20.0&

21.90&

1.80&

1.02\\

106944&

02:18:14.05&

-05:13:54.7&

22.07&

1.80&

1.32\\

113302&

02:17:14.07&

-05:11:34.2&

21.48&

1.80&

1.17\\

117258&

02:17:33.50&

-05:10:05.1&

21.36&

1.80&

1.95\\

119091&

02:17:51.06&

-05:09:26.1&

21.77&

1.80&

1.07\\

 \end{tabular}

 \end{table*}

 \newpage

   \begin{table*}

\begin{minipage}{5in}

\begin{center}

\textbf{Table D1. Continued}

\smallskip

\end{center}

\end{minipage}

 \begin{tabular}{ m{1.5cm}m{2.0cm}m{2.0cm}m{2.8cm}m{1.2cm}m{2.8cm}}

 \hline

 ID  &

 RA &

 DEC &

 H$_{160}$ Total Mag &

 $z_{phot}$ &

 Mass /$10^{11} M_{\sun}$  \\

 \hline

121062&

02:17:37.41&

-05:08:41.5&

21.69&

1.80&

2.14\\

109905&

02:18:10.75&

-05:12:43.2&

21.85&

1.82&

1.51\\

110645&

02:18:11.78&

-05:12:30.5&

21.01&

1.82&

3.72\\

114669&

02:17:13.84&

-05:11:06.2&

22.26&

1.82&

1.32\\

115841&

02:18:19.29&

-05:10:38.8&

22.12&

1.82&

1.02\\

117884&

02:17:37.16&

-05:09:53.9&

21.74&

1.82&

2.34\\

120014&

02:17:21.57&

-05:08:58.7&

21.13&

1.85&

1.74\\

100741&

02:17:25.11&

-05:16:17.8&

22.27&

1.87&

1.10\\

104404&

02:17:55.36&

-05:14:51.2&

22.35&

1.87&

1.29\\

105238&

02:17:54.64&

-05:14:30.5&

22.03&

1.87&

1.62\\

118954&

02:17:48.86&

-05:09:32.1&

21.75&

1.90&

1.86\\

102297&

02:17:23.47&

-05:15:40.3&

21.71&

1.97&

1.48\\

106298&

02:18:08.72&

-05:14:09.9&

21.95&

1.97&

1.66\\

110734&

02:17:05.00&

-05:12:28.3&

22.76&

1.97&

1.95\\

120314&

02:17:20.29&

-05:09:00.2&

22.34&

1.97&

1.41\\

120345&

02:17:20.77&

-05:08:56.4&

21.68&

1.97&

1.17\\

107080&

02:17:17.43&

-05:13:48.1&

22.46&

2.00&

1.17\\

121825&

02:18:03.95&

-05:08:25.9&

22.45&

2.00&

1.00\\

123330&

02:17:46.90&

-05:07:49.8&

22.67&

2.00&

1.29\\

123457&

02:17:04.19&

-05:07:46.7&

23.23&

2.00&

1.02\\

100934&

02:17:39.09&

-05:16:12.9&

22.00&

2.02&

1.15\\

107453&

02:18:05.43&

-05:13:43.3&

21.78&

2.02&

1.51\\

111656&

02:17:14.06&

-05:12:09.4&

22.91&

2.02&

1.15\\

113744&

02:18:16.84&

-05:11:27.7&

22.76&

2.02&

1.00\\

115054&

02:17:52.44&

-05:10:56.6&

22.95&

2.02&

2.09\\

119667&

02:17:56.45&

-05:09:15.1&

23.55&

2.02&

1.35\\

119944&

02:17:04.63&

-05:09:06.3&

22.05&

2.02&

1.45\\

120268&

02:17:19.69&

-05:08:56.6&

21.86&

2.02&

2.34\\

120920&

02:17:55.69&

-05:08:37.2&

21.41&

2.02&

1.70\\

102986&

02:17:50.41&

-05:15:27.2&

22.91&

2.05&

1.10\\

109891&

02:18:09.54&

-05:12:49.5&

22.40&

2.05&

1.05\\

111030&

02:17:31.66&

-05:12:24.2&

22.93&

2.05&

1.51\\

111336&

02:18:03.03&

-05:12:17.9&

22.17&

2.05&

1.17\\

114933&

02:17:26.10&

-05:10:58.2&

21.78&

2.05&

1.58\\

116891&

02:17:39.79&

-05:10:18.7&

23.48&

2.05&

1.12\\

118757&

02:17:05.22&

-05:09:36.2&

23.17&

2.05&

1.10\\

122721&

02:17:51.33&

-05:08:03.4&

22.81&

2.05&

1.26\\

103749&

02:17:04.68&

-05:15:09.7&

23.29&

2.07&

1.17\\

103751&

02:17:21.88&

-05:15:08.1&

22.03&

2.07&

1.29\\

107730&

02:17:06.71&

-05:13:38.3&

22.22&

2.07&

1.41\\

111461&

02:17:57.64&

-05:12:14.4&

22.44&

2.07&

1.38\\

111782&

02:17:50.68&

-05:12:04.6&

22.31&

2.07&

1.00\\

119679&

02:18:06.56&

-05:09:15.4&

22.52&

2.07&

1.00\\

123325&

02:17:24.79&

-05:07:51.3&

22.44&

2.07&

1.02\\

100894&

02:17:42.33&

-05:16:15.5&

23.02&

2.10&

1.07\\

107689&

02:17:06.93&

-05:13:35.9&

21.78&

2.10&

2.04\\

108249&

02:18:07.84&

-05:13:25.1&

22.45&

2.10&

1.07\\

108777&

02:17:20.80&

-05:13:16.0&

23.63&

2.10&

1.32\\

115620&

02:18:21.31&

-05:10:44.0&

22.23&

2.10&

1.48\\

117347&

02:17:13.48&

-05:10:05.6&

22.66&

2.10&

1.38\\

121641&

02:17:40.41&

-05:08:30.3&

22.72&

2.10&

1.05\\

100858&

02:17:30.39&

-05:16:16.6&

23.96&

2.12&

1.55\\

102168&

02:17:05.60&

-05:15:43.5&

21.20&

2.12&

2.63\\

104794&

02:17:19.60&

-05:14:43.0&

23.11&

2.12&

1.12\\

110029&

02:17:41.59&

-05:12:46.6&

23.15&

2.12&

1.12\\

116835&

02:17:31.35&

-05:10:18.3&

22.05&

2.12&

1.66\\

 \end{tabular}

 \end{table*}

 \newpage

  \begin{table*}

\begin{minipage}{5in}

\begin{center}

\textbf{Table D1. Continued}

\smallskip

\end{center}

\end{minipage}

 \begin{tabular}{ m{1.5cm}m{2.0cm}m{2.0cm}m{2.8cm}m{1.2cm}m{2.8cm}}

 \hline

 ID  &

 RA &

 DEC &

 H$_{160}$ Total Mag &

 $z_{phot}$ &

 Mass /$10^{11} M_{\sun}$  \\

 \hline

119583&

02:17:07.61&

-05:09:17.5&

23.25&

2.12&

1.35\\

121896&

02:18:03.20&

-05:08:23.1&

22.17&

2.12&

1.51\\

109051&

02:17:20.02&

-05:13:05.7&

22.66&

2.15&

2.34\\

112374&

02:17:32.56&

-05:11:56.3&

23.06&

2.15&

1.02\\

114727&

02:17:21.18&

-05:11:02.7&

21.81&

2.15&

3.72\\

102387&

02:18:03.40&

-05:15:41.3&

22.25&

2.17&

1.48\\

110626&

02:17:04.97&

-05:12:31.4&

22.38&

2.17&

1.20\\

116591&

02:17:35.58&

-05:10:23.1&

22.01&

2.17&

2.14\\

116644&

02:16:55.05&

-05:10:22.8&

22.27&

2.17&

1.38\\

101298&

02:17:19.82&

-05:16:04.5&

23.12&

2.20&

1.02\\

108854&

02:17:12.54&

-05:13:09.2&

22.39&

2.20&

1.45\\

111731&

02:17:27.41&

-05:12:08.0&

22.05&

2.20&

1.02\\

109877&

02:17:11.07&

-05:12:49.1&

22.32&

2.22&

1.17\\

111146&

02:17:07.97&

-05:12:21.6&

22.58&

2.22&

1.41\\

111909&

02:17:27.16&

-05:11:57.7&

21.18&

2.22&

2.40\\

123324&

02:17:43.95&

-05:07:51.3&

23.12&

2.22&

1.58\\

107752&

02:18:08.19&

-05:13:38.4&

22.09&

2.25&

1.10\\

111836&

02:17:41.80&

-05:12:06.7&

23.09&

2.25&

1.38\\

119585&

02:17:42.89&

-05:09:17.9&

22.71&

2.25&

1.32\\

103664&

02:17:57.56&

-05:15:08.6&

22.46&

2.27&

2.51\\

107610&

02:17:13.69&

-05:13:41.3&

22.54&

2.27&

1.45\\

100564&

02:17:25.97&

-05:16:21.3&

21.79&

2.30&

2.69\\

101313&

02:17:24.85&

-05:16:06.3&

22.91&

2.30&

1.38\\

101818&

02:17:44.98&

-05:15:51.0&

22.10&

2.30&

1.45\\

109082&

02:17:37.39&

-05:13:07.9&

22.44&

2.30&

1.86\\

109262&

02:18:11.09&

-05:13:04.4&

22.81&

2.30&

1.35\\

114138&

02:18:11.78&

-05:11:15.9&

22.21&

2.30&

1.41\\

104698&

02:17:17.29&

-05:14:44.6&

22.94&

2.32&

1.55\\

101714&

02:17:37.25&

-05:15:49.6&

22.30&

2.35&

1.86\\

103841&

02:17:51.76&

-05:15:07.0&

24.27&

2.35&

2.57\\

108887&

02:16:55.80&

-05:13:12.7&

22.50&

2.35&

1.95\\

108892&

02:17:18.39&

-05:13:10.7&

22.48&

2.35&

1.74\\

115739&

02:17:56.02&

-05:10:43.3&

24.37&

2.35&

1.00\\

113904&

02:17:03.66&

-05:11:22.2&

22.57&

2.40&

1.00\\

121971&

02:16:57.46&

-05:08:23.1&

22.50&

2.40&

2.82\\

110871&

02:17:25.20&

-05:12:29.7&

24.03&

2.43&

1.05\\

108716&

02:17:41.32&

-05:13:14.6&

23.36&

2.48&

1.15\\

104392&

02:17:43.16&

-05:14:51.3&

23.66&

2.50&

1.26\\

121395&

02:17:20.95&

-05:08:37.1&

22.95&

2.50&

2.34\\

117233&

02:17:35.90&

-05:10:09.4&

22.91&

2.55&

1.86\\

120369&

02:18:17.17&

-05:08:59.4&

21.88&

2.55&

2.09\\

114460&

02:17:34.76&

-05:11:11.1&

23.35&

2.58&

1.32\\

115338&

02:17:41.37&

-05:10:51.8&

23.86&

2.58&

1.74\\

101548&

02:16:54.85&

-05:16:01.1&

23.38&

2.60&

1.32\\

110846&

02:18:21.40&

-05:12:29.2&

22.56&

2.60&

1.35\\

101885&

02:17:09.17&

-05:15:45.4&

22.63&

2.63&

1.26\\

106767&

02:17:01.41&

-05:14:01.8&

24.16&

2.63&

3.24\\

116142&

02:17:13.11&

-05:10:32.5&

22.39&

2.65&

2.14\\

110731&

02:18:04.64&

-05:12:32.3&

22.96&

2.78&

1.15\\

122586&

02:18:06.38&

-05:08:09.7&

22.77&

2.98&

3.72\\

107762&

02:17:05.79&

-05:13:38.5&

23.46&

3.00&

1.07\\

 \hline

 \end{tabular}

 \caption{Table for the physical properties of each object, listed by ascending redshift.}

 \end{table*}

 \begin{table*}

\begin{minipage}{5in}

\begin{center}

\textbf{Table D2.}

Table of Best-Fit Parameters

\smallskip

\end{center}

\end{minipage}

 \begin{tabular}{ m{1cm}m{0.9cm}m{0.7cm}m{0.7cm}m{0.7cm}m{1.7cm}m{1.7cm}m{1.7cm}m{1.7cm}m{0.7cm}m{0.7cm}m{0.7cm}}

 \hline

 ID  &

 n &

 $r_{e}$ \newline /kpc &

 axial \newline ratio &

 psf \newline /\% &

 bulge $r_{e}$ \newline /kpc &

 disk $r_{e}$ \newline /kpc &

 bulge axial \newline ratio &

 disk axial \newline ratio &

 bulge \newline /\% &

 disk  \newline/\% &

 psf  \newline/\%\\

 \hline

104291&

 2.7&

 3.8&

 0.5&

  0.&

 4.6&

 3.3&

0.57&

0.45&

 74.&

 26.&

  0.\\

107814&

 0.8 $^{*}$&

 6.2&

 0.5&

  0.&

 $-^{*}$ &

 6.4&

 $-$ &

0.49&

  0.&

100.&

  0.\\

110641&

 2.8 $^{*}$&

 5.2&

 0.6&

  0.&

 3.5 $^{*}$&

 6.7&

0.43&

0.93&

 64.&

 36.&

  0.\\

117875&

11.7 $^{*}$&

 3.8&

 0.6&

  0.&

 1.2&

12.0&

0.42&

0.70&

 83.&

 17.&

  0.\\

104128&

 1.2&

 4.2&

 0.5&

 15.&

 0.8&

 4.5&

0.28&

0.56&

 28.&

 72.&

  0.\\

109330&

 5.7 $^{*}$&

 7.1&

 0.6&

  0.&

 7.1 $^{*}$&

 $-$ &

0.64&

 $-$ &

 87.&

  0.&

 13.\\

120725&

 1.9&

 2.9&

 0.4&

  0.&

 5.2&

 2.6&

0.47&

0.33&

 45.&

 55.&

  0.\\

121549&

 7.0&

 4.0&

 0.7&

  0.&

 6.2&

 1.1&

0.62&

0.69&

 68.&

 20.&

 12.\\

121600&

 4.6&

 1.5&

 0.7&

  0.&

 1.4&

 $-$ &

0.66&

 $-$ &

100.&

  0.&

  0.\\

107906&

 3.5&

 2.3&

 0.5&

  0.&

 2.5&

 2.0&

0.55&

0.27&

 86.&

 14.&

  0.\\

115478&

 2.7 $^{*}$&

 5.3&

 0.5&

  0.&

 3.7 $^{*}$&

 7.4&

0.31&

0.60&

 65.&

 35.&

  0.\\

107886&

 3.4 $^{*}$&

 1.6&

 0.5&

  0.&

 2.6&

 1.9&

0.61&

0.26&

 50.&

 35.&

 15.\\

111163&

 1.4&

 2.5&

 0.8&

 12.&

 8.6&

 2.6&

0.68&

0.69&

 11.&

 73.&

 16.\\

116097&

 1.4&

 3.0&

 0.7&

  0.&

 1.1&

 3.6&

0.30&

0.67&

 17.&

 83.&

  0.\\

116189&

 0.8&

 3.9&

 0.6&

 42.&

 0.2&

 3.8&

0.24&

0.64&

 40.&

 60.&

  0.\\

117976&

 3.4&

 2.8&

 0.8&

  0.&

 2.3&

 5.4&

0.75&

0.63&

 82.&

 18.&

  0.\\

108718&

 3.7&

 4.8&

 0.4&

  0.&

 8.7&

 2.9&

0.55&

0.17&

 62.&

 38.&

  0.\\

109018&

 2.7&

 2.9&

 0.5&

 19.&

 0.9&

 6.2&

0.59&

0.37&

 70.&

 30.&

  0.\\

105061&

 2.4 $^{*}$&

 3.3&

 0.5&

 27.&

 0.7 $^{*}$&

 5.2&

0.48&

0.47&

 62.&

 38.&

  0.\\

116928&

 4.5&

 1.8&

 0.9&

  0.&

 1.7&

 $-$ &

0.94&

 $-$ &

100.&

  0.&

  0.\\

117116&

 4.2&

 4.2&

 0.9&

  0.&

 4.0&

 $-$ &

0.87&

 $-$ &

100.&

  0.&

  0.\\

120336&

 1.1&

 3.7&

 0.8&

  0.&

 $-$ &

 3.7&

 $-$ &

0.75&

  0.&

100.&

  0.\\

102534&

 1.3 $^{*}$&

 5.8&

 0.9&

  0.&

 0.9 $^{*}$&

 6.2&

0.34&

0.88&

 11.&

 89.&

  0.\\

103000&

 1.8&

 4.4&

 1.0&

 11.&

 1.2&

 5.4&

0.98&

0.99&

 39.&

 61.&

  0.\\

108988&

 2.3&

 2.8&

 0.8&

 31.&

 0.6&

 4.7&

0.78&

0.76&

 68.&

 32.&

  0.\\

113554&

 1.8&

 2.7&

 0.3&

 22.&

 1.6&

 3.4&

0.46&

0.22&

 53.&

 35.&

 12.\\

102857&

 3.9&

 4.1&

 0.9&

  0.&

 4.1&

 $-$ &

0.95&

 $-$ &

100.&

  0.&

  0.\\

113491&

 3.5&

 4.9&

 0.8&

  0.&

 3.7&

 9.2&

0.92&

0.50&

 84.&

 16.&

  0.\\

116852&

 3.1 $^{*}$&

 3.9&

 0.7&

  0.&

 3.1 $^{*}$&

 6.7&

0.76&

0.59&

 82.&

 18.&

  0.\\

118791&

 3.2 $^{*}$&

 4.6&

 0.5&

  0.&

 3.1 $^{*}$&

 7.1&

0.40&

0.54&

 72.&

 28.&

  0.\\

104282&

 5.2&

 2.3&

 0.6&

  0.&

 1.3&

 7.0&

0.44&

0.64&

 76.&

 24.&

  0.\\

120093&

 0.4&

 5.2&

 0.3&

 17.&

 $-^{*}$ &

 5.1&

 $-$ &

0.32&

  0.&

 88.&

 12.\\

120134&

 0.7&

 6.5&

 0.4&

  0.&

 $-$ &

 7.0&

 $-$ &

0.36&

  0.&

100.&

  0.\\

112575&

 2.0&

 1.6&

 0.7&

 18.&

 3.3&

 1.7&

0.88&

0.54&

 23.&

 53.&

 24.\\

105017&

 4.5&

 2.6&

 1.0&

  0.&

 1.4&

 5.4&

0.90&

0.90&

 72.&

 28.&

  0.\\

109704&

 3.2&

 5.2&

 0.7&

  0.&

 5.1&

 6.1&

0.81&

0.42&

 81.&

 19.&

  0.\\

113419&

 1.8&

 2.4&

 0.7&

 11.&

 1.2&

 3.1&

0.65&

0.62&

 57.&

 43.&

  0.\\

113972&

 1.9&

 2.0&

 0.8&

  0.&

 2.3&

 2.3&

0.80&

0.54&

 47.&

 53.&

  0.\\

122843&

 2.3 $^{*}$&

 7.1&

 0.6&

  0.&

 7.0&

 5.9&

0.27&

0.92&

 44.&

 56.&

  0.\\

102704&

 2.4&

 4.5&

 0.3&

  0.&

 1.4&

 5.9&

0.65&

0.20&

 36.&

 64.&

  0.\\

104371&

 1.2&

 1.9&

 0.7&

 15.&

 0.8&

 2.2&

0.78&

0.60&

 46.&

 54.&

  0.\\

113151&

 1.0&

 3.1&

 0.8&

 17.&

 $-$ &

 3.1&

 $-$ &

0.77&

  0.&

 83.&

 17.\\

107026&

 3.5&

 3.2&

 0.8&

  0.&

 2.9&

 5.6&

0.85&

0.59&

 88.&

 12.&

  0.\\

107210&

 1.8&

 2.0&

 0.5&

  0.&

 1.6&

 2.4&

0.87&

0.41&

 44.&

 56.&

  0.\\

100222&

 2.3&

 2.2&

 0.5&

  0.&

 2.4&

 2.5&

0.58&

0.43&

 60.&

 40.&

  0.\\

107573&

 1.0&

 4.1&

 0.6&

  0.&

 $-$ &

 4.1&

 $-$ &

0.60&

  0.&

100.&

  0.\\

119019&

 1.7&

 3.0&

 0.2&

 22.&

 0.9&

 4.4&

0.26&

0.17&

 59.&

 41.&

  0.\\

102613&

 4.6&

 1.1&

 0.7&

  0.&

 2.0&

 0.7&

0.87&

0.40&

 64.&

 36.&

  0.\\

104240&

 1.2&

 3.4&

 0.8&

  0.&

12.4&

 3.5&

0.47&

0.67&

 17.&

 83.&

  0.\\

105024&

 3.5&

 1.5&

 0.8&

  0.&

 2.2&

 1.3&

0.93&

0.32&

 67.&

 33.&

  0.\\

112384&

 3.5&

 1.7&

 1.0&

  0.&

 1.7&

 3.2&

0.89&

0.50&

 90.&

 10.&

  0.\\

114942&

 1.6 $^{*}$&

 5.0&

 0.8&

  0.&

10.0 $^{*}$&

 5.2&

0.56&

0.67&

 36.&

 64.&

  0.\\

 \end{tabular}

 \end{table*}

 \newpage

  \begin{table*}

\begin{minipage}{5in}

\begin{center}

\textbf{Table D2. Continued}

\smallskip

\end{center}

\end{minipage}

 \begin{tabular}{ m{1cm}m{0.9cm}m{0.7cm}m{0.7cm}m{0.7cm}m{1.7cm}m{1.7cm}m{1.7cm}m{1.7cm}m{0.7cm}m{0.7cm}m{0.7cm}}

 \hline

 ID  &

 n &

 $r_{e}$ \newline /kpc &

 axial \newline ratio &

 psf \newline /\% &

 bulge $r_{e}$ \newline /kpc &

 disk $r_{e}$ \newline /kpc &

 bulge axial \newline ratio &

 disk axial \newline ratio &

 bulge \newline /\% &

 disk  \newline/\% &

 psf  \newline/\%\\

 \hline

122919&

 0.9&

 4.8&

 0.5&

 14.&

 $-$ &

 4.8&

 $-$ &

0.54&

  0.&

 87.&

 13.\\

105818&

 2.8&

 2.3&

 0.6&

 13.&

 1.1&

 4.3&

0.56&

0.64&

 74.&

 26.&

  0.\\

118545&

 2.8&

 3.2&

 0.7&

 16.&

 1.5&

 6.0&

0.64&

0.72&

 79.&

 21.&

  0.\\

112149&

 4.8&

 1.2&

 0.6&

  0.&

 1.2&

 $-$ &

0.56&

 $-$ &

100.&

  0.&

  0.\\

110670&

 3.2&

 2.1&

 0.8&

  0.&

 1.3&

 4.3&

0.78&

0.50&

 66.&

 34.&

  0.\\

100855&

 3.2&

 2.1&

 0.5&

  0.&

 2.4&

 2.0&

0.44&

0.36&

 81.&

 19.&

  0.\\

111783&

 3.1&

 2.1&

 0.6&

 12.&

 1.1&

 4.4&

0.46&

0.72&

 77.&

 23.&

  0.\\

118417&

 1.8&

 1.9&

 0.6&

  0.&

 1.4&

 2.4&

0.65&

0.50&

 39.&

 61.&

  0.\\

123058&

 3.2&

 0.7&

 0.9&

  0.&

 0.4&

 2.3&

0.84&

0.64&

 78.&

 22.&

  0.\\

109795&

 0.8&

 4.2&

 0.6&

  0.&

 $-$ &

 4.2&

 $-$ &

0.57&

  0.&

100.&

  0.\\

110261&

 3.7&

 1.3&

 0.7&

  0.&

 1.4&

 $-$ &

0.69&

 $-$ &

100.&

  0.&

  0.\\

121157&

 1.9 $^{*}$&

 6.1&

 0.3&

  0.&

 6.4&

 6.7&

0.39&

0.23&

 46.&

 54.&

  0.\\

102712&

 2.1&

 1.5&

 0.6&

 22.&

 1.0&

 2.0&

0.71&

0.41&

 51.&

 31.&

 19.\\

121682&

 3.2&

 4.4&

 0.7&

  0.&

 4.2&

 6.1&

0.73&

0.35&

 82.&

 18.&

  0.\\

102967&

 3.7&

 3.7&

 0.6&

  0.&

 1.8&

 7.1&

0.61&

0.53&

 66.&

 34.&

  0.\\

117838&

 1.9&

 3.6&

 0.3&

 15.&

 1.6&

 5.0&

0.23&

0.36&

 60.&

 40.&

  0.\\

118244&

 1.6 $^{*}$&

 3.9&

 0.7&

 17.&

 1.5 $^{*}$&

 4.9&

0.70&

0.71&

 52.&

 48.&

  0.\\

113309&

 3.6&

 5.1&

 0.4&

  0.&

 9.2&

 1.8&

0.36&

0.56&

 70.&

 30.&

  0.\\

115725&

 2.4&

 3.3&

 0.5&

  0.&

 2.4&

 4.7&

0.48&

0.37&

 59.&

 41.&

  0.\\

116275&

 2.5&

 1.0&

 0.7&

  0.&

 1.2&

 1.0&

0.91&

0.54&

 59.&

 41.&

  0.\\

121585&

 1.9&

 2.8&

 0.6&

 17.&

 0.7&

 4.2&

0.90&

0.43&

 56.&

 44.&

  0.\\

101385&

 0.8&

 1.4&

 0.6&

 24.&

 $-$ &

 1.3&

 $-$ &

0.58&

  0.&

 80.&

 20.\\

109022&

 4.1&

 3.8&

 0.6&

  0.&

 3.8&

 $-$ &

0.63&

 $-$ &

100.&

  0.&

  0.\\

110839&

 1.2&

 5.2&

 0.6&

 11.&

 0.4&

 5.4&

0.11&

0.58&

 16.&

 84.&

  0.\\

113066&

 2.6&

 1.9&

 0.8&

  0.&

 3.9&

 1.4&

0.80&

0.84&

 54.&

 46.&

  0.\\

105503&

 1.9&

 2.2&

 0.7&

  0.&

 2.5&

 2.3&

0.48&

0.70&

 47.&

 53.&

  0.\\

105929&

 1.1 $^{*}$&

 2.7&

 0.8&

 22.&

 $-^{*}$ &

 2.8&

 $-$ &

0.83&

  0.&

 76.&

 24.\\

110901&

 1.3&

 4.7&

 0.5&

 14.&

 0.6&

 5.2&

0.43&

0.51&

 26.&

 74.&

  0.\\

113549&

 1.7&

 1.3&

 0.5&

  0.&

 0.6&

 1.6&

0.42&

0.36&

 28.&

 72.&

  0.\\

117922&

 2.4&

 2.2&

 0.9&

  0.&

 1.2&

 3.2&

0.72&

0.74&

 42.&

 58.&

  0.\\

119123&

 4.1&

 3.5&

 0.9&

  0.&

 2.7&

 6.8&

0.89&

0.78&

 85.&

 15.&

  0.\\

120201&

 3.0&

 2.0&

 0.6&

  0.&

 2.2&

 2.3&

0.63&

0.34&

 80.&

 20.&

  0.\\

100592&

 2.1&

 1.1&

 0.3&

  0.&

 0.9&

 1.4&

0.43&

0.16&

 56.&

 44.&

  0.\\

111966&

 1.7&

 1.9&

 0.4&

 14.&

 0.8&

 2.6&

0.34&

0.35&

 53.&

 47.&

  0.\\

115630&

 3.9&

 1.4&

 0.4&

  0.&

 3.8&

 1.7&

0.60&

0.21&

 38.&

 42.&

 20.\\

116508&

 3.1&

 5.4&

 0.7&

 16.&

 1.7&

 7.7&

0.62&

0.91&

 68.&

 32.&

  0.\\

120574&

 2.0&

 5.8&

 0.3&

  0.&

 1.6&

 6.6&

0.20&

0.37&

 24.&

 76.&

  0.\\

120940&

 1.4 $^{*}$&

 6.1&

 0.7&

  0.&

 2.3&

 7.0&

0.59&

0.62&

 18.&

 82.&

  0.\\

121595&

 2.4&

 4.0&

 0.6&

 10.&

 1.8&

 5.2&

0.44&

0.78&

 58.&

 42.&

  0.\\

104918&

 1.2&

 4.0&

 0.7&

 26.&

 0.4&

 4.5&

0.44&

0.64&

 37.&

 63.&

  0.\\

110317&

 2.5&

 2.2&

 0.6&

  0.&

 1.2&

 3.1&

0.35&

0.69&

 45.&

 55.&

  0.\\

114574&

 1.7&

 2.4&

 0.7&

  0.&

 $-$ &

 2.7&

 $-$ &

0.71&

  0.&

 88.&

 12.\\

115661&

 3.7&

 1.0&

 0.8&

  0.&

 1.3&

 0.7&

0.76&

0.76&

 78.&

 22.&

  0.\\

117377&

 2.1&

 2.3&

 0.2&

 18.&

 0.9&

 3.6&

0.35&

0.16&

 68.&

 32.&

  0.\\

101558&

 0.9&

 4.9&

 0.9&

 16.&

 $-$ &

 4.9&

 $-$ &

0.87&

  0.&

 85.&

 15.\\

102867&

 2.5&

 1.9&

 0.9&

 14.&

 0.7&

 3.7&

0.57&

0.61&

 57.&

 43.&

  0.\\

110152&

 1.3 $^{*}$&

 5.0&

 0.7&

  0.&

 0.8 $^{*}$&

 5.6&

0.15&

0.69&

 13.&

 87.&

  0.\\

113470&

 0.9&

 4.8&

 0.9&

 12.&

 1.3&

 4.9&

0.22&

0.94&

 21.&

 79.&

  0.\\

117047&

 1.5&

 1.9&

 0.2&

  0.&

 $-$ &

 1.9&

 $-$ &

0.22&

  0.&

100.&

  0.\\

117332&

 2.4&

 2.4&

 0.4&

  0.&

 3.2&

 2.2&

0.64&

0.19&

 59.&

 41.&

  0.\\

122623&

 2.3&

 2.4&

 1.0&

  0.&

 1.9&

 3.5&

0.71&

0.80&

 56.&

 44.&

  0.\\

103252&

 0.7&

 4.1&

 0.3&

  0.&

 $-$ &

 4.1&

 $-$ &

0.34&

  0.&

100.&

  0.\\

106944&

 3.2&

 2.6&

 0.8&

  0.&

 1.5&

 4.1&

0.59&

0.83&

 56.&

 44.&

  0.\\

113302&

11.6&

 2.9&

 0.7&

 14.&

 0.8&

11.4&

0.70&

0.60&

 85.&

 15.&

  0.\\

117258&

 3.3&

 1.5&

 0.8&

  0.&

 2.0&

 1.6&

0.65&

0.36&

 80.&

 20.&

  0.\\

119091&

 3.5&

 5.0&

 0.5&

  0.&

 2.7&

 5.5&

0.29&

0.98&

 56.&

 44.&

  0.\\

 \end{tabular}

 \end{table*}

 \newpage

  \begin{table*}

\begin{minipage}{5in}

\begin{center}

\textbf{Table D2. Continued}

\smallskip

\end{center}

\end{minipage}

 \begin{tabular}{ m{1cm}m{0.9cm}m{0.7cm}m{0.7cm}m{0.7cm}m{1.7cm}m{1.7cm}m{1.7cm}m{1.7cm}m{0.7cm}m{0.7cm}m{0.7cm}}

 \hline

 ID  &

 n &

 $r_{e}$ \newline /kpc &

 axial \newline ratio &

 psf \newline /\% &

 bulge $r_{e}$ \newline /kpc &

 disk $r_{e}$ \newline /kpc &

 bulge axial \newline ratio &

 disk axial \newline ratio &

 bulge \newline /\% &

 disk  \newline/\% &

 psf  \newline/\%\\

 \hline

121062&

 4.8&

 1.7&

 1.0&

  0.&

 2.8&

 1.5&

0.79&

0.41&

 72.&

 18.&

 10.\\

109905&

 2.4&

 1.6&

 0.7&

  0.&

 1.3&

 2.1&

0.87&

0.55&

 63.&

 37.&

  0.\\

110645&

 2.1&

 3.1&

 0.7&

 12.&

 1.3&

 4.4&

0.69&

0.72&

 56.&

 44.&

  0.\\

114669&

 1.4&

 1.6&

 0.3&

 12.&

 0.4&

 2.0&

0.47&

0.23&

 36.&

 64.&

  0.\\

115841&

 3.9&

 1.1&

 0.7&

  0.&

 1.1&

 $-$ &

0.73&

 $-$ &

100.&

  0.&

  0.\\

117884&

 1.3&

 2.8&

 0.8&

 25.&

 0.4&

 3.3&

0.50&

0.80&

 39.&

 61.&

  0.\\

120014&

 7.6&

 9.9&

 0.8&

  0.&

 1.8&

10.9&

0.69&

0.74&

 68.&

 32.&

  0.\\

100741&

 1.7&

 2.7&

 0.5&

  0.&

 $-$ &

 2.5&

 $-$ &

0.52&

  0.&

100.&

  0.\\

104404&

 6.3&

10.4&

 0.9&

  0.&

 5.3&

 $-$ &

0.91&

 $-$ &

100.&

  0.&

  0.\\

105238&

 1.9&

 3.1&

 0.7&

  0.&

 5.2&

 2.8&

0.55&

0.69&

 49.&

 51.&

  0.\\

118954&

 1.8&

 1.0&

 0.4&

  0.&

 1.0&

 1.1&

0.46&

0.38&

 37.&

 63.&

  0.\\

102297&

 2.7&

 2.7&

 0.8&

  0.&

 5.2&

 2.1&

0.71&

0.43&

 67.&

 33.&

  0.\\

106298&

 1.1&

 2.1&

 0.9&

 17.&

 $-$ &

 2.2&

 $-$ &

0.86&

  0.&

 82.&

 18.\\

110734&

 5.1&

17.1&

 0.4&

  0.&

11.5&

 $-$ &

0.42&

 $-$ &

100.&

  0.&

  0.\\

120314&

 4.2&

 1.1&

 0.8&

  0.&

 1.1&

 $-$ &

0.82&

 $-$ &

100.&

  0.&

  0.\\

120345&

 4.9&

 1.9&

 0.8&

  0.&

 2.3&

 0.6&

0.85&

0.06&

 87.&

 13.&

  0.\\

107080&

 0.9&

 3.8&

 0.9&

 17.&

 0.4&

 4.0&

0.36&

0.85&

 21.&

 79.&

  0.\\

121825&

 3.5&

 2.4&

 1.0&

  0.&

 1.2&

 4.0&

0.84&

0.89&

 60.&

 40.&

  0.\\

123330&

 1.2&

 4.8&

 0.5&

  0.&

 $-$ &

 4.6&

 $-$ &

0.51&

  0.&

100.&

  0.\\

123457&

 0.6&

 4.7&

 0.6&

  0.&

 $-$ &

 5.0&

 $-$ &

0.65&

  0.&

100.&

  0.\\

100934&

15.3&

 6.3&

 0.5&

  0.&

 1.3&

10.5&

0.30&

0.47&

 63.&

 25.&

 12.\\

107453&

 0.8&

 3.2&

 0.7&

 23.&

 $-$ &

 3.2&

 $-$ &

0.68&

  0.&

 79.&

 21.\\

111656&

 8.4&

 2.6&

 0.5&

  0.&

 2.5&

 $-$ &

0.50&

 $-$ &

 86.&

  0.&

 14.\\

113744&

 2.1&

 1.0&

 0.3&

  0.&

 0.6&

 1.4&

0.46&

0.07&

 49.&

 51.&

  0.\\

115054&

 1.3&

 4.4&

 0.8&

 17.&

 0.4&

 5.1&

0.48&

0.74&

 27.&

 73.&

  0.\\

119667&

 1.6&

 4.2&

 0.5&

  0.&

 $-$ &

 3.7&

 $-$ &

0.50&

  0.&

100.&

  0.\\

119944&

 1.0&

 5.4&

 0.7&

  0.&

 $-$ &

 5.5&

 $-$ &

0.70&

  0.&

100.&

  0.\\

120268&

 2.7&

 8.0&

 0.4&

  0.&

 4.5&

 8.3&

0.16&

0.48&

 42.&

 58.&

  0.\\

120920&

 4.6&

 2.3&

 0.8&

  0.&

 1.4&

 5.9&

0.87&

0.69&

 81.&

 19.&

  0.\\

102986&

 5.8&

 1.8&

 0.8&

  0.&

 0.9&

 6.3&

0.73&

0.72&

 74.&

 26.&

  0.\\

109891&

 8.7&

 2.3&

 0.7&

  0.&

 4.6&

 0.5&

0.65&

0.67&

 63.&

 37.&

  0.\\

111030&

 1.6&

 7.0&

 0.3&

  0.&

 $-$ &

 6.2&

 $-$ &

0.30&

  0.&

100.&

  0.\\

111336&

 1.7&

 2.2&

 0.6&

  0.&

 1.2&

 2.9&

0.43&

0.51&

 28.&

 72.&

  0.\\

114933&

 1.6&

 4.7&

 0.9&

  0.&

 $-$ &

 4.2&

 $-$ &

0.88&

  0.&

100.&

  0.\\

116891&

 0.6&

 3.1&

 0.5&

 21.&

 $-$ &

 3.0&

 $-$ &

0.50&

  0.&

 84.&

 16.\\

118757&

 0.7&

 6.4&

 0.7&

  0.&

 $-$ &

 7.0&

 $-$ &

0.71&

  0.&

100.&

  0.\\

122721&

 2.5&

 1.3&

 0.5&

  0.&

 1.5&

 $-$ &

0.51&

 $-$ &

100.&

  0.&

  0.\\

103749&

 3.4&

 1.5&

 0.8&

  0.&

 1.0&

 4.1&

0.66&

0.33&

 69.&

 31.&

  0.\\

103751&

 2.1&

 1.4&

 0.8&

  0.&

 0.7&

 2.1&

0.81&

0.64&

 46.&

 54.&

  0.\\

107730&

 4.0&

 1.2&

 0.8&

  0.&

 1.2&

 $-$ &

0.78&

 $-$ &

100.&

  0.&

  0.\\

111461&

 1.3&

 2.4&

 0.8&

 15.&

 0.8&

 3.0&

0.32&

0.80&

 36.&

 64.&

  0.\\

111782&

 0.7&

 4.7&

 0.5&

  0.&

 $-$ &

 5.0&

 $-$ &

0.46&

  0.&

100.&

  0.\\

119679&

 3.6&

 0.8&

 0.8&

  0.&

 0.8&

 $-$ &

0.80&

 $-$ &

100.&

  0.&

  0.\\

123325&

 1.2&

 1.1&

 0.9&

 26.&

 $-$ &

 1.2&

 $-$ &

0.88&

  0.&

 70.&

 30.\\

100894&

 0.7&

 4.0&

 0.7&

  0.&

 $-$ &

 4.4&

 $-$ &

0.66&

  0.&

100.&

  0.\\

107689&

 2.9&

 2.8&

 0.8&

  0.&

 2.1&

 3.8&

0.90&

0.59&

 67.&

 33.&

  0.\\

108249&

20.0&

 1.6&

 0.7&

  0.&

 0.4&

19.5&

0.76&

0.31&

 82.&

 18.&

  0.\\

108777&

 2.1&

 5.0&

 0.5&

  0.&

 8.8&

 $-$ &

0.49&

 $-$ &

100.&

  0.&

  0.\\

115620&

 0.7&

 7.8&

 0.4&

  0.&

 $-$ &

 8.4&

 $-$ &

0.38&

  0.&

100.&

  0.\\

117347&

 0.7&

 3.5&

 0.8&

 17.&

 $-$ &

 3.6&

 $-$ &

0.73&

  0.&

 85.&

 15.\\

121641&

 1.6&

 2.0&

 0.8&

  0.&

 0.8&

 2.5&

0.30&

0.79&

 20.&

 80.&

  0.\\

100858&

 0.5&

 4.2&

 0.6&

  0.&

 $-$ &

 4.5&

 $-$ &

0.67&

  0.&

100.&

  0.\\

102168&

 3.8&

 1.8&

 0.6&

  0.&

 1.3&

 4.4&

0.65&

0.52&

 82.&

 18.&

  0.\\

104794&

 1.5&

 3.8&

 0.4&

  0.&

 $-$ &

 3.5&

 $-$ &

0.44&

  0.&

100.&

  0.\\

110029&

 0.3&

 4.6&

 0.6&

  0.&

 $-$ &

 5.3&

 $-$ &

0.66&

  0.&

100.&

  0.\\

116835&

 2.5&

 0.9&

 0.9&

  0.&

 0.6&

 1.7&

0.66&

0.71&

 59.&

 41.&

  0.\\

 \end{tabular}

 \end{table*}

 \newpage

  \begin{table*}

\begin{minipage}{5in}

\begin{center}

\textbf{Table D2. Continued}

\smallskip

\end{center}

\end{minipage}

 \begin{tabular}{ m{1cm}m{0.9cm}m{0.7cm}m{0.7cm}m{0.7cm}m{1.7cm}m{1.7cm}m{1.7cm}m{1.7cm}m{0.7cm}m{0.7cm}m{0.7cm}}

 \hline

 ID  &

 n &

 $r_{e}$ \newline /kpc &

 axial \newline ratio &

 psf \newline /\% &

 bulge $r_{e}$ \newline /kpc &

 disk $r_{e}$ \newline /kpc &

 bulge axial \newline ratio &

 disk axial \newline ratio &

 bulge \newline /\% &

 disk  \newline/\% &

 psf  \newline/\%\\

 \hline

119583&

 0.5&

 3.0&

 0.9&

  0.&

 $-$ &

 3.4&

 $-$ &

0.90&

  0.&

100.&

  0.\\

121896&

 2.7&

 1.1&

 0.5&

  0.&

 1.8&

 0.9&

0.40&

0.57&

 58.&

 42.&

  0.\\

109051&

 2.1&

 6.0&

 0.6&

  0.&

 4.7&

 5.2&

0.18&

0.89&

 30.&

 70.&

  0.\\

112374&

 0.9&

 1.5&

 0.9&

  0.&

 $-$ &

 1.5&

 $-$ &

0.93&

  0.&

100.&

  0.\\

114727&

 2.1&

 2.1&

 0.3&

  0.&

 1.9&

 2.5&

0.54&

0.19&

 55.&

 45.&

  0.\\

102387&

 2.8&

 0.7&

 0.6&

  0.&

 1.3&

 0.6&

0.56&

0.54&

 42.&

 58.&

  0.\\

110626&

 1.2&

 4.4&

 0.8&

  0.&

 $-$ &

 4.1&

 $-$ &

0.78&

  0.&

100.&

  0.\\

116591&

 1.5&

 1.8&

 0.8&

  0.&

 0.6&

 2.0&

0.14&

0.80&

 17.&

 83.&

  0.\\

116644&

 1.5&

 1.2&

 0.6&

 17.&

 $-$ &

 1.4&

 $-$ &

0.65&

  0.&

 75.&

 25.\\

101298&

 7.8&

 2.5&

 0.8&

  0.&

 2.7&

 $-$ &

0.83&

 $-$ &

 84.&

  0.&

 16.\\

108854&

 4.2&

11.2&

 0.4&

 16.&

 0.9&

10.6&

0.60&

0.34&

 50.&

 50.&

  0.\\

111731&

19.9&

 0.7&

 0.6&

  0.&

 4.7&

 0.5&

0.46&

0.01&

 42.&

 58.&

  0.\\

109877&

 1.9&

 1.9&

 0.9&

  0.&

 1.1&

 2.4&

0.73&

0.83&

 37.&

 63.&

  0.\\

111146&

 5.5&

 1.0&

 0.9&

  0.&

 0.7&

 5.2&

0.88&

0.53&

 83.&

 17.&

  0.\\

111909&

 1.4 $^{*}$&

 5.9&

 0.8&

  0.&

 4.8&

 6.9&

0.35&

0.67&

 22.&

 78.&

  0.\\

123324&

 1.1&

 2.8&

 0.8&

 10.&

 $-$ &

 2.8&

 $-$ &

0.81&

  0.&

 89.&

 11.\\

107752&

 1.1&

 1.6&

 0.8&

  0.&

 $-$ &

 1.6&

 $-$ &

0.75&

  0.&

100.&

  0.\\

111836&

 0.9&

 2.5&

 0.9&

  0.&

 $-$ &

 2.5&

 $-$ &

0.94&

  0.&

100.&

  0.\\

119585&

 1.9&

 1.4&

 0.8&

 13.&

 1.1&

 2.0&

0.63&

0.56&

 70.&

 30.&

  0.\\

103664&

 1.3 $^{*}$&

 4.2&

 0.7&

  0.&

 $-^{*}$ &

 4.1&

 $-$ &

0.69&

  0.&

100.&

  0.\\

107610&

 0.3&

 4.8&

 0.4&

  0.&

 $-$ &

 5.5&

 $-$ &

0.33&

  0.&

100.&

  0.\\

100564&

 2.6&

 1.9&

 0.9&

 21.&

 1.5&

 3.4&

0.75&

0.59&

 67.&

 18.&

 14.\\

101313&

 1.7&

 1.1&

 0.5&

  0.&

 $-$ &

 1.3&

 $-$ &

0.47&

  0.&

 87.&

 13.\\

101818&

 1.1&

 2.5&

 0.8&

  0.&

 $-$ &

 2.5&

 $-$ &

0.85&

  0.&

100.&

  0.\\

109082&

 2.8&

 0.9&

 0.9&

  0.&

 1.0&

 1.0&

0.85&

0.70&

 76.&

 24.&

  0.\\

109262&

 1.2&

 2.4&

 0.7&

 14.&

 $-$ &

 2.4&

 $-$ &

0.67&

  0.&

 84.&

 16.\\

114138&

 2.2&

 3.1&

 0.8&

  0.&

 1.7&

 4.3&

0.80&

0.60&

 41.&

 59.&

  0.\\

104698&

 2.0&

 2.9&

 0.7&

  0.&

 7.3&

 1.9&

0.51&

0.81&

 50.&

 50.&

  0.\\

101714&

 4.0&

 1.1&

 0.6&

  0.&

 1.1&

 $-$ &

0.64&

 $-$ &

100.&

  0.&

  0.\\

103841&

 0.9 $^{*}$&

68.7&

 0.1&

 36.&

 $-$ &

 0.0&

 $-$ &

0.05&

  0.&

100.&

  0.\\

108887&

 3.9&

 1.0&

 0.4&

  0.&

 1.0&

 $-$ &

0.39&

 $-$ &

100.&

  0.&

  0.\\

108892&

 9.7&

 5.3&

 0.8&

  0.&

 0.9&

 6.4&

0.60&

0.88&

 65.&

 35.&

  0.\\

115739&

 0.1&

 4.0&

 0.6&

  0.&

 $-$ &

 5.0&

 $-$ &

0.52&

  0.&

100.&

  0.\\

113904&

 1.2&

 3.4&

 0.7&

 20.&

 $-$ &

 3.4&

 $-$ &

0.71&

  0.&

 78.&

 22.\\

121971&

 2.2&

 1.4&

 0.5&

  0.&

 1.5&

 1.7&

0.78&

0.25&

 59.&

 41.&

  0.\\

110871&

 0.8&

 4.2&

 0.6&

  0.&

 $-$ &

 4.4&

 $-$ &

0.60&

  0.&

100.&

  0.\\

108716&

19.7&

 1.0&

 0.6&

  0.&

 0.6&

 $-$ &

0.59&

 $-$ &

100.&

  0.&

  0.\\

104392&

 0.6&

 5.7&

 0.3&

  0.&

 $-$ &

 6.4&

 $-$ &

0.27&

  0.&

100.&

  0.\\

121395&

 4.1 $^{*}$&

 0.0&

 0.9&

  0.&

 0.0&

 $-$ &

1.00&

 $-$ &

100.&

  0.&

  0.\\

117233&

 1.0&

 4.9&

 0.5&

 12.&

 0.6&

 5.2&

0.23&

0.55&

 18.&

 82.&

  0.\\

120369&

 2.4&

 0.9&

 0.4&

  0.&

 5.3&

 1.2&

0.17&

0.41&

 11.&

 67.&

 22.\\

114460&

19.3 $^{*}$&

96.6&

 0.4&

  0.&

 1.3&

23.0&

0.48&

0.29&

 68.&

 32.&

  0.\\

115338&

 0.8&

 3.4&

 0.6&

 10.&

 $-$ &

 2.8&

 $-$ &

0.64&

  0.&

100.&

  0.\\

101548&

 0.7&

 2.6&

 0.7&

  0.&

 $-$ &

 2.6&

 $-$ &

0.74&

  0.&

100.&

  0.\\

110846&

 3.4&

 1.1&

 0.7&

  0.&

 1.5&

 0.9&

0.55&

0.03&

 85.&

 15.&

  0.\\

101885&

20.0&

 2.4&

 1.0&

  0.&

10.0&

 $-$ &

0.51&

 $-$ &

 51.&

  0.&

 49.\\

106767&

 3.2&

 1.6&

 0.3&

  0.&

 1.8&

 $-$ &

0.31&

 $-$ &

100.&

  0.&

  0.\\

116142&

 1.1&

 4.7&

 0.6&

 14.&

 0.4&

 5.1&

0.35&

0.66&

 21.&

 79.&

  0.\\

110731&

 1.9&

 1.2&

 0.5&

 30.&

 0.7&

 $-$ &

0.49&

 $-$ &

100.&

  0.&

  0.\\

122586&

 2.6&

 1.3&

 0.4&

  0.&

 1.0&

 1.9&

0.28&

0.44&

 66.&

 34.&

  0.\\

107762&

 0.9&

 1.8&

 0.8&

 10.&

 $-$ &

 1.5&

 $-$ &

0.76&

  0.&

100.&

  0.\\

 \hline

 \end{tabular}

 \caption{Table showing the fitted parameters for each object. Objects with an unacceptable single component model have been flagged with an asterisks in the S\'{e}rsic index column, while objects with unacceptable double component models are flagged similarly in the bulge effective radius column.    }

\end{table*}

\label{lastpage}

\end{document}